\documentstyle[preprint,aps,eqsecnum,floats,epsf]{revtex}

\tighten
\begin{document}

\def\beq{\begin{equation}}
\def\eeq{\end{equation}}

\def\beqn{\begin{eqnarray}}
\def\eeqn{\end{eqnarray}}
\def\nn{\nonumber\\}

\def\Re{\,{\rm Re}\,}
\def\Im{\,{\rm Im}\,}
\def\O{{\cal O}}
\def\etal{{\it et al.}}

\def\vec#1{{\mbox{\boldmath $\bf #1$}}}

\def\e{\vec\epsilon}
\def\E{\vec\epsilon^{\prime *}}
\def\s{\vec s}
\def\S{\vec s^{\prime *}}
\def\k{\hat{\vec k}}
\def\K{\hat{\vec k}'}

\def\See{\e\cdot\E}
\def\Sss{\s\cdot\S}
\def\Ses{\s\cdot\E}
\def\Sse{\e\cdot\S}

\def\Sk{\vec\sigma\cdot\vec k}
\def\SK{\vec\sigma\cdot\vec k'}

\def\Vee{\vec\sigma\cdot\E\times\e}
\def\Vss{\vec\sigma\cdot\S\times\s}

\title{Deuteron Compton scattering \\
 below pion photoproduction threshold}

\author{M.I. Levchuk$^a$ and A.I. L'vov$^b$}

\address
{ ~
\\ $^a$B.I. Stepanov Institute of Physics,
   Belarus National Academy of Sciences,
\\ F. Scaryna prospect 70, Minsk 220072, Belarus
\\  \sf (levchuk@dragon.bas-net.by)}

\address
{ ~
\\ $^b$P.N. Lebedev Physical Institute, Russian Academy of Sciences,
\\ Leninsky prospect 53, Moscow 117924, Russia
\\ \sf (lvov@x4u.lebedev.ru)}

\maketitle

\begin{abstract}
Deuteron Compton scattering below pion photoproduction threshold is
considered in the framework of the nonrelativistic diagrammatic
approach with the Bonn OBE potential.  A complete gauge-invariant set
of diagrams is taken into account which includes resonance diagrams
without and with $NN$-rescattering and diagrams with one- and two-body
seagulls. The seagull operators are analyzed in detail, and their
relations with free- and bound-nucleon polarizabilities are discussed.
It is found that both dipole and higher-order polarizabilities
of the nucleon are required for a quantitative description of
recent experimental data. An estimate of the isospin-averaged dipole
electromagnetic polarizabilities of the nucleon and the
polarizabilities of the neutron is obtained from the data.
\end{abstract}



\newpage
\section{Introduction}
\label{sec:intro}

Elastic photon, or Compton, scattering is a powerful tool
for probing the structure of hadrons and nuclei.  A deformation of the
system's ground state caused by an incoming electromagnetic wave and
encoded into electromagnetic polarizabilities of the system contributes
to radiation of outgoing photons and thus shows itself in such
observables as the differential cross section of Compton scattering.
A particular example is forward Compton scattering. The corresponding
spin-averaged amplitude at sufficiently low energies $\omega$
has the form%
\footnote{The factor of $4\pi$ in Eqs.\ (\ref{LEX}) and (\ref{Vpol})
below stands because we use Heaviside's units for the electric charges
and electromagnetic fields (e.g., $e^2=4\pi/137$) but, for historical
reasons, Gaussian units for the polarizabilities themselves.}
\beq
\label{LEX}
  T(\omega) = \See \Big( -\frac{Z^2e^2}{M_t} +
     4\pi (\bar\alpha + \bar\beta) \omega^2 + \ldots \Big).
\eeq
Here $\e$ and $\e'$ are polarizations of the initial and final photons,
$Ze$ and $M_t$ are the electric charge and the mass of the target,
and $\bar\alpha$ and $\bar\beta$ are the electric and magnetic
dipole polarizabilities of the target.
Many efforts have been spent to measure the polarizabilities of the
nucleon, $\bar\alpha_N$ and $\bar\beta_N$ (as well as polarizabilities
of other hadrons and nuclei), and to understand them theoretically.
For a review and further references see
\cite{petr81,aren86,lvov93,bern95,chpt97,huet00}.

The polarizabilities of the proton have been successfully found in
a series of experiments on $\gamma p$ scattering
\cite{gold60,bara74,fede91,zieg92,hall93,macg95,tonn98} which
ultimately yielded quite an accurate result,
\beq
\label{alpha-p-exp}
  \bar\alpha_p = 12.1 \pm 0.8 \pm 0.5, \quad
  \bar\beta _p =  2.1 \mp 0.8 \mp 0.5
\eeq
(in the units of $10^{-4}~\rm fm^3$ used for the dipole
polarizabilities throughout the paper).
The values (\ref{alpha-p-exp}) quoted here \cite{macg95} have been
extracted from data of a few experiments of 90's performed at energies
below pion photoproduction threshold under the theoretical constraint
of the Baldin sum rule \cite{bald60,lapi63}:
\beq
\label{Baldin}
   \bar\alpha + \bar\beta =
   \int_0^\infty \sigma_{\rm tot}(\omega)\,
   \frac{d\omega}{2\pi^2 \omega^2},
\eeq
where $\sigma_{\rm tot}(\omega)$ is the total photoabsorption cross
section.  Recently, the result (\ref{alpha-p-exp}) has been confirmed by
a more comprehensive analysis of a larger data base \cite{bara99}.

Meantime, studies of polarizabilities of the neutron which began even
earlier than those for the proton (see a book \cite{alex92} which
summarizes a long history of these studies) achieved a knowledge of
$\bar\alpha_n$ and $\bar\beta_n$ far less satisfactory.  Most of the
experiments performed for measuring the polarizabilities of the neutron
had deals with neutron transmission in the substance. The long-range
polarization interaction
\beq
\label{Vpol}
    V_{\rm pol}(r) = - \frac12 4\pi
    [\bar\alpha \vec E^2(r) + \bar\beta \vec H^2(r)]
\eeq
of the neutrons with the electromagnetic (actually, electric) field
near the edge of nuclei in the substance creates a small but detectable
contribution to the total cross section due to its anomalous energy
dependence $\propto \sqrt E$ \cite{alex92,thal59,lvov87}.  The best
results for the electric polarizability of the neutron, $\bar\alpha_n$,
obtained from these studies read \cite{schm91,koes95}
\beq
\label{alpha-n-exp}
   \bar\alpha_n = 12.6 \pm 1.5 \pm 2.0 ~\cite{schm91}, \quad
   \bar\alpha_n =  0.6 \pm  5 ~\cite{koes95},
\eeq
where a small relativistic correction $=+0.62$ \cite{lvov93}
missing in a fully nonrelativistic formalism used in the original works
\cite{thal59,schm91,koes95} is included.%
\footnote{Recently, the relation between the polarizability
$\bar\alpha_n$ standing in the $\gamma n$ scattering amplitude and the
so-called static polarizability $\alpha_n$ determined in the neutron
transmission experiments was re-derived in Ref.\ \cite{bawi97}.
In essence, the main conclusion of that analysis agrees with our own
findings \cite{lvov93,lvov87}.  This agreement, however, might be
difficult to see from the paper \cite{bawi97}, because its authors
erroneously claim that there is a difference between $\bar\alpha$,
which is understood as a parameter standing in an effective
relativistic Hamiltonian, and the polarizability $\alpha_s$ (using
their notation) determined through Compton scattering.}
Since the two values in Eq.\ (\ref{alpha-n-exp}) seem to contradict to
each other, the current situation with knowing the polarizability
$\bar\alpha_n$ is hardly satisfactory. Moreover, there is an argument
\cite{enik97} that the systematic (in fact, theoretical) uncertainty,
which is a very delicate problem for those experiments, may be strongly
underestimated in Ref.\ \cite{schm91}.
Furthermore, the neutron transmission experiments do not constrain
the magnetic polarizability of the neutron at all, although
$\bar\beta_n$ can be theoretically derived from $\bar\alpha_n$ by using
the Baldin sum rule (\ref{Baldin}).
For all these reasons there is a need for searching for other ways of
finding the polarizabilities of the neutron, e.g., by using real photons.

There are several reasons why experimental studies of neutron Compton
scattering and a further extraction of the neutron polarizabilities are
more difficult than those for the proton.  First, because of the
absence of dense free-neutron targets, actual measurements of $\gamma n$
scattering are forced to have a deal with neutrons bound in nuclei
and hence to take into account effects of the nuclear environment.
Second, due to vanishing the neutron Thomson scattering amplitude (viz.\
the amplitude of photon scattering off the electric charge of the
neutron which is zero), the contribution of polarizabilities of the
neutron to the differential cross section at low energies ($\alt 100$
MeV) turns out to be rather small.  It is of order $\O(\omega^4)$ in
the low-energy expansion over the photon energy $\omega$ vs.\ order
$\O(\omega^2)$ in the proton case.  Third, the $\O(\omega^4)$
contribution of the neutron polarizabilities is accompanied with other
terms $\O(\omega^4)$ which come from the spin-dependent part of the
scattering amplitude; these additional terms are determined by the
so-called spin polarizabilities and they cannot be isolated in a
model-independent way \cite{lvov85,babu98}.  Therefore, a use of
further assumptions, like those constituting the dispersion theory of
Compton scattering \cite{guia78,filk81,lvov81,lvov85,lvov97,drec99},
for evaluating the unknown pieces becomes unavoidable. All that
introduces larger theoretical uncertainties to the obtained
polarizabilities which are at least $\pm 2$ even without the
nuclear-environment corrections.

The first attempt to measure low-energy $\gamma n$ scattering and to
extract polarizabilities of the neutron has been done by the
G\"ottingen--Mainz group \cite{rose90} which followed an earlier
theoretical suggestion \cite{levc94} to exploit the reaction $\gamma d
\to \gamma n p$ in the quasi-free kinematics. The result of this very
first experiment,
\beq
\label{alpha-n-Goett}
        \bar\alpha_n = 10.7{\,}^{~+3.3}_{-10.7} \, ,
\eeq
is not yet so accurate.  However, with the use of a wider range of
photon energies (up to 200--250 MeV), further improvements are quite
feasible \cite{lvov85,levc94,wiss98}. A high accuracy of the underlying
dispersion calculations of $\gamma n$ scattering \cite{lvov85,lvov97}
is crucial for finding the polarizabilities from Compton scattering
data taken at ``high" energies, viz.\ those above pion photoproduction
threshold. Therefore, a determination of $\bar\alpha_n$ and
$\bar\beta_n$ from such data also assumes a careful check and
normalization of pion photoproduction off the neutron which is used in
the dispersion calculations as a crucial input.  Fortunately, such a
check can be done in parallel with Compton scattering measurements,
because the $\gamma(n,\pi)$ reaction can be learned from
$\gamma(d,\pi)$ in the quasi-free kinematics as well \cite{levc96}.

In the present work we analyze another possibility for probing the
polarizabilities of the neutron which requires no strong assumptions on
the ``high-energy" behavior of $\gamma N$ interactions.  This
possibility mentioned already in Ref.\ \cite{bald60} is elastic $\gamma
d$ scattering below pion photoproduction threshold. The presence of the
proton next to the neutron and the coherence of the proton and neutron
contributions makes two advantages.  First, the $\O(\omega^2)$
contribution of the neutron polarizabilities to the scattering
amplitude can interfere with the $\O(1)$ contribution from proton
Thomson scattering, so that a sensitivity of the differential cross
section with respect to the polarizabilities is enhanced.  Second, the
largest contribution to the spin polarizabilities of nucleons which
comes from the $t$-channel $\pi^0$-exchange does not contribute to
$\gamma d$ scattering at all (due to isospin), so that $\O(\omega^4)$
corrections, which are not small for individual nucleons, especially
for the neutron, are more suppressed in the considered case.
Nevertheless, various binding corrections, including meson-exchange
currents (MEC) and meson-exchange seagulls (MES), are rather important
and have to be introduced and carefully evaluated.  Their analysis is
the central subject of the present paper.

Theoretical studies of deuteron Compton scattering have been started by
Bethe and Peierls \cite{beth35} who considered this process in the
dipole $E1$ approximation. After then a number of calculations has been
performed in 50's and 60's, mostly based on the impulse approximation
\cite{50s-60s}.  A higher level of art, with an explicit consideration
of MEC and of their influence on the so-called resonance and seagull
amplitudes of Compton scattering was introduced by Weyrauch and
Arenh\"ovel \cite{weyr83}.  They directly calculated the seagull
contribution from the pion exchange and developed an approximate scheme
based on dispersion relations in the long wave-length limit to find the
resonance amplitude of Compton scattering from a theoretically known
deuteron photodisintegration amplitude.  Later on, direct calculations
of the resonance amplitude free from the approximations of the
oversimplified dispersion relations have been performed using a simple
separable $NN$ potential \cite{weyr90}.  More recently, this
consideration was further improved \cite{wilb95} by using realistic
$NN$ potentials for evaluation of $NN$ rescattering in the intermediate
state and by taking into account leading relativistic corrections and
MEC beyond the Siegert approximation.  A similar (but technically
different) approach was presented by us in Ref.\  \cite{levc95}, in
which MEC and two-body seagull effects were evaluated using two methods:
through a procedure of the minimal substitution in the $NN$ potential
and through a direct diagrammatic computation in the framework of the
Bonn OBE picture of the $NN$ interactions.

It should be said that despite a resemblance of many physical
ingredients of Refs.\ \cite{weyr83,weyr90,wilb95,levc95}, the results
of these works are sometimes rather different, what may indicate
unnoticed computational errors or unjustified approximations.
For instance, the differential cross
section at the forward angle and the photon energy $\omega=100$ MeV
found in Ref.\  \cite{wilb95} is only 2/3 of that found in Refs.\
\cite{weyr90,levc95} (in this comparison, polarizabilities of the
nucleon are omitted).  There are large discrepancies at the backward
angle either.  Very recently, one more calculation in the approach
close to that of Ref.\ \cite{wilb95} was presented \cite{kara99}.
Their results differ from our and previous predictions too,
especially at energies $\omega \sim 100$ MeV. Possible reasons for that
are discussed in Section~\ref{sec:results} below.

Recently methods of effective field theories have been applied to
deuteron Compton scattering as well \cite{chen98,chen99,bean99}. In
such an approach, a model-independent part of the low-energy scattering
amplitude which dominates in the chiral limit of $m_\pi\to 0$ (but
still $m_\pi^2 \gg M\Delta$, where $\Delta=2.2246$ MeV is the deuteron
binding energy and $M$ is the nucleon mass) was found in a closed
analytical form \cite{chen98}.
Generally, the results of both calculations are similar to those
obtained by virtue of the ``standard-nuclear-theory" technique.
The advantage of the calculations \cite{chen98,chen99,bean99} is
that they naturally include important non-static effects
in the pion propagation which is an outside feature
for the ``standard" theory based on the notion of the $NN$ potential.
A disadvantage also exists which is related with unavoidable
truncation of the expansion series leading to a lost of contributions
important for a determination of the neutron polarizabilities.
The $\Delta$-isobar is one example. See Section~\ref{sec:results} for a
further discussion.

In the present paper we complete the calculation with the
nonrelativistic Bonn OBE $NN$ potential (OBEPR) started earlier
\cite{levc95}. Technically, this is done in the framework of the
diagrammatic approach which relies on an explicit consideration of
relevant Feynman diagrams of the reaction in the momentum
representation. It avoids Siegert-like transformations and it is rather
convenient for incorporating non-static and relativistic corrections
\cite{lage84,levc95a}. Because of inherent restrictions of the
potential picture, our analysis covers energies below pion
photoproduction threshold.

Our model is essentially nonrelativistic. However, we include a few
most important relativistic corrections (like the spin-orbit
interaction) into the one-body electromagnetic current and seagull.
After a brief description of the notion of the seagull given in the
next Section, we introduce the Hamiltonian of the model and analyze
one- and two-body electromagnetic operators.  Then we calculate the
Compton scattering amplitude and discuss the obtained results.

\section{Hamiltonians, currents and seagulls}

In the framework of the time-ordered perturbation theory, a computation
of the photon scattering amplitude starts with a specification of
system's effective degrees of freedom and the system's Hamiltonian
$H[A]$, including its dependence on the external electromagnetic vector
potential $A_\mu$.  We need both linear and quadratic terms in the
expansion of $H[A]$ in powers of $A_\mu$ which determine operators of
the electromagnetic current $j_\mu(x)$ and the electromagnetic seagull
$S_{\mu\nu}(x,y)$ for the system and, correspondingly, the so-called
resonance $R$ and seagull $S$ parts of the Compton scattering
amplitude. Simplifying a real situation, we write
\beqn
\label{H:j&S}
    H[A](t) &=& H(t)
   + \int \Big[ j_\mu(x) A^\mu(x) \Big]_{x_0=t} \,d^3x
\nn && \qquad {}
     - \frac12  \int \!\!\! \int
     \Big[ S_{\mu\nu}(x,y) A^\mu(x) A^\nu(y) \Big]_{x_0=y_0=t}
    \,d^3x \,d^3y  + \ldots ~,
\eeqn
where $S_{\mu\nu}(x,y) = S_{\nu\mu}(y,x)$ is assumed to be a symmetric
function of its arguments.  Accordingly, the photon scattering
amplitude of
\beq
   |i \rangle + \gamma(\epsilon,k) \to
   |f \rangle + \gamma'(\epsilon',k')
\eeq
reads
\beq
\label{R&S}
   T(\omega,\theta) = R(\omega,\theta) + S(\omega,\theta)
\eeq
to leading order $\O(e^2)$, where
\beq
\label{R}
    R(\omega,\theta) = \sum_n \frac{
   \langle f | \epsilon^{\prime *\mu} j_\mu(-k') | n \rangle
   \langle n | \epsilon^\nu j_\nu(k) | i \rangle }
   {E_n - E_i - \omega -i0 }
           + ( \epsilon \leftrightarrow \epsilon^{\prime *},
           k \leftrightarrow -k' )
\eeq
and
\beq
\label{S}
    S(\omega,\theta) = \langle f | \epsilon^{\prime *\mu} \epsilon^\nu
        S_{\mu\nu}(-k',k) | i \rangle .
\eeq
Here $\omega=k_0$ is the photon energy, $\theta$ is the scattering angle,
$E_n$ are energies of eigen states $|n\rangle$ of the system,
$j_\mu(k)$ means a Fourier component of the current density, i.e.\
\beq
\label{j-k}
   j_\mu(k) = \int \Big[ j_\mu(x)
    \,e^{-ik\cdot x} \Big]_{x_0=0} \,d^3x,
\eeq
and
\beq
\label{S-kk}
   S_{\mu\nu}(-k',k) = \int\!\!\!\int
    \Big[ S_{\mu\nu}(x,y) \,e^{ik'\cdot x - ik\cdot y}
    \Big]_{x_0=y_0=0} \,d^3x \,d^3y.
\eeq
In a more general situation, the Hamiltonian (\ref{H:j&S}) can depend
on time derivatives of the vector potential too (e.g.\ owing to a
presence of terms dependent on the electric field).  Nothing changes then
in Eqs.\ (\ref{R})--(\ref{S}) with the except that the Fourier
components of the current and seagull become dependent on both the space
and time components of the photon momenta.

As is well known, the structure of the effective Hamiltonian (and thus
that of the current and seagull too) is closely related with a choice
of the effective degrees of freedom.  In the present context of
low-energy deuteron Compton scattering, we consider nonrelativistic
nucleons as the only dynamical variables of the system, whereas all
mesons, anti-nucleons and other degrees of freedom are encoded into the
internal structure and effective interactions of the nucleons
themselves.  Such an approach is certainly applicable at energies below
pion threshold.

With this choice, the resonance amplitude $R$ comes from low-lying
(two-nucleon) intermediate excitations $n$ of the deuteron, including
the deuteron itself.  It corresponds to two-step scattering via photon
absorption followed by photon emission and vice versa.  This piece can
have and generally has the imaginary part.  Meanwhile, the seagull
amplitude $S$ is real and corresponds to processes, in which the photon
absorption and emission happens at indistinguishable time moments, as
seen at the considered energy scale.  Among such processes are
excitations of heavier intermediate states like $\pi NN$, which
describe, in particular, meson exchanges between photon interaction
points and an internal structure of the nucleon related with
intermediate-meson production.  Considering meson-exchange processes in
general as instantaneous (unretarded) and neglecting dependence of
meson propagators on the photon energy, we retain a retardation
correction for the pion exchange which is known to be quite important
in the seagull amplitude $S$ and to lead to effective modifications of
nucleon polarizabilities in nuclei \cite{huet96}. As for contributions
to $S$ due to the nucleon internal structure, it is taken into account
by introducing polarizabilities of the nucleon.

Both the current operator $j_\mu(x)$ and the seagull operator
$S_{\mu\nu}(x,y)$ have to be consistent with the nuclear Hamiltonian
$H(t)$ of the system and satisfy conditions of the gauge invariance.
Generally, these conditions take the form of the conservation of the
electromagnetic current $j_\mu[A](x)$ found in the presence of the
external vector potential $A_\mu$:
\beq
\label{j[A]-conservation}
   0 = \partial^\mu j_\mu[A](x) = \partial^k j_k[A](x) +
     i \Big[ H[A](t),j_0[A](x) \,\Big] + \partial_A^0 j_0[A](x).
\eeq
Here the Latin index $k$ runs over the space components and the time
derivative $\partial_A^0$ acts only on the external potential $A_\mu$.
In the simplest case of the time-local Hamiltonian (\ref{H:j&S}) the
current $j_\mu[A](x)$ is given by the three-dimensional variational
derivative of $H[A]$,
\beq
\label{j[A]}
   j_\mu[A](x) = \frac{\delta H[A](t)}{\delta A^\mu(x)}
     = j_\mu(x) - \int S_{\mu\nu}(x,y) A^\nu(y)\,d^3y  + \ldots ~.
\eeq
In this case the term with $\partial_A^0$ in Eq.\
(\ref{j[A]-conservation}) must vanish, since this is the only piece
which depends on the time derivative of $A_\mu$.  Therefore, the
charge density $j_0[A](x)$ is $A$-independent, and the
following consistency equations emerge \cite{brow66,aren80}:
\beq
\label{j-conservation-x}
   [j_0(x),H(t)] = -i\,\frac{\partial j^k (x)}{\partial x^k}
\eeq
and
\beq
\label{Schwinger-x}
   [j_0(x), j^\mu(y)] =
     i\,\frac{\partial S^{k\mu}(x,y)}{\partial x^k},
   \quad S^{k0}(x,y) = S^{00}(x,y) = 0.
\eeq
Here all operators are taken at the same time moment $t=x_0=y_0$.

In the nonrelativistic approximation, which will be used in the following
consideration of {\em two-body} effects, the charge density $j_0(x)$
is not affected by meson exchanges (Siegert's theorem) and therefore
coincides with the one-body charge density of the two nucleons $i=1,2$~:
\beq
\label{j0-nonrel}
   j_0(\vec x) = j_0^{[1]}(\vec x) =
      \sum_{i=1,2} eZ_i \delta(\vec x - \vec r_i), \qquad
     Z_i = \frac{1 + \tau_i^z}{2}.
\eeq
Then Eqs.\ (\ref{j-conservation-x}) and (\ref{Schwinger-x}) give
relations between the nuclear (two-body) potential $V$ standing in the
nuclear Hamiltonian $V$ and the two-body parts of the current and
seagull, $j_\mu^{[2]}$ and $S_{\mu\nu}^{[2]}$~:
\beq
\label{j2-conservation-x}
   \Big[j_0^{[1]}(\vec x), V \Big] = -i\vec\nabla\cdot\vec j^{[2]}(\vec x),
\eeq
and
\beq
\label{Schwinger2-x}
   \Big[j_0^{[1]}(\vec x), j_l^{[2]}(\vec y) \Big] =
     i\,\frac{\partial S_{kl}^{[2]}(\vec x,\vec y)}{\partial x_k}.
\eeq

The resonance and seagull contributions at zero energy are constrained
by the Thirring's low-energy theorem. Within the nonrelativistic
accuracy we have
\beq
\label{LET}
   T(0,\theta) = R(0,\theta) + S^{[1]}(0,\theta) + S^{[2]}(0,\theta)
        = - \frac{Z^2 e^2}{AM} \,\See,
\eeq
where $M$ is the nucleon mass, $Z=1$ and $A=2$ for the deuteron,
and the radiation gauge
\beq
\label{gauge}
   \epsilon_0  = \vec k \cdot\e  = 0, \quad
   \epsilon_0' = \vec k'\cdot\e' = 0
\eeq
is assumed for the photon polarization vectors.  In the absence of the
two-body currents and seagulls, the model-independent relation
(\ref{LET}) is fulfilled due to a balance between the one-body seagull
contribution,
\beq
\label{S1(0)}
   S^{[1]}(0,\theta) = -\frac{Ze^2}{M} \,\See,
\eeq
and the resonance amplitude
$R(0,\theta) = (NZe^2/AM)\See$. Here $N=A-Z$.  The presence of
the two-body currents results in an enhancement of the resonance
amplitude, viz.\ $R(0,\theta) \to (NZe^2/AM)(1+\kappa)\See$ for
spinless nuclei, where $\kappa$ is the enhancement parameter standing
in the modified Thomas-Reiche-Kuhn sum rule (see, e.g., the review
\cite{huet00} for a discussion and further references).
Then, in order to support the
balance suggested by the low-energy theorem (\ref{LET}), a two-body
seagull contribution is required.  It is $S^{[2]}(0,\theta) =
-(NZe^2/AM)\kappa\See$ for the spinless nucleus.  For a general case of
a nucleus of spin $S \ge 1$, the two-body seagull amplitude is
characterized by the scalar and tensor enhancement parameters, $\kappa$
and $\kappa_T$:
\beq
\label{S2(0)}
   S^{[2]}(0,\theta) = - \frac{NZ}{AM}  e^2\epsilon^{\prime *}_i \epsilon_j
     \Big\{ \kappa \delta_{ij}  + \kappa_T
    \Big[ S_i S_j + S_j S_i - \frac23 S(S+1)\delta_{ij} \Big] \Big\}.
\eeq

Now we proceed with a consideration of free and interacting nucleons.

\section{Hamiltonian for a single polarizable nucleon}
\label{sec:one-body}

\subsection{Leading-order effects}

Phenomenologically, the dipole polarizabilities $\bar\alpha$ and
$\bar\beta$ are defined as low-energy parameters determining the
quadratic-in-the-field energy shift $V_{\rm pol}$, Eq.\ (\ref{Vpol}).
This shift has to be added to a ``bare" Hamiltonian $H_0[A]$
which is linear in the electromagnetic field, describes an ``unpolarizable"
nucleon with the electric charge $eZ$ and anomalous magnetic moment
$e\kappa/2M$ and produces the so-called Born contribution to the
Compton scattering amplitude.  In the relativistic phenomenology, the
standard choice for $H_0[A]$ and hence the standard {\em
definition} of the unpolarizable nucleon is given by the Dirac--Pauli
Hamiltonian
\beq
\label{H-DP}
   H_0[A] = eZ A_0 + \vec\alpha\cdot(\vec p -eZ \vec A)
   + \beta M + \frac{e\kappa}{4M} \beta\sigma_{\mu\nu} F^{\mu\nu}
\eeq
with $F_{\mu\nu} = \partial_\mu A_\nu - \partial_\nu A_\mu$.
Actually, the given form of $H_0[A]$ is valid only for the nucleon
interacting with real photons.  This is all we need in the present
paper. In the more general case, additional derivatives of the
electromagnetic field appear in $H_0[A]$ as well \cite{lvov93}. They
account for electromagnetic form factors of the nucleon, i.e.\ its
finite size.

Polarizabilities manifest themselves in low-energy Compton scattering
as an $\O(\omega^2)$ addition to the Born amplitude, the latter
becoming the Thomson scattering amplitude $-(e^2 Z^2/M)\,\See$ at zero
energy. In order to correctly identify the contribution of the
polarizabilities, $\O(\omega^2)$ terms in the Born amplitude have to be
retained as well. Since some of them are of order $\O(\omega^2/M^3)$,
an effective low-energy Hamiltonian covering all the $\O(\omega^2)$
terms has to include relativistic corrections up to order $\O(M^{-3})$.

A nonrelativistic reduction of the Dirac--Pauli Hamiltonian valid to
order required was found in Ref.\ \cite{lvov87}. Using the
Foldy--Wouthuysen method \cite{gole86,sche94}
or expelling lower components and higher derivatives
as described in Refs.\ \cite{lvov87,akhi65},
a lengthy but straightforward computation gives%
\footnote{We give the answer in the form obtained in Ref.\ \cite{lvov87}.
In Ref.\ \cite{gole86} the anomalous magnetic moment is not
considered and the final result contains a sign mistake.
In Ref.\ \cite{sche94} terms of order $\O(e^n)$, $n \le 2$ were only
retained.}
\beqn
\label{H-DP-nonrel}
   H_0[A] &=& eZ A_0 + \frac{\vec\pi^2}{2M}
     - \frac{(\vec\pi^2)^2}{8M^3}
     -\frac{e(Z+\kappa)}{2M} \vec\sigma\cdot\vec H
\nn && {}
     -\frac{e(Z+2\kappa)}{8M^2} \Big[\vec\nabla\cdot\vec E
    + \vec\sigma\cdot(\vec E\times\vec\pi - \vec\pi\times\vec E) \Big]
\nn && {}
   + \frac{eZ}{8M^3}\{ \vec\pi^2, \vec\sigma\cdot\vec H \}
   + \frac{e\kappa}{8M^3}\{ \vec\sigma\cdot\vec\pi,
        \vec\pi\cdot\vec H \}
\nn && {}
   + \frac{e\kappa}{16M^3} \Big[ \{ \vec\pi,
           \vec\nabla\times\vec H - \dot{\vec E} \}
   +  (\vec\sigma\times\vec\nabla) \cdot
      (\vec\nabla\times\vec H - \dot{\vec E} ) \Big]
\nn && {}
   + \frac{e^2}{8M^3} \Big[
    (Z^2+Z\kappa+\kappa^2)\vec E^2 - Z^2\vec H^2 \Big]
   + \O(M^{-4}).
\eeqn
Here $\vec\pi = \vec p-eZ\vec A$ is a covariant momentum, $\{A,B\}$
denotes the symmetrized product $AB+BA$, and $\dot{\vec E}$ means the
time derivative of the electric field.  In the region lying outside any
sources $J_\mu$ of the electromagnetic field, the combinations
$\vec\nabla\cdot\vec E = J_0$ and $\vec\nabla\times\vec H - \dot{\vec E}
= \vec J$ vanish, so that the above equation turns out even simpler.

When anti-nucleon degrees of freedom are removed and absorbed into new
effective interactions, the resulting effective Hamiltonian
(\ref{H-DP-nonrel}) becomes non-linear in the electromagnetic field.
In particular, it contains polarizability-like parts which have to be
kept in computations using nonrelativistic variables alone.  Among
these parts is the term $\kappa^2\vec E^2$ which imitates a negative
electric polarizability of the neutron and which is known due to
Foldy \cite{fold59}.

One can easily check that the Hamiltonian (\ref{H-DP-nonrel}) exactly
reproduces the Born amplitude of nucleon Compton scattering to order
$\O(\omega^2)$ which is explicitly given, e.g., in Ref.\ \cite{lvov93}.
Moreover, all the $\O(\omega^2)$ terms in the scattering amplitude are
retained when the kinetic energy in the nucleon propagator is
calculated to leading order $\O(M^{-1})$ (i.e.\ nonrelativistically),
the electromagnetic current is taken to order $\O(M^{-2})$ (i.e.\ with
the spin-orbit correction), and the full electromagnetic seagull up to
order $\O(M^{-3})$ is taken as it stands in Eq.\ (\ref{H-DP-nonrel}).

In the present work we adopt a few further simplifications to Eq.\
(\ref{H-DP-nonrel}). First, we neglect those parts of the Hamiltonian
which do not contribute to the $\O(\omega^2)$ terms at all. These are
the $\O(M^{-3})$ parts of the kinetic energy and the current.
Second, we neglect the $\O(M^{-3})$ part of the spin-dependent seagull
which gives an $\O(\omega^2)$ contribution to the Compton scattering
amplitude but does not contribute to the differential cross section of
Compton scattering to order $\O(\omega^2)$ with {\em unpolarized}
nucleons. Third, omitting the $\O(M^{-3})$ component of the kinetic
energy, we omit also a $\sim e^2 Z^2/M^3$ part of the seagull standing
in $-\vec\pi^4/8M^3$; moreover, we omit such a part in the coefficients
of the fields squared. Thus, we use the following effective Hamiltonian
for a single nucleon which interacts with real photons:
\beqn
\label{H1}
   H^{[1]}[A] &=& eZ A_0 + \frac{\vec\pi^2}{2M}
     -\frac{e(Z+\kappa)}{2M} \vec\sigma\cdot\vec H
     -\frac{e(Z+2\kappa)}{8M^2}
     \vec\sigma\cdot(\vec E\times\vec\pi - \vec\pi\times\vec E)
\nn && {} \quad
   - \frac12 4\pi (\bar\alpha + \delta\alpha_0) \vec E^2
   - \frac12 4\pi \bar\beta \vec H^2,
\eeqn
where%
\footnote{The correction $\delta\alpha_0$ has a direct relation with
the difference mentioned in Section~\ref{sec:intro} between
$\bar\alpha_n$ and the ``static polarizability" $\alpha_n$ found in the
neutron transmission experiments \cite{schm91,koes95}.
In fact, in the formalism used in
these works $\alpha_n$ denotes the coefficient of $\vec E^2$ in the
effective nonrelativistic Hamiltonian.  In order to get warning
against wrong generalizations note, however, that the coefficient
of $\vec E^2$ in the proton case is not equal
to the static polarizability $\alpha_p$
which differs from $\bar\alpha_p$ by a term containing the electric
radius of the proton \cite{petr81,lvov93}.}
\beq
\label{delta-a0}
   \delta\alpha_0 =  -\frac{e^2}{4\pi}\,
            \frac{\kappa^2 + Z\kappa}{4M^3}
   = \cases{ -0.85, & proton,  \cr  -0.62, & neutron.}
\eeq
The corresponding electromagnetic vertices, i.e.\ matrix elements of the
one-body current and seagull in the momentum representation, read
\beqn
\label{j1}
   \epsilon^\mu j_\mu^{[1]}(k; p',p) &=&
   -\frac{eZ}{2M} \e\cdot(\vec p+\vec p')
   -\frac{e}{2M}(Z+\kappa) \,i\omega\vec\sigma\cdot\s
\nn && \qquad {}
   -\frac{e}{8M^2}(Z+2\kappa)
        \,i\omega\vec\sigma\cdot\e\times(\vec p+\vec p')
\eeqn
and
\beqn
\label{S1}
    \epsilon^{\prime * \mu} \epsilon^\nu
       S_{\mu\nu}^{[1]}(-k',k) &=&
     -\frac{e^2 Z^2}{M} \,\See
     +\frac{e^2 Z}{4M^2}(Z+2\kappa) (\omega + \omega') \,i\Vee
\nn && \qquad {}
     + 4\pi\omega\omega' (\bar\alpha + \delta\alpha_0) \,\See
     + 4\pi\omega\omega' \bar\beta \,\Sss,
\eeqn
where $\omega$ and $\omega'$ are the initial and final photon energies,
\beq
\label{s}
   \vec s=\k\times\e, \quad \vec s'=\k'\times\e'
\eeq
are the magnetic field vectors, and we have used the radiation gauge
(\ref{gauge}) for the photon polarization vectors.
It is worth mentioning that the absence of the kinetic term $-p^4/8M^3$
in the Hamiltonian (\ref{H1}) allows us to use self-consistently
nonrelativistic phenomenological potentials developed for a
description of $NN$ interactions at low energies.

The Hamiltonian (\ref{H1}) with the leading relativistic corrections
included possesses an accuracy of about
$e^2/16\pi M^3 = 0.17\times 10^{-4}~\rm fm^3$ for
treating the leading-order effects of the polarizabilities.  For example,
being used in the lab frame, the Hamiltonian $H^{[1]}[A]$ and the
vertices (\ref{j1})--(\ref{S1}) generate the following $\gamma N$
scattering amplitude at the forward angle,
\beqn
\label{Tlab-0}
   T_{\rm lab}(\theta=0^\circ) &=&
   \Big( -\frac{e^2 Z^2}{M} + 4\pi\omega^2
    (\bar\alpha + \bar\beta) \Big) \,\See
\nn && \qquad {}
      -\frac{e^2 \kappa^2}{2M^2} \,i\omega\Vee  + \O(\omega^3).
\eeqn
The $\delta\alpha_0$ and spin-orbit contributions of the seagull
properly correct $\omega$-dependent terms coming from the resonance
amplitude $R$ and bring the resulting amplitude (\ref{Tlab-0}) into an
exact agreement with a known low-energy expansion of $T$ (given, e.g.,
in Ref.\ \cite{babu98}).  At the backward angle, the amplitude found with the
Hamiltonian (\ref{H1}) reads
\beqn
\label{Tlab-180}
   T_{\rm lab}(\theta=180^\circ) &=&
   \Big( -\frac{e^2 Z^2}{M} + 4\pi\omega\omega'
     (\bar\alpha - \bar\beta) \Big) \,\See
\nn && \qquad {}
      +\frac{e^2 (\kappa^2 + 4Z\kappa + 2Z^2) }{4M^2}
   \,i(\omega+\omega') \Vee + \O(\omega^3).
\eeqn
This time an exact result is slightly different. It contains an
additional term $(e^2 Z^2 / 2M^3) \,\omega\omega'\See$ which comes from
a recoil $\O(\omega^2)$-correction to the Thomson amplitude and which
is lost in Eq.\ (\ref{Tlab-180}) because of omitting the $e^2 Z^2/M^3$
pieces of the seagull.%
\footnote{Making such a comparison of the two amplitudes, one has to
take into account a different normalization of the nucleon states.  It
is one particle per unit volume in the present paper and $2E$ particles
in Ref.\ \cite{babu98}.}

\subsection{Polarizabilities and the Baldin sum rule}

In the case of $\gamma d$ scattering, the seagull amplitudes (\ref{S})
for the proton and neutron contribute coherently and dominate the
scattering amplitude (\ref{R&S}) at energies of a few tens MeV.  Their
joint result depends only on the isospin-averaged polarizabilities of
the nucleon, viz.\ $\bar\alpha_N = \frac12(\bar\alpha_p +
\bar\alpha_n)$ and $\bar\beta_N = \frac12(\bar\beta_p + \bar\beta_n)$.

In the following we will consider the difference
$\bar\alpha_N - \bar\beta_N$ as the only free parameter of the nucleon
structure. It is hard to reliably predict this difference, because it
can be affected by $t$-channel exchanges with poorly known couplings
(like the $\sigma$-meson exchange) -- see, e.g., Refs.\
\cite{petr81,lvov93}.  Meanwhile the sum $\bar\alpha_N + \bar\beta_N$
can be safely found from the well-convergent Baldin sum rule
(\ref{Baldin}).  This is quite sufficient in the present context.

There are several evaluations of the dispersion integral in Eq.\
(\ref{Baldin}). Earlier calculations gave
$\bar\alpha_p + \bar\beta_p = 14.2 \pm (0.3{-}0.5)$
\cite{bara74,dama70,lvov79,schr80} and
$\bar\alpha_n + \bar\beta_n = 15.8 \pm 0.5$ \cite{lvov79}
(we comment on the other result,
$\bar\alpha_n + \bar\beta_n = 13.3$ \cite{schr80} below).
A recent re-analysis \cite{babu98a} gave lower values:
\begin{mathletters}
\label{(a+b)matone}
\beqn
\label{(a+b)matone-p}
   \bar\alpha_p + \bar\beta_p &=& 13.69 \pm 0.14,
\\
\label{(a+b)matone-n}
   \bar\alpha_n + \bar\beta_n &=& 14.40 \pm 0.66.
\eeqn
\end{mathletters}
Doing our own calculations with modern sets of photoabsorption data,
we also obtain somewhat lower values than those found in 70's.
However, they are not so low as those in Ref.\ \cite{babu98a},
especially for the neutron.

Specifically, finding $\sigma_{\rm tot}(\omega)$ through the set of
pion photoproduction amplitudes of Ref.\ \cite{hans98} at energies
below 400 MeV, taking total photoabsorption cross sections from
Refs.\ \cite{arms72,macc97} at energies 0.5--1.5 GeV, making a smooth
mixture of the ``theoretical" \cite{hans98} and experimental
\cite{arms72,macc97} cross sections in between, and using a Regge
parameterization of $\sigma_{\rm tot}(\omega)$ at energies
$\omega > 1.5$ GeV (the same as in Refs.\ \cite{petr81,lvov79}), we obtain
\begin{mathletters}
\label{(a+b)lvov}
\beqn
\label{(a+b)lvov-p}
   \bar\alpha_p + \bar\beta_p = 14.0,
\\
\label{(a+b)lvov-n}
   \bar\alpha_n + \bar\beta_n = 15.2.
\eeqn
\end{mathletters}
Uncertainties in these numbers mainly originate from the region of
$\omega \alt 500$ MeV which essentially saturates the dispersion
integral. They can be again conservatively estimated as
$\pm (0.3{-}0.5)$.  For example, we obtain very close results
(13.8 and 15.2, respectively) using in this computation photo-pion
amplitudes from the code SAID \cite{arnd96} (as of beginning of 1999)
instead of the amplitudes from Ref.\ \cite{hans98}.  The lower value for
$\bar\alpha_p + \bar\beta_p$, which follows from the SAID amplitudes,
is mainly caused by that the pion photoproduction multipole
$E_{0+}(\pi^+n)$, as given by that partial-wave analysis close to pion
threshold, is by $\sim 12\%$ too low \cite{drec98} in comparison with
predictions of independent analyses like \cite{hans98} and with
predictions of chiral perturbation theory \cite{bern96}.%
\footnote{Actually, the SAID authors gave an explicit warning against
using the SAID amplitudes very close to pion threshold \cite{arnd96}.
{\it Addition written after the first submission of the present paper:}
Very recently a new set of the SAID amplitudes appeared on the SAID
web-site (solution SM99K). This set is free from most above-mentioned
near-threshold problems, and its use in the evaluation of the Baldin
sum rule gives results which perfectly agree with the estimates
(\ref{(a+b)lvov}).
}
In accordance with (\ref{(a+b)lvov}), we accept the following number
for the isospin-averaged sum of the dipole polarizabilities of the nucleon:
\beq
\label{a+b}
   \bar\alpha_N + \bar\beta_N = 14.6 \, .
\eeq

There are several reasons why we prefer to rely on our own findings
(\ref{(a+b)lvov}) both for the proton and the neutron rather than on
the quoted recent results (\ref{(a+b)matone}).  For the proton case, the
central number for the sum of the polarizabilities obtained by the
authors of Ref.\ \cite{babu98a} is shifted down by their use of the
SAID amplitudes very close to pion threshold. This shift almost
explains the difference between (\ref{(a+b)matone-p}) and
(\ref{(a+b)lvov-p}).  It is worth saying that the tiny uncertainty
$\pm 0.14$ ascribed to $\bar\alpha_p + \bar\beta_p$ in Eq.\
(\ref{(a+b)matone-p}) represents {\em only} statistical errors in the
experimental data on $\sigma_{\rm tot}$. It does not include systematic
errors which are equal to $2{-}3\%$ in $\sigma_{\rm tot}$ and hence
produce the uncertainty of at least $\pm 0.3$ in
$\bar\alpha_p + \bar\beta_p$.

For the neutron, we are even more far from reproducing the very low value
obtained in Ref.\ \cite{babu98a};
we are also far from the result of Ref.\ \cite{schr80},
where the number obtained was even lower.
The reason might be in a different
use of the (indirect) data \cite{arms72} on the neutron total
photoabsorption cross section $\sigma_{\rm tot}(\gamma n)$.  Close to
the $\Delta$-resonance energy, the cross section
$\sigma_{\rm tot}(\gamma n)$ given in Ref.\ \cite{arms72}
is by $\sim 20\%$ (!) lower than predictions
of all modern partial-wave analyses of pion photoproduction.
The procedure used in Ref.\ \cite{arms72} to extract
$\sigma_{\rm tot}(\gamma n)$ from the primary cross section
$\sigma_{\rm tot}(\gamma d)$ obtained with the deuteron target is not
so clear in the $\Delta$-resonance region, in which medium
corrections are large.  That is why we believe that the results of
Ref.\ \cite{arms72} for the neutron should not be taken seriously at
energies below 400 MeV.  As was already said, in our own evaluation of
Eq.\ (\ref{(a+b)lvov-n}) we have found $\sigma_{\rm tot}(\gamma n)$
below 400 MeV through the partial-wave analyses \cite{hans98,arnd96}.

\subsection{Higher-order corrections}

It is clear that higher-order kinematical corrections neglected in Eq.\
(\ref{H1}) are suppressed by powers of $\omega/M$ and therefore they are
small below pion threshold.%
\footnote{The suppression in $T$-even observables like the differential
cross section is in fact $(\omega/M)^2$.}
An actual accuracy of the effective Hamiltonian (\ref{H1}) is
determined by dynamical effects which originate from the pion and
$\Delta$-isobar structure of the nucleon and give corrections of the
relative order $(\omega/m_\pi)^2$. They become important at energies
$\agt 70$ MeV.  To next-to-leading order, such effects are
parametrized by eight structure constants of the nucleon,
viz.\ the quadrupole ($\alpha_{E2}$, $\beta_{M2}$), dispersion
($\alpha_{E\nu}$, $\beta_{M\nu}$) and spin ($\gamma_{E1}$,
$\gamma_{M1}$, $\gamma_{E2}$, $\gamma_{M2}$) polarizabilities of the
nucleon \cite{guia79,levc85,ragu93,babu98}, as represented by the
following effective Hamiltonian \cite{babu98}:
\beqn
\label{H1add}
 \delta H^{[1]}[A] &=&  -\frac12 4\pi
    (\alpha_{E\nu} \vec{\dot E}^2 + \beta_{M\nu} \vec{\dot H}^2)
    -\frac1{12} 4\pi (\alpha_{E2} E_{ij}^2 + \beta_{M2} H_{ij}^2)
\nn && {} \qquad
     -\frac12 4\pi \Big(
     \gamma_{E1} \vec\sigma \cdot \vec E \times \vec{\dot E}
   + \gamma_{M1} \vec\sigma \cdot \vec H \times \vec{\dot H}
\nn && {} \qquad\qquad
     -2 \gamma_{E2} E_{ij}\sigma_i H_j
     +2 \gamma_{M2} H_{ij}\sigma_i E_j \Big).
\eeqn
Here
\beq
  E_{ij} = \frac12 (\nabla_i E_j + \nabla_j E_i), \quad
  H_{ij} = \frac12 (\nabla_i H_j + \nabla_j H_i)
\eeq
are quadrupole strengths of the electric and magnetic fields.  Such an
effective interaction contributes to the seagull amplitude of $\gamma
N$ scattering which gets an addition
\beqn
\label{S1add}
   && \epsilon^{\prime * \mu} \epsilon^\nu
       \delta S_{\mu\nu}^{[1]}(-k',k) =
      4\pi\omega\omega'
    \Big[ \See \delta\alpha + \Sss \delta\beta
\nn && {} \qquad
   + \frac i2 (\omega+\omega')
       \Big( \Vee \,(\gamma_{M2}-\gamma_{E1})
      + \Vss \,(\gamma_{E2}-\gamma_{M1}) \Big)
\nn && \qquad {}
      - i(\Sk\,\Sse - \SK\,\Ses) \gamma_{E2}
      - i(\SK\,\Sse - \Sk\,\Ses) \gamma_{M2} \Big].
\eeqn
The functions $\delta\alpha$, $\delta\beta$ which depend on the photon
energies and on the cosine $z=\k\cdot\K$ of the scattering angle,
\beqn
\label{delta-ab}
   \delta\alpha &=& \omega\omega' \Big( \alpha_{E\nu}
    + \frac{z}{6}\alpha_{E2} - \frac{1}{12}\beta_{M2} \Big),
\nn
   \delta\beta  &=& \omega\omega' \Big( \beta_{M\nu}
    + \frac{z}{6}\beta_{M2} - \frac{1}{12}\alpha_{E2} \Big),
\eeqn
can be handled as dynamical corrections to the dipole polarizabilities
$\bar\alpha$, $\bar\beta$ standing in Eq.\ (\ref{S1}).

Using estimates for the isospin-averaged polarizabilities of the
nucleon obtained through fixed-$t$ dispersion relations \cite{babu98},
\beq
\label{alphas}
  (\alpha_{E\nu})_N \simeq  -3.1, \quad
  (\beta _{M\nu})_N \simeq   9.1, \quad
  (\alpha_{E2})_N   \simeq  27.3, \quad
  (\beta _{M2})_N   \simeq -23.0
\eeq
(units are $10^{-4}~\rm fm^5$), we find, e.g., that the contribution of
$\delta\alpha$, $\delta\beta$ increases the backward-angle differential
cross section of $\gamma d$ scattering and makes the same change as a
shift of $\bar\alpha_N - \bar\beta_N$ by $-1$, $-2$ and $-4$ at 50 MeV,
70 MeV and 100 MeV, respectively.

It is not easy to estimate a model uncertainty in the above numbers
(\ref{alphas}). In particular, they are sensitive to the so-called
asymptotic contribution $A_1^{\rm as}$ to the Compton scattering
amplitude $A_1$ which was represented by the $\sigma$-exchange of the
effective mass $m_\sigma=500{-}600$ MeV in the framework of Refs.\
\cite{babu98,lvov97}.  Recently, an alternative dispersion approach was
presented \cite{drec99,hols99} which allowed to avoid an explicit
introduction of the $\sigma$-exchange and to calculate the amplitude
$A_1$ and the quadrupole polarizabilities of the nucleon under
reasonable assumptions on the reactions $\gamma\gamma\to\pi\pi$ and
$\pi\pi\to N\bar N$.%
\footnote{There are many predecessors of this approach, of which we
would like to indicate Refs.\ \cite{guia78,filk81,hols94} as most
recent works, in which further references can be found.}
Specific numbers obtained in Ref.\ \cite{hols99} for the higher-order
polarizabilities (given for the proton case only) are rather close to
those obtained earlier \cite{babu98}.  Their use for evaluating the
deviation of the backward Compton scattering amplitude from the
low-energy expansion of order $\O(\omega^2)$, Eq.\ (\ref{Tlab-180}),
gives only a 3\% bigger effect than that obtained with our numbers
(\ref{alphas}). More cautiously, we could state that the effect of the
quadrupole and dispersion polarizabilities is known within 20\%, where
the last number is obtained by a reasonable variation of the effective
$\sigma$-meson mass.

In order to evaluate the spin-dependent contribution in Eq.\
(\ref{S1add}), we use spin polarizabilities of the nucleon as found
through the dispersion relations too \cite{babu98,drec98,lvov99}:
\beq
\label{gammas}
    (\gamma_{E1})_N \simeq -3.7, \quad
    (\gamma_{M1})_N \simeq  2.3, \quad
    (\gamma_{E2})_N \simeq  1.4, \quad
    (\gamma_{M2})_N \simeq  0.6
\eeq
(units are $10^{-4}~\rm fm^4$). Writing Eq.\ (\ref{gammas}), we have
corrected predictions for $\gamma$'s of Refs.\ \cite{babu98,drec98}
which include a poorly constrained asymptotic contribution $A_2^{\rm
as}$ arising in the fixed-$t$ dispersion relation for the invariant
amplitude $A_2$ of nucleon Compton scattering. Since $A_2^{\rm as}$
determines the backward spin polarizability of the nucleon, $\gamma_\pi
= -\gamma_{E1} + \gamma_{M1} + \gamma_{E2} - \gamma_{M2}$, which was
recently re-evaluated through a more reliable backward dispersion
relation \cite{lvov99}, we have introduced the appropriate changes to
$\gamma$'s.  Specifically, they are $\delta(\gamma_{E1})_N =
-\delta(\gamma_{M1})_N = -\delta(\gamma_{E2})_N = \delta(\gamma_{M2})_N
= -\frac14 \delta(\gamma_\pi)_N$, where $\delta(\gamma_\pi)_N \simeq
-4$ is a correction to the previous estimate \cite{babu98,drec98} of
$(\gamma_\pi)_N$. About one half of that corrections stems from the
$\eta$ and $\eta'$ exchanges.
At backward angles, the spin effects of order
$\O(\omega^3)$ result in enhancing the coefficient $(e^2/4M^2)(\kappa^2
+ 4Z\kappa + 2Z^2)$ in Eq.\ (\ref{Tlab-180}) by
$2\pi\omega\omega'\gamma_\pi$ with $(\gamma_\pi)_N \simeq 7$ and make
an increase in the differential cross section of backward-angle $\gamma
d$ scattering which is about one third of what the
$(\delta\alpha,\delta\beta)$ correction does.

Recently, there was a controversy on the value of $\gamma_\pi$, at
least for the proton. It was experimentally found \cite{tonn98} that
$(\gamma_\pi)_p = -27.1 \pm 2.2\, ^{+2.8}_{-2.4}$ (in units of
$10^{-4}~\rm fm^4$) and, therefore, it largely deviates from
theoretical predictions \cite{babu98,drec98,lvov99,hemm98} which give
$(\gamma_\pi)_p = -36.7$ to $-39.5$ and $(\gamma_\pi)_n = 50.3$ to
$52.5$.  Taking this deviation seriously and assuming that the
isospin-averaged backward spin polarizability $(\gamma_\pi)_N$ may
differ from the above-accepted value of $(\gamma_\pi)_N=7$ by as much
as $+10$, we should conclude that the theoretical differential cross
section of $\gamma d$ scattering at backward angles is visibly enhanced
due to the effect of $(\gamma_\pi)_N$.
Accordingly, the value of $\bar\alpha_N - \bar\beta_N$ extracted, for
instance, from 100 MeV data on $\gamma d$ scattering should be shifted
up by as much as $+2\times 10^{-4}~\rm fm^3$, what is not negligible!
On the other hand, recent Mainz measurements of backward-angle proton
Compton scattering \cite{wiss99} done with the deuterium target give
the differential cross section which is lower than that obtained at
LEGS \cite{tonn98}, and these newer cross sections assume that the
theoretical value of $(\gamma_\pi)_p \approx -37$ is fully compatible
with the data.  Moreover, recent Mainz measurements of proton Compton
scattering data done with the hydrogen target \cite{schu00} seem to
fully exclude the previous finding \cite{tonn98} as well and to
completely agree with the quoted theoretical predictions. In view of
that, we rely our following analysis on the theoretical values of the
spin polarizabilities of the nucleon specified by Eq.\ (\ref{gammas})
and assume that the related theoretical uncertainties in extracting
$\bar\alpha_N - \bar\beta_N$ are less than $\approx 0.5\times
10^{-4}~\rm fm^3$.

At forward angles, all the higher-order corrections (\ref{S1add}) are
less important.

\section{Two-body currents and seagulls}
\label{sec:MES}

\subsection{Potential-induced electromagnetic currents and seagulls}

The remaining part of the Hamiltonian $H[A]$ is related with two-body
interactions of the nucleons. In the absence of the electromagnetic
fields, such interactions can be represented by a (generally non-local)
$NN$-potential $V$ which has to accurately describe differential cross
sections and polarization observables in $NN$ scattering at energies
below pion threshold, as well as the deuteron binding energy.  There
are several phenomenological potentials of that sort in the literature.
We have chosen to use the Bonn potential (specifically, its
nonrelativistic version OBEPR) \cite{mach87,mach89}, because it
implies a very simple physical picture of one-boson exchanges (OBE)
mediating the $NN$ interaction and, in the framework of this picture,
allows constructing the meson-exchange current $j_\mu^{[2]}$ and
the meson-exchange seagull $S_{\mu\nu}^{[2]}$ directly from the
corresponding Feynman diagrams.  Of course, the OBE picture cannot be
true in all detail.  However, at least, it takes fully into account the
most important long-range contribution, which is the one-pion exchange.

\begin{figure}[htb]
\epsfxsize=\textwidth
\centerline{\epsfbox[10 670 585 810]{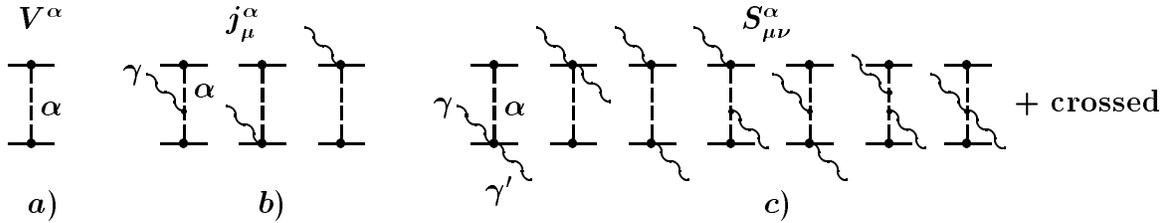}}
\caption{The one-boson-exchange potential (a), and the corresponding
meson-exchange current (b) and meson-exchange seagull (c).
Strong and radiative $\alpha NN$ vertices include antinucleons
(due to PS couplings and/or relativistic corrections) and form factors.}
\label{fig:V-j-S}
\end{figure}

In the momentum representation, the OBEPR potential has the form
\beq
   V (\vec p'_1, \vec p'_2; \vec p_1, \vec p_2)
    = \sum_{\alpha = \pi,\eta,\delta,\sigma,\omega,\rho}
     V^\alpha (\vec p'_1, \vec p'_2; \vec p_1, \vec p_2),
\eeq
where $\vec p_i$ and $\vec p'_i$ are the initial and final momenta of
the $i$-th nucleon subject to the constraint $\vec p_1 + \vec p_2 = \vec
p'_1 + \vec p'_2$, and $V^\alpha$ are potentials stemming
from the exchanges with the specified mesons $\alpha = \pi$, $\eta$,
$\delta$ (which is $a_0(980)$ in the modern notation), $\sigma$, $\omega$
and $\rho$ (see Fig.~\ref{fig:V-j-S}{\it a}).

Let us consider in some details the pion exchange which determines the
long-range part of $V$ and gives the biggest contribution to the
matrix elements of MEC and MES relevant to low-energy $\gamma d$
scattering.  The potential $V^\pi$ is velocity-independent,
i.e.\ it depends only on the momentum transfer $\vec q$:
\beq
\label{V-pi}
   V^\pi(\vec q)
    = -\frac{g_\pi^2}{4M^2} \,
     \vec\sigma_1 \cdot \vec q \, \vec\sigma_2 \cdot \vec q \,
     \vec\tau_1\cdot\vec\tau_2 \, G_\pi(\vec q).
\eeq
Here $g_\pi$ is the $\pi NN$ coupling constant, and the function $G_\pi$,
\beq
\label{G-pi}
   G_\pi(\vec q) = \frac{F_\pi^2(\vec q)}{\vec q^2 + m_\pi^2},
\eeq
contains the pion propagator and the $\pi NN$ vertex form factor
of the monopole form,
\beq
\label{F-pi}
   F_\pi(\vec q) = \frac{\Lambda_\pi^2 - m_\pi^2}
        {\Lambda_\pi^2 + \vec q^2},
\eeq
given by the isospin-averaged mass of the pion, $m_\pi$, and the
cutoff parameter $\Lambda_\pi$.

The pion-exchange current $j_\mu^\pi$ can be obtained by attaching
the photon line to the exchange pion and to the $\pi NN$ vertices as
shown in Fig.~\ref{fig:V-j-S}{\it b}, in which the electromagnetic
$\gamma\pi NN$ vertex arises due to a momentum dependence of the
$\pi NN$ coupling.  This momentum dependence comes partly from the
derivative, or the factor of $\vec q$, standing in the $\pi NN$ vertex
(or, equivalently, from a contribution of antinucleons in the formalism
of the pseudo-scalar coupling adopted in Refs.\ \cite{mach87,mach89}).
Then the minimal substitution
\beq
\label{min-subst-pi}
  \vec\nabla (\vec\tau\cdot\vec\pi) \to
  \vec\nabla (\vec\tau\cdot\vec\pi)
    - ie \vec A \, \Big[ \frac{\tau^z}{2}, \vec\tau\cdot\vec\pi \Big]
\eeq
in the effective $\pi NN$ Lagrangian generates the well-known
Kroll--Ruderman component of the $\gamma\pi NN$ vertex.  An additional
momentum dependence is introduced by the vertex form factor, Eq.\
(\ref{F-pi}), and it should also be taken into account.

\begin{figure}[htb]
\epsfxsize=12cm
\centerline{\epsfbox[40 670 450 810]{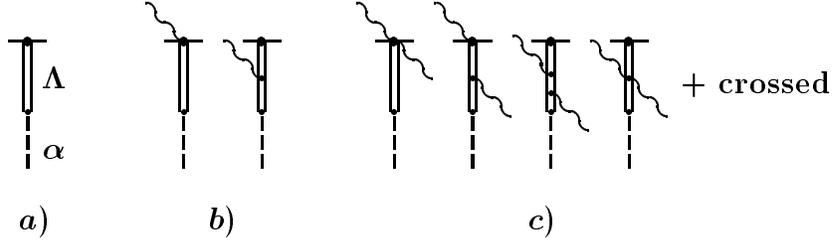}}
\caption{Diagram $a$: the meson-nucleon form factor $F_\alpha(\vec q)$
viewed as resulting from an intermediate heavy-boson exchange $\Lambda$.
Diagrams $b$ and $c$: corresponding contributions to the electromagnetic
meson-nucleon vertices.}
\label{fig:Lambda}
\end{figure}

\begin{figure}[htb]
\epsfxsize=0.8\textwidth
\centerline{\epsfbox[50 680 450 830]{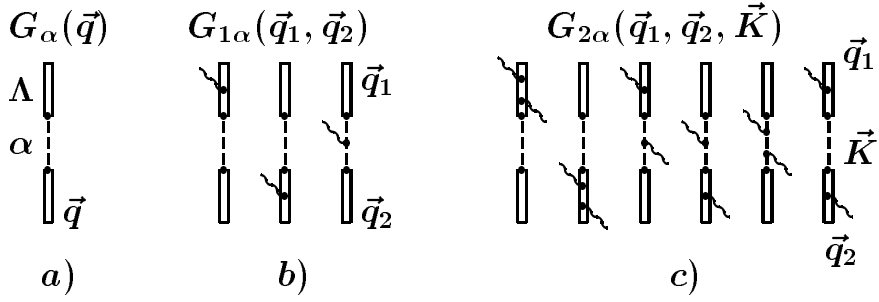}}
\caption{Modifications of a light-boson propagator by intermediate
heavy bosons $\Lambda$.  Shown are the cases with zero, one, and two
photon vertices on the line.}
\label{fig:G012}
\end{figure}

Without knowing the dynamical nature of $F_\pi(q)$, there is no unique
way to restore the electromagnetic current associated with the form
factor. Different prescriptions proposed in the literature
for maintaining gauge invariance in such cases (see, e.g., Refs.\
\cite{aren79,korc84,risk84,math84,buch85,lvov86,adam89,ohta89,schm89})
give different answers, especially in the region of high momenta $q$.
Fortunately, at low momenta $q \ll \Lambda_\pi$ which are only relevant
to the present consideration, such ambiguities are expected to be
small, as is suggested by the Siegert's theorem.  In the following we
choose a simple phenomenological way explicitly formulated
by Riska \cite{risk84} and Mathiot \cite{math84}.%
\footnote{In essence, the solution given by Riska and Mathiot was
earlier obtained by Arenh\"ovel \cite{aren79}, who used the minimal
substitution (\ref{min-subst-pi}) and considered the specific case when
the monopole form factor (\ref{F-pi}) appears to the first power in the
potential $V^\pi(\vec q)$.  A straightforward generalization to the
case when this form factor is squared, as in Eq.\ (\ref{G-pi}), is
easily derived through a differentiation with respect to $\Lambda_\pi$
\cite{schm89}.  After this differentiation, the Arenh\"ovel's
prescription \cite{aren79} becomes identical to that proposed by Riska
and Mathiot \cite{risk84,math84}.}
That is, we assume that the vertex form factor (\ref{F-pi}) results
from a propagation of a fictitious particle $\Lambda$ of the mass
$\Lambda_\pi$, which has the same quantum numbers as the pion and
mediates the pion interaction with the nucleon (see
Figs.~\ref{fig:Lambda}{\it a} and \ref{fig:G012}{\it b}).  Accordingly,
the photon interacts with the particle $\Lambda$ as well
(Fig.~\ref{fig:Lambda}{\it b}), and this gives an additional
contribution to the vertex $\gamma\pi NN$ which restores the
fulfillment of the generalized Ward-Takahashi identities
\cite{bogoliubov} for the transition amplitude of $\gamma N \to \pi N$
and restores the electromagnetic current conservation in the
meson-nucleon system.

Evaluating the diagrams shown in Fig.\ \ref{fig:V-j-S}{\it b} with the
vertices shown in Fig.\ \ref{fig:Lambda}{\it b} and with
the {\em static} propagators of all particles,%
\footnote{Non-static, i.e.\ retardation, corrections will be
considered in the next subsection.}
we obtain the well-known result for the pion MEC:
\beqn
\label{j-pi}
   \vec j^\pi(\vec k; \vec p'_1, \vec p'_2; \vec p_1, \vec p_2)
   &=& -ie (\vec\tau_1 \times \vec\tau_2)^z
   \frac{g_\pi^2}{4M^2} \, \Big[
   \vec\sigma_1 \,(\vec\sigma_2 \cdot \vec q_2) \, G_\pi(\vec q_2)
   - (1 \leftrightarrow 2) \Big]
\nn && {}
   + ie (\vec q_1 - \vec q_2) \, (\vec\tau_1 \times \vec\tau_2)^z
   \frac{g_\pi^2}{4M^2} \,
   \vec\sigma_1 \cdot \vec q_1 \, \vec\sigma_2 \cdot \vec q_2
    \, G_{1\pi}(\vec q_1,\vec q_2) .
\eeqn
Here we introduced the function (cf.\ Refs.\ \cite{risk84,math84,schm89})
\beq
\label{G1-pi}
   G_{1\pi}(\vec q_1,\vec q_2) =
   \frac{F_\pi(\vec q_1) \, F_\pi(\vec q_2)}
    {(\vec q_1^2 + m_\pi^2) \, (\vec q_2^2 + m_\pi^2)}  \Bigg[ 1
   + \frac{\vec q_1^2 + m_\pi^2}{\vec q_2^2 + \Lambda_\pi^2}
   + \frac{\vec q_2^2 + m_\pi^2}{\vec q_1^2 + \Lambda_\pi^2} \Bigg] ,
\eeq
which provides a combination of propagators of the exchanged pion and
the particle $\Lambda$ as they appear in the case of a line with one
electromagnetic vertex (see Fig.~\ref{fig:G012}{\it b}). The vectors
\beq
\label{q_i}
   \vec q_1 = \vec p_1' - \vec p_1, \quad
   \vec q_2 = \vec p_2' - \vec p_2
\eeq
are the momenta transferred to the nucleons.  These momenta
are subject to the constraint $\vec q_1 + \vec q_2 = \vec k$,
where $\vec k$ is the momentum of the incoming photon.
Using the identity
\beq
\label{G-G1}
    G_\pi(\vec q_1) - G_\pi(\vec q_2) =
     (\vec q_2^2 - \vec q_1^2) \, G_{1\pi}(\vec q_1, \vec q_2),
\eeq
one can easily check that the pion-exchange current (\ref{j-pi}) satisfies
Eq.\ (\ref{j2-conservation-x}), which has the following form
in the momentum representation:
\beqn
\label{j2-conservation-p}
  \vec k\cdot\vec j^\pi
     (\vec k; \vec p'_1, \vec p'_2; \vec p_1, \vec p_2)
  &=& e V^\pi
      (\vec p'_1, \vec p'_2; \vec p_1+\vec k, \vec p_2) Z_1
\nn && {}
  - e Z_1 V^\pi
      (\vec p'_1-\vec k, \vec p'_2; \vec p_1, \vec p_2)
     + (1 \leftrightarrow 2).
\eeqn
Here $Z_i$ ($i=1,2$) are the electric charges of the first and second
nucleon, Eq.\ (\ref{j0-nonrel}).

The pion-exchange seagull $S_{\mu\nu}^\pi$ is determined by the
diagrams shown in Fig.\ \ref{fig:V-j-S}{\it c}. There, the
electromagnetic meson-nucleon vertices include again the form factors
generated by the fictitious particle $\Lambda$ through the mechanism
shown in Figs.\ \ref{fig:Lambda}{\it b} and \ref{fig:Lambda}{\it c}.
In the case of the pion exchange the very first diagram of Fig.\
\ref{fig:Lambda}{\it c} is absent.  However, it contributes when the
strong meson-nucleon vertex of the OBE potential has a quadratic
dependence on the particle momenta. This is the case for $\sigma$,
$\delta$, $\omega$ and $\rho$ exchanges (see Ref.\ \cite{mach89}, Appendix
A.3).  Evaluating the diagrams \ref{fig:V-j-S}{\it c}, we obtain
\beqn
\label{S-pi}
  && \epsilon_i^{\prime *} \epsilon_j S_{ij}^\pi
  (-\vec k',\vec k; \vec p'_1, \vec p'_2; \vec p_1, \vec p_2) =
   \frac{e^2 g_\pi^2}{4M^2} \, T_{12} \,
   \Big[ \vec\sigma_1\cdot\e \, \vec\sigma_2\cdot\E
     \, G_\pi(\vec K_1) + (1 \leftrightarrow 2) \Big]
\nn && \qquad\qquad {}
   - \frac {e^2 g^2_\pi}{4M^2} \, T_{12} \, \Big\{ \Big[
    \vec\sigma_1\cdot\e \, \vec\sigma_2\cdot\vec q_2
    \, (\vec q_2 - \vec K_1) \cdot\E \, G_{1\pi}(\vec K_1, \vec q_2)
   + (1 \leftrightarrow 2) \Big]
\nn && \hspace*{12em} {}
   + (\e \leftrightarrow \E,~ \vec K_1 \leftrightarrow -\vec K_2) \Big\}
\nn && \qquad\qquad {}
   + \frac {e^2 g^2_\pi}{4M^2} \, T_{12} \,
   \vec\sigma_1\cdot\vec q_1 \, \vec\sigma_2\cdot\vec q_2
    \, D_\pi (\vec q_1, \vec q_2, \vec K_1, \vec K_2).
\eeqn
Here $\vec q_1$ and $\vec q_2$ are again given by Eq.\ (\ref{q_i}), and
the vectors $\vec K_1$ and $\vec K_2$ are defined as
\beqn
\label{K_i}
   \vec K_1 &=& \vec q_1 - \vec k = -\vec q_2 - \vec k',
\nn
   \vec K_2 &=& \vec q_2 - \vec k = -\vec q_1 - \vec k'.
\eeqn
The isotopic factor $T_{12}$ is equal to
\beq
\label{T12}
   T_{12} = T_{21} = \vec\tau_1\cdot\vec\tau_2 - \tau_1^z\tau_2^z.
\eeq
The function $D_\pi$ is proportional to the amplitude of pion Compton
scattering modified by the form factor corrections. It reads
\beqn
\label{D-pi}
    && D_\pi (\vec q_1, \vec q_2, \vec K_1, \vec K_2) =
   2\e\cdot\E \, G_{1\pi}(\vec q_1, \vec q_2)
\nn && \qquad {} +
    \Big[ (\vec q_1 + \vec K_1)\cdot\e \; (\vec q_2 - \vec K_1)\cdot\E \,
    G_{2\pi}(\vec q_1, \vec q_2, \vec K_1) + (1 \leftrightarrow 2) \Big].
\eeqn
Here the function
\beqn
\label{G2-pi}
   && G_{2\pi}(\vec q_1, \vec q_2, \vec K) =
   \frac{F_\pi(\vec q_1) \, F_\pi(\vec q_2)}
   {(\vec q_1^2 + m_\pi^2) \,(\vec q_2^2 + m_\pi^2) \,(\vec K^2 + m_\pi^2) }
\nn && \qquad {} \times  \Bigg[
   \Bigg(1 + \frac{\vec q_1^2 + m_\pi^2} {\vec K^2 + \Lambda_\pi^2} \Bigg)
   \Bigg(1 + \frac{\vec q_2^2 + m_\pi^2} {\vec K^2 + \Lambda_\pi^2} \Bigg)
   + \Bigg(
    \frac{\vec q_1^2 + m_\pi^2}{\vec q_2^2 + \Lambda_\pi^2}
  + \frac{\vec q_2^2 + m_\pi^2}{\vec q_1^2 + \Lambda_\pi^2} \Bigg)  \,
    \frac{\vec K^2   + m_\pi^2}{\vec K^2   + \Lambda_\pi^2} \Bigg]
\eeqn
provides a combination of propagators of the exchanged pion and
the particle $\Lambda$ as they appear in the case of a line with two
electromagnetic vertices (see Fig.~\ref{fig:G012}{\it c}).
Writing Eq.\ (\ref{S-pi}), we did not assume any special gauge for the
photon polarizations. Therefore, that equation specifies all individual
components of the tensor $S_{ij}^\pi$. Using Eq.\ (\ref{G-G1}) and
the identity
\beq
\label{G1-G2}
    G_{1\pi}(\vec q_1,\vec q_2) - G_{1\pi}(\vec K,\vec q_2) =
     (\vec K^2 - \vec q_1^2) \, G_{2\pi}(\vec q_1, \vec q_2, \vec K),
\eeq
one can check that the obtained MES satisfies Eq.\ (\ref{Schwinger2-x}).
In the momentum representation,
\beqn
\label{S2-conservation-p}
   -k_j S_{ij}^\pi
   (-\vec k',\vec k; \vec p'_1, \vec p'_2; \vec p_1, \vec p_2) &=&
   e j_i^\pi (-\vec k';
       \vec p'_1, \vec p'_2; \vec p_1+\vec k, \vec p_2) Z_1
\nn && {}
   - e Z_1  j_i^\pi (-\vec k';
       \vec p'_1-\vec k, \vec p'_2; \vec p_1, \vec p_2)
     + (1 \leftrightarrow 2).
\eeqn
We may note that formulas for the seagull $S_{ij}^\pi$
(in the $\vec r$-space) were derived long ago in Refs.\
\cite{fria76,weyr83} by considering the appropriate Feynman diagrams,
and in Ref.\ \cite{aren80} by using the minimal substitution.  Neither
of those considerations, however, takes into account the $\pi NN$
vertex form factor. Therefore, in order to achieve a consistency with the
pion-exchange potential (\ref{V-pi}), we do need Eq.\ (\ref{S-pi}).

Meson-exchange currents $\vec j^\alpha$ and seagulls $S_{ij}^\alpha$
related with other bosons $\alpha$ of the OBE potential can be derived
in a similar way.  Formulas for $\vec j^\alpha$ were already obtained
in Ref.\ \cite{levc95a}. Newer results for $S_{ij}^\alpha$ are given in
Appendix~\ref{sec:MES-other}.

All the considered MECs and MESs can be called potential-induced,
because they contain only those pieces which are intimately related
with the OBE potential itself and which are needed to fulfil the
electromagnetic current conservation in the $NN$ system as given by
Eqs.\ (\ref{j[A]-conservation}), (\ref{j2-conservation-x}) and
(\ref{Schwinger2-x}).  Non-potential contributions to $\vec j^{[2]}$
and $S_{ij}^{[2]}$ also exist, and now we proceed with a consideration
of them.

\subsection{Non-potential contributions}

The most important degree of freedom explicitly missing in the
OBE-potential picture of the $NN$ interaction at low energies is an
excitation of the intermediate $\Delta$-isobar. Nevertheless, within
the purely hadron sector (viz.\ $NN$) effects of the
$\Delta$-excitation are indirectly included owing to the use of fitted
parameters adjusted to the experimental data on $NN$ scattering.  Then,
in accordance with the Siegert's theorem, the electric contributions to
MEC and MES found with such parameters take the effects of the $\Delta$
into account as well, at least in the long wave-length approximation.

\begin{figure}[htb]
\epsfxsize=\textwidth
\epsfbox[20 680 590 820]{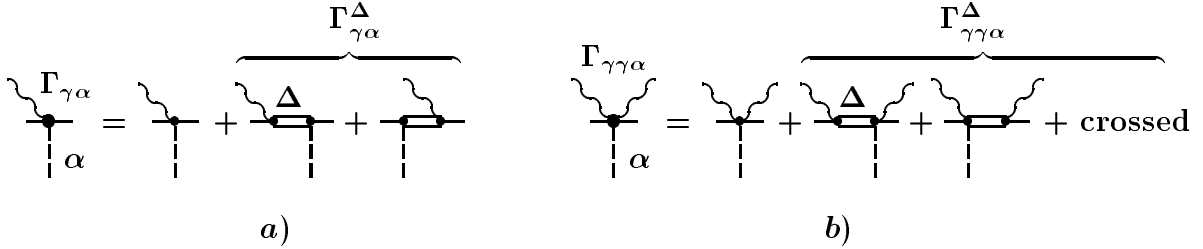}
\caption{The effective $\gamma\alpha NN$ and $\gamma\gamma\alpha NN$
vertices. For $\alpha=\pi$ they take into account the
$\Delta$-resonance contribution.}
\label{fig:g-ga}
\end{figure}

This mechanism, however, does not work for magnetic contributions to
MEC and MES which have to be added independently.  The dominating
(long-range) parts of such contributions come from the one-pion
exchange, and they appear through modifications
$\Gamma_{\gamma\pi}^\Delta$, $\Gamma_{\gamma\gamma\pi}^\Delta$, of the
effective $\gamma\pi NN$ and $\gamma\gamma\pi NN$ vertices caused by
the $\gamma N\Delta$ transition (see Fig.\ \ref{fig:g-ga}).  We do not
include similar modifications for the case of the $\rho$ meson, because
they are completely negligible in the present context.
It is worth noticing that the vertex $\Gamma_{\gamma\gamma\pi}^\Delta$
appears due to the momentum dependence of the $\pi N\Delta$ coupling,
and it is needed to maintain the gauge-invariance of the resulting
Compton scattering amplitude.

Being used for evaluation of the diagrams in Fig.\ \ref{fig:V-j-S}{\it b},
the $\Gamma_{\gamma\pi}^\Delta$ component of the $\gamma\pi NN$ vertex
gives the following contribution to MEC:
\beqn
\label{j-pi-Delta}
  &&  \vec j^{\pi\Delta}
   (\vec k; \vec p'_1, \vec p'_2; \vec p_1, \vec p_2) =
    \frac{ ig_\pi g_\gamma^\Delta g_\pi^\Delta}{36M^2} \Big[
    \frac{ \vec\sigma_2\cdot\vec q_2 \, G_\pi(\vec q_2)}
           {M_\Delta - M - \omega}
      \Big( 2\tau_2^z - i(\vec\tau_1\times\vec\tau_2)^z \Big)
\nn && \qquad {} \times
       (2 \vec q_2 - i\vec\sigma_1\times\vec q_2 ) \times\vec k
    + (\vec\tau_1\to -\vec\tau_1, ~\vec\sigma_1\to -\vec\sigma_1,
             ~\omega\to -\omega) \Big]
    + (1 \leftrightarrow 2) .
\eeqn
Writing this equation, we assumed that the form factor of the
$\pi N\Delta$ vertex was equal to that of the $\pi NN$ vertex.
The mass and couplings of the $\Delta$ are taken to be \cite{blom77}
\beq
   M_\Delta=1225~\mbox{MeV}, \quad
   g_\gamma^\Delta = 0.282 e \, \frac{M_\Delta + M}{m_\pi}, \quad
   g_\pi^\Delta = \frac{2.18}{m_\pi}.
\eeq
Actually, the crossed term in (\ref{j-pi-Delta}), i.e.\ the term having
$M_\Delta - M + \omega$ in the denominator, vanishes when the operator
$\vec j^{\pi\Delta}$ acts upon the deuteron state which has the isospin
$I=0$.

\begin{figure}[htb]
\epsfxsize=0.6\textwidth
\centerline{\epsfbox[80 670 410 820]{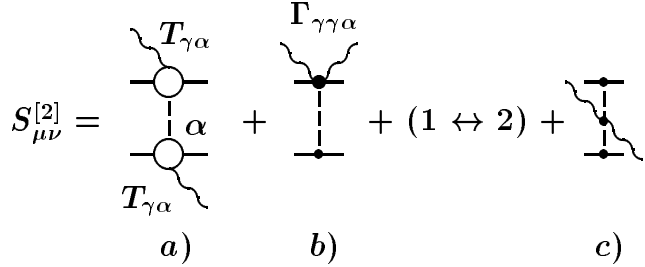}}
\caption{The diagrammatic representation of the two-body seagull $S^{[2]}$.
The content of the pion photoproduction amplitude $T_{\gamma\pi}$ is
explained in Fig.\ \protect\ref{fig:T-gpi}.}
\label{fig:S2}
\end{figure}

\begin{figure}[htb]
\epsfxsize=0.4\textwidth
\centerline{\epsfbox[20 710 220 810]{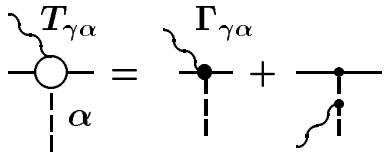}}
\caption{The amplitude of pion photoproduction.}
\label{fig:T-gpi}
\end{figure}

The $\Delta$-isobar contributes to the seagull operator too.  This
happens in two ways. First, the $\Gamma_{\gamma\pi}^\Delta$ component
of the $\gamma\pi NN$ vertex works in the diagrams with one or two
contact single-photon vertices shown in Fig.\ \ref{fig:V-j-S}{\it c}.
This gives a contribution which, together with pieces without $\Delta$,
can be written through (off-shell) pion photoproduction amplitudes
as
\beq
\label{S-pi+piDelta}
   \epsilon^{\prime *}_i \epsilon_j S_{ij}^{\pi + \pi\Delta}
   (-\vec k',\vec k; \vec p'_1, \vec p'_2; \vec p_1, \vec p_2) =
   \frac{ T(\gamma N_1 \to \pi^a N'_1) T(\pi^a N_2 \to \gamma' N'_2)}
    {\vec K_1^2 + m_\pi^2} + (1 \leftrightarrow 2),
\eeq
where $\vec K_1$ is given by Eq.\ (\ref{K_i}), and the sum over the
pion's isospin index $a$ is assumed (see Figs.\ \ref{fig:S2}{\it a}
and \ref{fig:T-gpi}).
Second, the $\Gamma_{\gamma\gamma\pi}^\Delta$ component of the
$\gamma\gamma\pi NN$ vertex works in the first two diagrams of Fig.\
\ref{fig:V-j-S}{\it c} with the contact two-photon vertex. This gives
the contribution shown in Fig.\ \ref{fig:S2}{\it b}:
\beqn
\label{S-piDelta-c}
   && \epsilon^{\prime *}_i \epsilon_j S_{ij}^{\pi\Delta (c)}
   (-\vec k',\vec k; \vec p'_1, \vec p'_2; \vec p_1, \vec p_2) =
     iT_{12} \frac{ g_\pi g_\gamma^\Delta g_\pi^\Delta}{36M^2} \,
    \Big\{ \frac{\vec\sigma_2\cdot\vec q_2 \, G_\pi(\vec q_2)}
           {M_\Delta - M - \omega}
\nn && \qquad {} \times
   \Big[ 2\omega \E_\Lambda\cdot\s
       + 2\omega'\e_\Lambda\cdot\S
     -i\omega \vec\sigma_1\cdot\E_\Lambda \times\s
     +i\omega'\vec\sigma_1\cdot\e_\Lambda \times\S \Big]
\nn && \qquad\qquad\qquad {}
     + (1 \leftrightarrow 2) \Big\}
   + (\e \leftrightarrow \E,~ \vec k \leftrightarrow -\vec k',~
      \omega \to -\omega').
\eeqn
Here $\vec s$ and $\vec s'$ are given by Eq.\ (\ref{s}).
The meson-baryon form factor and the appropriate electromagnetic
coupling with the fictitious $\Lambda$-boson (Fig.\
\ref{fig:Lambda}{\it b}) are included into Eq.\ (\ref{S-piDelta-c}) via
the use of the vertex function $G_\pi(\vec q_2)$ and the modified
polarization vectors
\beq
    \e_\Lambda = \e - 2\vec K_1 \,
        \frac{\e\cdot\vec K_1}{\vec K_1^2 + \Lambda_\pi^2}
\eeq
(and the same for $\e'_\Lambda$; under the crossing,
$\vec K_1 \to -\vec K_2$ and $\s \leftrightarrow \S$).

Another feature missing in the potential picture of the $NN$
interaction is that the exchange-boson fields are generally non-static.
Non-static, or retardation, effects are most important for the pion
exchange due to its large range.  It was recently emphasized by H\"utt
and Milstein in their studies of Compton scattering by heavy nuclei
\cite{huet96} that the retardation correction gives a noticeable
contribution to $\pi$-MES and to the Compton scattering amplitude.  In
the framework of our formalism, we take this correction into account by
using the retarded pion propagator in Eq.\ (\ref{S-pi+piDelta}):
\beq
\label{retard}
  \frac{1}{\vec K^2 + m_\pi^2} \to
  \frac{1}{\vec K^2 + m_\pi^2 - \omega^2} \simeq
  \frac{1}{\vec K^2 + m_\pi^2} +
  \frac{\omega^2}{(\vec K^2 + m_\pi^2)^2}.
\eeq
Here we neglect the energy carried by the nucleons and replace the pion
energy by the photon one, $\omega$.  The adopted procedure is not fully
self-consistent, because we neglect retardation corrections to the
$\pi$-MEC and to the pion-exchange potential $V^\pi$.  However, the
omitted corrections are expected to be less significant than the
retardation correction to the seagull amplitude (cf.\ Refs.\
\cite{huet96,huet98}).

\section{Computation of the amplitudes and cross sections}
\label{sec:computation}

We do actual computations of the scattering amplitude
$T(E_\gamma,\Theta_\gamma)$ in the center-of-mass frame of the $\gamma
d$ system. Accordingly, $\omega=\omega'$ and $\vec k$, $\vec k'$ mean
the energy and momenta of the photons in the CM frame; also,
$\Theta_\gamma$ means the CM scattering angle.  The symbol $E_\gamma$
is reserved to denote the photon beam energy in the Lab frame.  To
specify polarizations of the particles, we introduce the
{\em helicities} of the photons
(viz.\ $\lambda$ and $\lambda'$) and the spin {\em projections}
of the deuteron and nucleons to the beam direction $\vec e_z$
(viz.\ $m$, $m_1$ and $m_2$). In the radiation gauge (\ref{gauge}),
\beq
    \vec\epsilon  = -\frac{1}{\sqrt 2}(\lambda \vec e_x + i\vec e_y), \quad
    \vec\epsilon' = -\frac{1}{\sqrt 2}(\lambda'\vec e_{x'} + i\vec e_y),
\eeq
where the axis $x'$ is orthogonal to $\vec k'$ and lies in the
reaction plane $xz$.

\begin{figure}[htb]
\epsfxsize=0.7\textwidth
\centerline{\epsfbox[30 680 415 810]{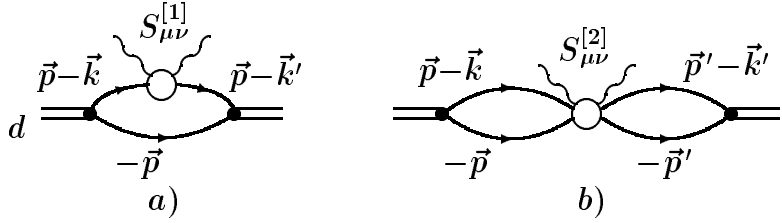}}
\caption{The one- and two-body seagull amplitudes of $\gamma d$
scattering.}
\label{fig:S}
\end{figure}

Using the seagull operator $S^{[1]}_{1\,\mu\nu}(-k',k)$ specified in
Section~\ref{sec:one-body}, we obtain the one-body seagull amplitude
$S^{[1]}(E_\gamma,\Theta_\gamma)$ through a loop integral in the
momentum space (see Fig.\ \ref{fig:S}{\it a}):
\beqn
\label{S[1]}
  &&  \langle \lambda', m' | S^{[1]}(E_\gamma,\Theta_\gamma)
         | \lambda, m \rangle =
   \int \frac{d\vec p}{(2\pi)^3}
    \Psi_{m_1' m_2}^{m'*} \Big( \vec p - \frac12\vec k' \Big) \,
    \Psi_{m_1  m_2}^{m}   \Big( \vec p - \frac12\vec k  \Big)
\nn && \qquad\qquad {} \times
   \langle m_1', m_2 | \epsilon^{\prime *}_i \epsilon_j
      S^{[1]}_{1\, ij}(-k',k) | m_1, m_2 \rangle
   + (1 \leftrightarrow 2).
\eeqn
The notation here is that the state-vectors like $| m_1, m_2 \rangle$
are used to designate only spin variables of the particles. The
momentum variables, if any, are indicated as arguments of the operators.
The sum is always taken over spin projections of intermediate nucleons.
The subscript 1 in $S^{[1]}_{1\,\mu\nu}(-k',k)$ points out that this
operator acts on the nucleon 1.

The deuteron wave function $\Psi_{m_1 m_2}^{m}(\vec p)$ depends on the
relative momentum $\vec p$ of the nucleons. For the nonrelativistic
Bonn potential, one can use analytical parameterizations of
$\Psi_{m_1 m_2}^{m}(\vec p)$ given in Ref.\ \cite{levc95a}.
Note that the authors of the Bonn potential published three
nonrelativistic versions of that potential which we label OBEPR(A),
OBEPR(B) and simply OBEPR.  The only difference between these versions
is in boson's masses, couplings and form factors used.  The parameters
of OBEPR are specified in Ref.\ \cite{mach87}, Table 14.  Parameters of
OBEPR(A) and OBEPR(B) are given in Ref.\ \cite{mach89}, Table A.3, part
A and part B, respectively. We always give our predictions obtained
with the OBEPR version unless other stated explicitly.

Similarly to Eq.\ (\ref{S[1]}), the two-body seagull amplitude
$S^{[2]}(E_\gamma,\Theta_\gamma)$ is given by a two-loop integral over
the nucleon's momenta (see Fig.\ \ref{fig:S}{\it b}):
\beqn
\label{S[2]}
  && \langle \lambda', m' | S^{[2]}(E_\gamma,\Theta_\gamma)
    | \lambda, m \rangle =
   \int\!\!\!\int \frac{d\vec p\,d\vec p'}{(2\pi)^6}
    \Psi_{m_1' m_2'}^{m'*} \Big( \vec p' - \frac12\vec k' \Big) \,
    \Psi_{m_1  m_2 }^{m }  \Big( \vec p  - \frac12\vec k  \Big)
\nn && \qquad {} \times
    \langle m_1', m_2' | \epsilon^{\prime *}_i \epsilon_j
      S^{[2]}_{ij}
   (-\vec k',\vec k; \vec p'-\vec k',-\vec p'; \vec p-\vec k,-\vec p)
     | m_1, m_2 \rangle.
\eeqn
To evaluate this amplitude, we perform integrations over $\vec p$,
$\vec p'$ and sum over spin variables numerically.  Some details of the
integration procedure are given in Ref.\ \cite{levc95a}.

\begin{figure}[htb]
\epsfxsize=0.6\textwidth
\centerline{\epsfbox[85 680 450 780]{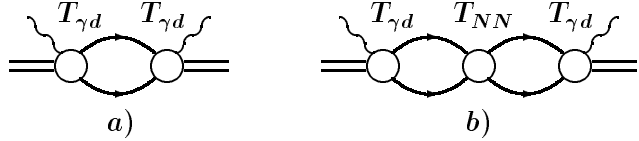}}
\caption{The resonance contribution $R$. Shown are terms without and
with $NN$-rescattering.}
\label{fig:R}
\end{figure}

\begin{figure}[htb]
\epsfxsize=0.7\textwidth
\centerline{\epsfbox[35 700 440 815]{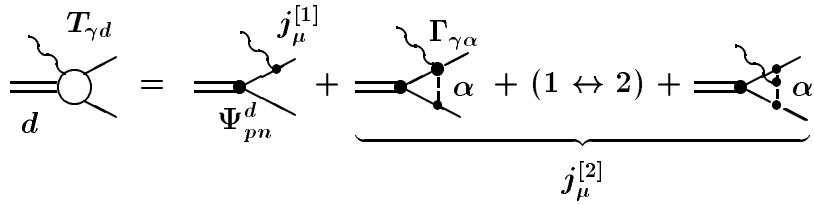}}
\caption{Structure of the $\gamma d \to NN$ vertex.}
\label{fig:T-gd}
\end{figure}

In order to calculate the resonance amplitude
$R(E_\gamma,\Theta_\gamma)$, we introduce the off-shell $T$-matrix of
$NN$ scattering, $T_{NN}(E)$, and write the propagator
$G(E)=(E-H+i0)^{-1}$ standing in Eq.\ (\ref{R}) in the form
\beq
   G(E)=G_0(E) + G_0(E)\, T_{NN}(E) \, G_0(E),
\eeq
where $G_0(E)=(E-H_0+i0)^{-1}$ is the propagator of free nucleons.  Then
$R(E_\gamma,\Theta_\gamma)$ turns out to be the sum of
two terms, without and with $NN$ rescattering in the intermediate state
(see Fig.\ \ref{fig:R}; cf.\ Ref.\ \cite{weyr90}).  The term without
rescattering reads
\beqn
\label{R1}
  && \langle \lambda', m' | R^{\rm no\,rescat}(E_\gamma,\Theta_\gamma)
    | \lambda, m \rangle =
   \int \frac{d\vec p}{(2\pi)^3}
   \, \Big( \frac{\vec p^2}{M} - E_k -i0 \Big)^{-1}
\nn && \qquad\qquad {} \times
    \langle \lambda', m' | T_{\gamma d}
         (-k'; \vec p; \vec p_d+\vec k) | m_1,m_2 \rangle \,
    \langle m_1,m_2 | T_{\gamma d}
         ( k ; \vec p; \vec p_d+\vec k) | \lambda, m \rangle
\nn && \qquad\qquad\qquad\qquad {}
    + (\e \leftrightarrow \E,~ k \leftrightarrow -k').
\eeqn
Here $T_{\gamma d}(k; \vec p, \vec P)$ denotes the amplitude of deuteron
photodisintegration without the final-state interaction (see Fig.\
\ref{fig:T-gd}) at the relative momentum $\vec p$ of the
intermediate nucleons and the total momentum $\vec P$.
The energy $E_k$ in Eq.\ (\ref{R1}) is equal to
\beq
\label{energy-loop}
  E_k = \omega + \frac{\vec p_d^2}{4M}
    - \Delta - \frac{(\vec p_d + \vec k)^2}{4M},
\eeq
where $\Delta$ is the deuteron binding energy. The deuteron momentum
$\vec p_d$ is equal to $-\vec k$ in the CM frame, and it is unchanged
when the crossing transformation $k \leftrightarrow -k'$ is applied.
The procedure of a computation of $T_{\gamma d}$ was the same as in
Ref.\ \cite{levc95a}, and we refer to this paper for further details and
comments.  Here we note only that the two-body contribution to
$T_{\gamma d}$,
\beqn
\label{T-gd[2]}
  &&  \langle m_1,m_2 | T_{\gamma d}^{[2]}
          ( k; \vec p; \vec 0) | \lambda, m \rangle =
   \int \frac{d\vec p'}{(2\pi)^3} \,
    \Psi_{m_1' m_2'}^{m} \Big( \vec p' - \frac12\vec k \Big)
\nn && \qquad\qquad {} \times
    \langle m_1,m_2 | \e\cdot\vec j^{[2]}
      (\vec k; \vec p, -\vec p; \vec p'-\vec k, -\vec p')
       | m_1',m_2' \rangle,
\eeqn
contains a loop integral over $\vec p'$.  Since we do not carry out
angular integrations analytically, the evaluation of Eq.\ (\ref{R1})
actually involves a 9-dimensional numerical integration.  Such a
computational work was hard but it was done with reasonable computer
resources.  We have carefully controlled that the number of chosen
nodes of integrations was sufficient to predict the observables of
$\gamma d$ scattering like the differential cross section with the
numerical accuracy better than 1\%.

The resonance amplitude with $NN$ rescattering has the form
\beqn
\label{R2}
  && \langle \lambda', m' | R^{\rm rescat}(E_\gamma,\Theta_\gamma)
    | \lambda, m \rangle =
   - \int \frac{d\vec p \, d\vec p'}{(2\pi)^6}
   \Big( \frac{\vec p^2}{M} - E_k -i0 \Big)^{-1}
   \Big( \frac{\vec p^{\prime 2}}{M} - E_k -i0 \Big)^{-1}
\nn && \qquad\qquad {} \times
    \langle \lambda', m' | T_{\gamma d}
         (-k'; \vec p'; \vec p_d+\vec k) | m_1',m_2' \rangle \,
    \langle m_1',m_2' | T_{NN}(E_k; \vec p',\vec p) | m_1,m_2 \rangle
\nn && \qquad\qquad\qquad\qquad {} \times
    \langle m_1,m_2 | T_{\gamma d}
         ( k; \vec p; \vec p_d+\vec k) | \lambda, m \rangle
    + (\e \leftrightarrow \E,~ k \leftrightarrow -k').
\eeqn
Here the $NN$ scattering $T$-matrix is determined by the $NN$ potential
$V$ through the Lippmann-Schwinger equation
$T_{NN}(E)=V + VG_0(E)\,T_{NN}(E)$.  It is difficult to evaluate
Eq.\ (\ref{R2}) straightforwardly. In order to simplify the
computation, we used a separable approximation to $T_{NN}$. Actually,
we took $T_{NN}$ from Ref.\ \cite{haid84}, in which the separable
$T$-matrix was built for the Paris potential \cite{laco75}%
\footnote{To our knowledge, a separable approximation to the $T$-matrix
of OBEPR does not exist in the literature.}
(see Ref.\ \cite{levc94} for an explicit form of $T_{NN}$ given in our
notation and normalization).

Since the off-shell properties of $T_{NN}$ for the Paris and Bonn
potentials are not the same, the use of the ``Paris" $T$-matrix
in Eq.\ (\ref{R2}) spoils the self-consistency of the whole calculation
and even violates the gauge invariance of the resulting Compton
scattering amplitude.  A small mismatch appears between the resonance
and seagull amplitudes at all energies, so that the balance (\ref{LET})
prescribed by the low-energy theorem is not exactly fulfilled. At zero
energy and forward scattering angle we get $T_{\rm s.a.}(0,0) = -0.47$
for the spin-averaged (s.a.) amplitude of deuteron Compton scattering
instead of the correct value of $-0.50$ (it is given in units of
$e^2/M$ used here and below for the amplitude).  For the spin-flip
(s.f.) $\gamma d \to \gamma d$ transition $|1,-1\rangle \to
|{-1},1 \rangle$ we get $T_{\rm s.f.}(0,0) = -0.02$ instead of zero.

We believe, however, that the use of the ``Paris" $T$-matrix in Eq.\
(\ref{R2}) does not lead to practical problems at energies of a few
tens MeV, because the rescattering contribution decreases with the
energy.  For example, at $E_\gamma=50$ MeV and $\Theta_\gamma=0$ the
amplitude (\ref{R2}) is equal to
$R^{\rm rescat}_{\rm s.a.} = 0.009 - i0.076$, what should be compared
with the rest amplitude
$T^{\rm no\,rescat}_{\rm s.a.} = -1.097 +i0.315$ (found without
nucleon-polarizability corrections).  At larger angles
the role of the rescattering contribution is even less important,
as will be illustrated in the next Section.

Given the Compton scattering amplitude
$\langle \lambda', m' | T | \lambda, m \rangle$,
we find the differential cross section of $\gamma d$ scattering in the
CM frame as
\beq
   \frac{d\sigma}{d\Omega_\gamma} =
     \frac16 \left( \frac{E_d}{4\pi W} \right)^2
    \sum_{\lambda'm',\lambda m}
  |\langle \lambda', m' | T | \lambda, m \rangle|^2,
\eeq
where $E_d=\sqrt{(2M-\Delta)^2 + \omega^2}$ and $W=E_d + \omega$ are
the deuteron and total energies in the CM frame. The photon beam
asymmetry is
\beq
\label{Sigma}
   \Sigma = \frac{d\sigma^\perp - d\sigma^\parallel}
                 {d\sigma^\perp + d\sigma^\parallel} =
  \frac{2\Re \sum_{\lambda'm',m} \,
     \langle \lambda', m' | T |   1 , m \rangle \,
     \langle \lambda', m' | T | {-1}, m \rangle^*}
   {\sum_{\lambda'm',\lambda m}
     |\langle \lambda', m' | T | \lambda, m \rangle|^2 },
\eeq
where $d\sigma^\perp$ and $d\sigma^\parallel$ are the differential
cross sections for the incoming photons polarized perpendicular or
parallel to the reaction plane $xz$, respectively.

\section{Results and discussion}
\label{sec:results}

\subsection{Zero-range limit}

Before discussing results of the full model, let us consider the
limiting case of a very weak binding of the deuteron.  Specifically, let
us assume that the typical nucleon momentum $\alpha = \sqrt{M\Delta}$,
which is 45.7 MeV in reality, is much less than the pion mass.  Said
differently, we assume that the inter-nucleon distance
$r\sim\alpha^{-1}$ in the deuteron is much larger than the
$NN$-potential range.  Moreover, we assume that the photon energy is
also small, $\omega = \O(\alpha)$, so that all effects related with
recoil corrections $1/M$ can be safely neglected too.  In this limit
the two-body contributions to the electromagnetic current and seagull
become negligible, and so does the rescattering contribution
(\ref{R2}). Therefore, the Compton scattering amplitude is determined
by the one-loop diagrams involving the operators $S^{[1]}$,
$\vec j^{[1]}$ and the asymptotic wave function of the deuteron,
\beqn
\label{Psi0}
   \Psi^m_{m_1 m_2}(\vec p) \simeq
   \frac{\sqrt{8\pi\alpha}} {\vec p^2+\alpha^2} \,
  C^{1m}_{\frac12 m_1, \frac12 m_2}  \, .
\eeqn
Here $C^{1m_d}_{\frac12 m_1, \frac12 m_2}$ is the Clebsch-Gordon
coefficient.  Keeping the terms of leading order over $1/M$ in the
electromagnetic operators and in the energy (\ref{energy-loop}) of the
intermediate nucleons, and calculating analytically the integrals
(\ref{S[1]}) and (\ref{R1}), one arrives at the following $\gamma d$
scattering amplitude \cite{beth35,chen98}:%
\footnote{Bethe and Peierls \cite{beth35} who gave the very first
analysis of $\gamma d$ scattering considered so low energies $\omega
\sim \Delta \ll \alpha$, at which the dipole $E1$ approximation is
applicable. Accordingly, they had the form factor $F_0(q)$ to be equal
to 1. Equation (\ref{BP}) as it is written here was given by Chen
\etal\ \cite{chen98}.}
\beq
\label{BP}
   T(E_\gamma,\Theta_\gamma) \simeq \frac{e^2}{M}
   \See \Big\{ {-}F_0(q) - \frac{4}{3w^2} +
   \frac{2}{3w^2} \Big[ (1-w-i0)^{3/2} + (1+w)^{3/2} \Big] \Big\} ,
\eeq
where
\beq
\label{F0}
    F_0(q)= \frac{4\alpha}{q} \, \arctan\frac{q}{4\alpha}
\eeq
is the deuteron form factor in the considered zero-range limit,
$q=|\vec k - \vec k'|$, and $w=E_\gamma / \Delta$.  The term with
$F_0(q)$ in Eq.\ (\ref{BP}) represents the seagull contribution,
whereas other pieces come from the resonance amplitude
$R(E_\gamma,\Theta_\gamma)$.

As a by-product of this computation, the total cross section
$\sigma ^{\gamma d\to pn}_{\rm tot}(E_\gamma)$ of deuteron
photodisintegration can be derived through the imaginary part of the
(spin-averaged) forward scattering amplitude (\ref{BP}):
\beq
\label{sigma-tot-BP}
   \Im T_{\rm s.a.}(E_\gamma,0)
  = E_\gamma \sigma^{\gamma d\to pn}_{\rm tot}(E_\gamma)
  \simeq \frac{2e^2}{3Mw^2} (w-1)^{3/2}.
\eeq

When the photon energy becomes much higher than the deuteron binding
energy $\Delta$, the amplitude (\ref{BP}) becomes equal to the proton
Thomson scattering amplitude times the deuteron form factor $F_0(q)$.
This is just the seagull contribution $S(E_\gamma,\Theta_\gamma)$ to
the amplitude (\ref{BP}).  The rest ($w$-dependent) terms in Eq.\
(\ref{BP}) give the resonance amplitude $R(E_\gamma,\Theta_\gamma)$
which vanishes in the limit of $E_\gamma \gg \Delta$.  An instructive
feature of Eq.\ (\ref{BP}) is however that this vanishing is rather
slow, like $\propto E_\gamma^{-1/2}$.  Therefore, the resonance
amplitude can give a $10{-}20\%$ contribution to the differential cross
section of $\gamma d$ scattering at $E_\gamma \sim 100$ MeV.

In the opposite limit of very low energies, the binding corrections
become large, and the amplitude (\ref{BP}) is equal only one half of
the proton Thomson scattering contribution:
\beq
\label{T(E=0)}
  T(0,\Theta_\gamma) = -\frac{e^2}{2M} \See,
\eeq
in exact agreement with the low-energy theorem for photon-nucleus
scattering, Eq.\ (\ref{LET}).

The deviation of the {\em nuclear} amplitude (\ref{T(E=0)}) from the
{\em nucleon} one, is related with the resonance contribution
$R(0,\Theta_\gamma)$ which is equal to
    $\displaystyle +\frac{e^2}{2M}\See\rule[-2ex]{0ex}{4ex}$
in the considered zero-range approximation.  In the real case of the
$NN$ interaction of a finite range, both the seagull and resonance
amplitudes get considerable modifications.  For example, the resonance
contribution $R^{[1]}$ from the one-body electromagnetic current
becomes smaller than $e^2/2M$ at zero energy \cite{aren83}.  Moreover,
it depends on the deuteron spin.  Omitting the rescattering correction
(\ref{R2}) and evaluating Eq.\ (\ref{R1}), we have found
\beq
  \langle {\pm 1}, 1| R^{[1]\,\rm no\,rescat}(0,0)| {\pm 1}, 1\rangle =
  \langle {\pm 1},-1| R^{[1]\,\rm no\,rescat}(0,0)| {\pm 1},-1\rangle =
          0.448 \frac{e^2}{M},
\eeq
but
\beq
  \langle {\pm 1}, 0| R^{[1]\,\rm no\,rescat}(0,0)| {\pm 1}, 0\rangle =
          0.392 \frac{e^2}{M}
\eeq
for the Bonn potential OBEPR.

In contrast to $R^{[1]}$, the total resonance amplitude
$R(0,\Theta_\gamma)$ is greater that
    $\displaystyle \frac{e^2}{2M}\See\rule[-2ex]{0ex}{4ex}$
but it is also spin dependent. These features are easily seen from the
relation (\ref{LET}) and from the explicit expressions (\ref{S1(0)})
and (\ref{S2(0)}) for the one- and two-body seagull amplitudes at zero
energy.

\subsection{Two-body seagull amplitude: low-energy behavior}

The two-body seagull contribution $S^{[2]}$ to the total Compton
scattering amplitude dominates the binding corrections at energies of a
few tens MeV, although the resonance contribution $R$ is not negligible
either.  One can get more insight into a physical meaning of $S^{[2]}$
considering its low-energy behavior.  Note that $S^{[2]}$ is a regular
function of the photon energy below pion threshold, and it can be
expanded in powers of $\omega$.  We have found that keeping terms up to
order $\O(\omega^2)$ is generally sufficient for getting quite an
accurate approximation to results obtained through a numerical
evaluation of Eq.\ (\ref{S[2]}) at all energies up to 100 MeV.  The
only exception is the contribution of the $\Delta$-isobar, which
requires also a $\O(\omega^3)$ term linear in the deuteron spin $S$.

The spin-averaged part of the seagull amplitude $S^{[2]}$ at the
considered energies can be described by the following most general form
containing terms up to $\O(\omega^2)$ and compatible with the discrete
symmetries of the Compton scattering amplitude:
\beqn
\label{S[2]-LEX}
   (S^{[2]}_{ij})_{\rm s.a.}   & \;\overcirc{=}\; &
    - \frac{NZ}{AM} e^2 \kappa
    \,\Big( 1 - \frac{\langle r^2_\kappa \rangle}{6}
       q^2 \Big) \,\delta_{ij}
\nn && \qquad {}
    + 4\pi A\,\Delta\alpha \,\omega^2 \delta_{ij}
    + 4\pi A\,\Delta\beta  \,\omega^2
      (\k\cdot\K \delta_{ij} - \hat k_i \hat k'_j).
\eeqn
Here, in our case, the number of neutrons, protons, and nucleons
is equal to $N=Z=1$ and $A=2$, respectively. The sign $\overcirc{=}$ is
used in order to indicate that we have omitted pieces vanishing in the
radiation gauge (\ref{gauge}).  The four coefficients entering Eq.\
(\ref{S[2]-LEX}) were found numerically to be equal to
\beq
\label{kappa-num}
   \kappa  = 0.47, \quad    \langle r^2_\kappa \rangle = 0.49~{\rm fm}^2
\eeq
and
\beq
\label{Delta-alpha-num}
      \Delta\alpha = -0.72\times 10^{-4}\,{\rm fm}^3, \quad
      \Delta\beta  =  0.27\times 10^{-4}\,{\rm fm}^3.
\eeq
The coefficient $\kappa$ which determines the two-body seagull
amplitude at zero energy is the same quantity which characterizes
enhancement in the well-known electric-dipole photonuclear sum rule (by
Thomas--Reihe--Kuhn, or TRK).  Experimental data on the total cross
section of deuteron photodisintegration seem to suggest somewhat lower
value for $\kappa$ than that found from the Bonn potential, Eq.\
(\ref{kappa-num}), namely $\kappa = 0.35 \pm 0.10$ (see, e.g., Ref.\
\cite{gian85}).  For the OBEPR(B) version of the Bonn potential the
difference would be even bigger since then $\kappa$ is predicted to be
equal to $\kappa=0.50$.

Discussing a comparison with the experiment, it is worth to say that
$\kappa$ as defined by Eq.\ (\ref{S[2]-LEX}) is not a direct observable
but rather a (important) theoretical quantity which appears in the
theoretical formalism after eliminating explicit meson degrees of
freedom.  Moreover, $\kappa$ may depend on the formalism specifically
chosen, because the very separation of the total amplitude $T$ into the
resonance and seagull parts is subject to ambiguities.  This, in turn,
is because the separation of the Hamiltonian (\ref{H:j&S}) into pieces
of different order in the vector potential $A_\mu$ generally depends on
the representation chosen and it is subject to unitary ambiguities
(including off-shell ambiguities).  In this respect, $\kappa$ in Eq.\
(\ref{S[2]-LEX}) is as ill-defined and representation-dependent
quantity as, for example, meson-baryon form factors which,
nevertheless, are useful and meaningful theoretical objects.

From a practical point of view, the representation ambiguity in the
resonance and seagull amplitudes is probably not crucial.  An
often-used and convenient choice for fixing the resonance amplitude is
to have $R(\omega,\theta)$ to vanish at high energies, as it is hold in
the dispersion theory of nuclear Compton scattering \cite{huet00}.
Yet, such vanishing is not exactly the case for our theory  --  in
part, owing to the spin-orbit contribution and also because our theory
is not relativistic. We refer to Ref.\ \cite{huet00} for further
details concerning the relevance of the latter point.

As for the above-quoted ``experimental" estimate of $\kappa$, it
appears after two theoretical assumptions:  (i) the validity of the
so-called Gerasimov's argument \cite{gera64} which makes a connection
between the TRK and Gell-Mann--Goldberger--Thirring (GGT) dispersion
sum rules having deal with the unretarded $E1$ and retarded total cross
sections, respectively, and (ii) a little bit arbitrary cutoff at about
140 MeV in the GGT sum rule.  Both assumptions are not strict.  For
instance, owing to a disbalance between corrections of higher order in
$v/c$ (viz., retardation and higher-multipole contributions), the
Gerasimov's argument is actually violated in many models (see, e.g.,
Refs.\ \cite{fria75,aren93,huet00} for further detail and references).
The quoted uncertainties in the ``experimental" estimate for $\kappa$
represent only the cutoff dependence of the GGT dispersion integral,
and one should take this into account when compares the ``experimental"
and theoretical predictions for $\kappa$.

The presence of the parameter $\langle r^2_\kappa \rangle$ in Eq.\
(\ref{S[2]-LEX}) implies that the energy-independent part of the
seagull has generally a $q$-dependent form factor, which is reduced to
a linear function of the momentum transfer squared at low energies.
Numerically, $\kappa$ is almost fully determined by the pion exchange,
Eq.\ (\ref{S-pi}), which gives $\kappa=0.44$, thus leaving only
$\kappa^{\rm HM}=+0.03$ for the contribution from heavier mesons of the
Bonn potential.  In contrast, the pure pion-exchange leads to a very
small radius, $\langle r^2_\kappa \rangle=0.13~{\rm fm}^2$. Actually,
the most part of $\langle r^2_\kappa \rangle$ comes from the pion
exchange accompanied with the $\Delta$-resonance excitation, as
described by Eqs.\ (\ref{S-pi+piDelta}) and (\ref{S-piDelta-c}).

The parameters $\Delta\alpha$ and $\Delta\beta$ in Eq.\
(\ref{S[2]-LEX}) determine the energy-dependent part of the seagull
amplitude. Compared with Eq.\ (\ref{S1}), they can be loosely
interpreted as medium modifications to the electric and magnetic
polarizabilities of the bound nucleon in the deuteron due to
meson-exchange effects.  Such quantities were introduced in this
context by H\"utt and Milstein \cite{huet96} (following the previous
works \cite{rosa85,schu88}) and analyzed for spinless nuclei.  See also
Ref.\ \cite{huet00}, in which a review of a related experimental work
is given.  Similarly to $\kappa$, the parameters $\Delta\alpha$ and
$\Delta\beta$ are representation dependent and not direct observables.
They are rather useful theoretical quantities which appear in the
formalism with eliminated meson degrees of freedom.

There is some distinction in the way how the quantities $\Delta\alpha$
and $\Delta\beta$ and the free-nucleon polarizabilities $\bar\alpha_N$
and $\bar\beta_N$ enter to the $\gamma d$ scattering amplitude.  The
medium modifications to the polarizabilities clearly have a non-local,
i.e.\ two-body (and generally, many-body) origin, and they are expected
to be accompanied with a ``two-body" form factor $F_2(q)$ which
describes a distribution of the center of the relevant nucleon pairs in
the nucleus (in the deuteron we would expect $F_2(q)=1$).
The form factor $F_2(q)$ should be
different from the usual ``one-body" form factor $F(q)$ describing the
distribution of single nucleons in the nucleus and accompanying the
contributions of the free-nucleon polarizabilities.

In the deuteron case the difference between $F_2(q)$ and $F(q)$ is
especially large since the radius squared of the one-body form factor
(averaged over the deuteron spin) is quite large:
$\langle r^2 \rangle = 3.9~{\rm fm}^2$.  See Refs.\
\cite{huet98,huet00} for a more quantitative analysis of this
difference in the case of heavy nuclei.%
\footnote{The presence of the radius $\langle r_\kappa^2 \rangle$ in
the two-body contribution to the seagull amplitude (\ref{S[2]-LEX})
implies that there is no universal two-body form factor which
accompanies both the energy-independent and energy-dependent part of
the seagull.  Due to the term with $\langle r_\kappa^2 \rangle >0$, the
form factor which multiplies the energy-independent part (i.e.\
$\kappa$) has a larger radius than that multiplying the
energy-dependent part. This feature was also first found in Ref.\
\cite{huet98} for heavy nuclei.}
Numerically, the values (\ref{Delta-alpha-num}) are dominated by the
(retarded) pion exchange which gives alone $\Delta\alpha = -0.99$ and
$\Delta\beta = 0.36$ (in units of $10^{-4}\,{\rm fm}^3$).  The
retardation effects incorporated through Eq.\ (\ref{retard}) are very
important here, and they give alone $\Delta\alpha^{(\rm ret)} = -0.84$.
The values (\ref{Delta-alpha-num}) are similar but essentially larger,
especially for $\Delta\alpha$, than estimates obtained in Refs.\
\cite{huet98,huet00} for the lightest even-even nucleus $^4$He on the
basis of the correlated Fermi-gas approximation which is suitable for
heavy nuclei.

Considering the spin-dependent part of the seagull amplitude in the
similar way, one has to introduce a few more parameters.  We will not
discuss all of them here and mention only two, the tensor enhancement
parameter $\kappa_T$ and the tensor modification of the electric
polarizability $\Delta\alpha_T$.  They appear in the low-energy
expansion of the tensor part of $S^{[2]}_{ij}$:\,%
\footnote{A complete basis for representing spin-dependent Compton
scattering amplitudes in the general case of spin $S \ge 1$ was found
by Pais \cite{pais68}.}
\beqn
\label{S[2]T-LEX}
   (S^{[2]}_{ij})_T  & \;\overcirc{=}\; &
    \Bigg( {-}\frac{NZ}{AM} e^2 \kappa_T
    \,\Big( 1 - \frac{\langle r^2_{\kappa T} \rangle}{6} q^2 \Big)
    + 4\pi A\,\Delta\alpha_T \,\omega^2 \Bigg)
\nn && \qquad {}\times
        \Big[ S_i S_j + S_j S_i - \frac23 S(S+1)\delta_{ij} \Big] + \ldots
\eeqn
We have found numerically that $\kappa_T = 0.24$, so that the seagull
amplitude at low energies has a strong spin dependence.  This number is
again dominated by the pion exchange.  The heavier mesons of the Bonn
potential OBEPR give only $\kappa_T^{\rm HM} = -0.03$.  As for
$\Delta\alpha_T$, it gets the largest contribution
from the retardation effects
in the exchange-pion propagator which give alone
$\Delta\alpha_T^{\rm (ret)} = -1.3 \times 10^{-4}~\rm fm^3$.
Therefore, in contrast to the case of heavy nuclei considered
by H\"utt and Milstein \cite{huet96}, the
meson-exchange-induced modification of the electric polarizability of
the bound nucleon is essentially deuteron-spin dependent.

The retardation correction which manifests itself in the parameters
$\Delta\alpha$ and $\Delta\alpha_T$ increases noticeably the
differential cross section. For example, this increase is equal to
$5{-}7\%$ at 100 MeV at all scattering angles.

\begin{figure}[hbt]
\centerline{
\leavevmode \epsfxsize=0.35\textwidth\epsfbox{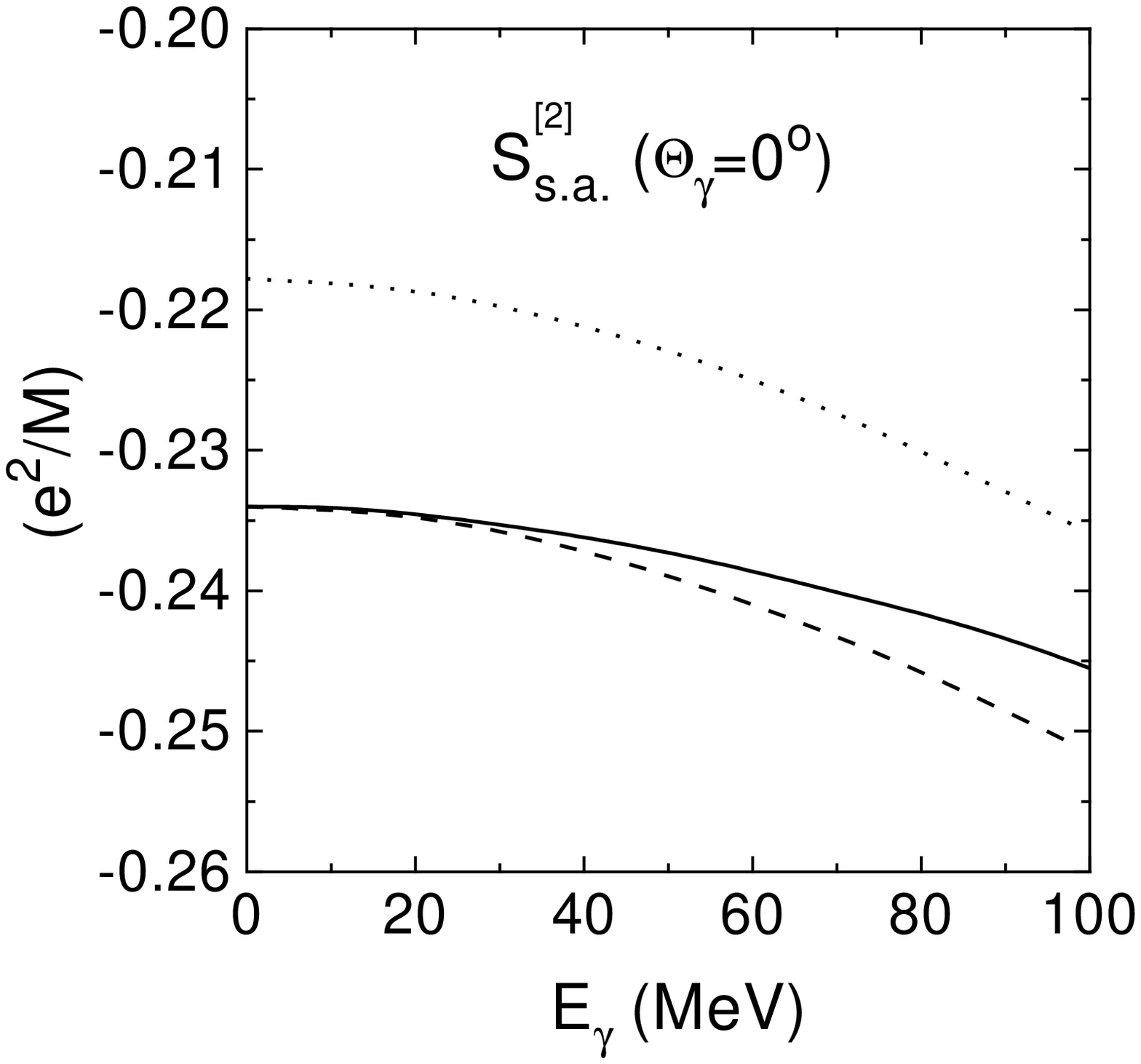}
            \epsfxsize=0.35\textwidth\epsfbox{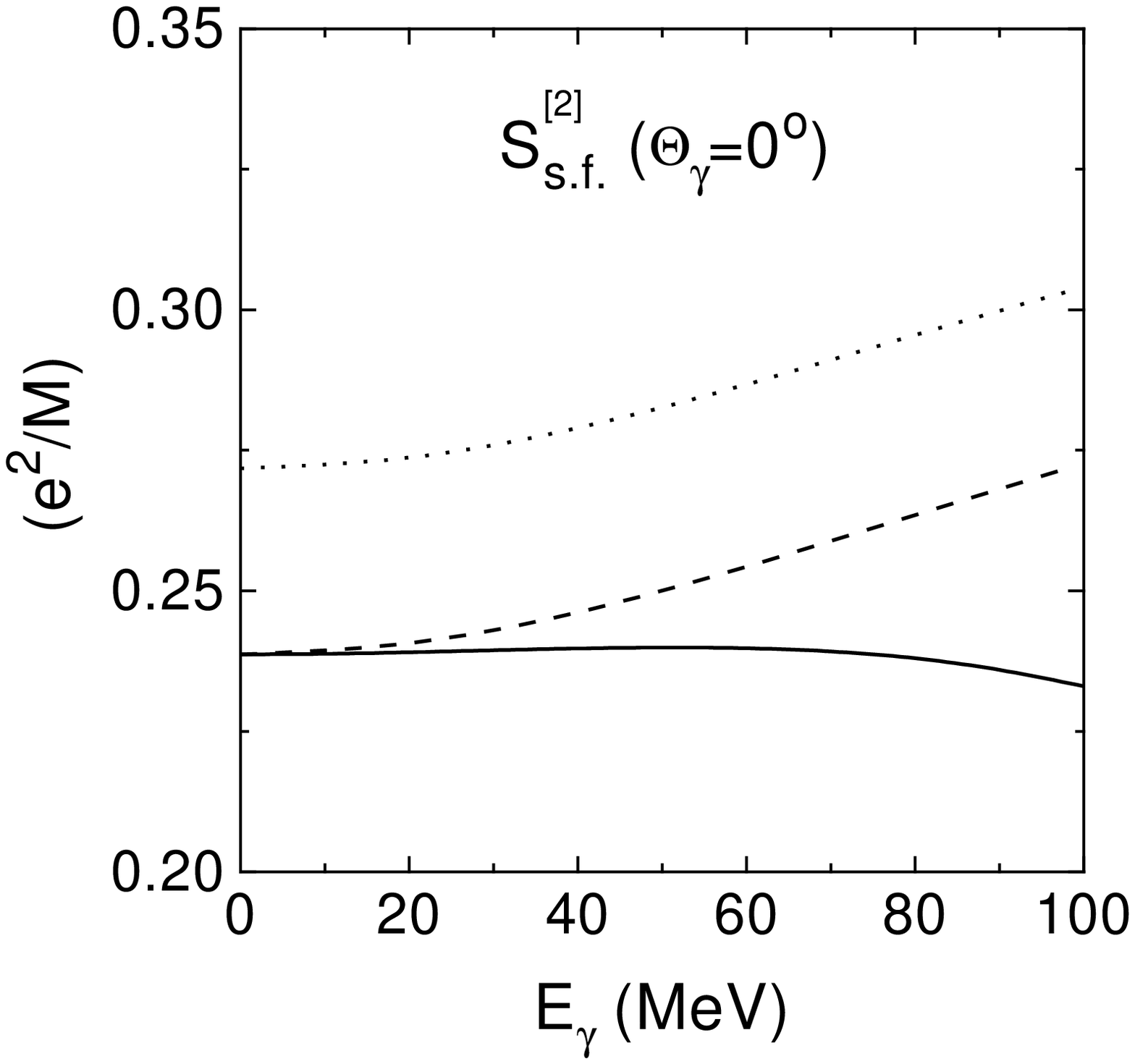}}
\centerline{
\leavevmode \epsfxsize=0.35\textwidth\epsfbox{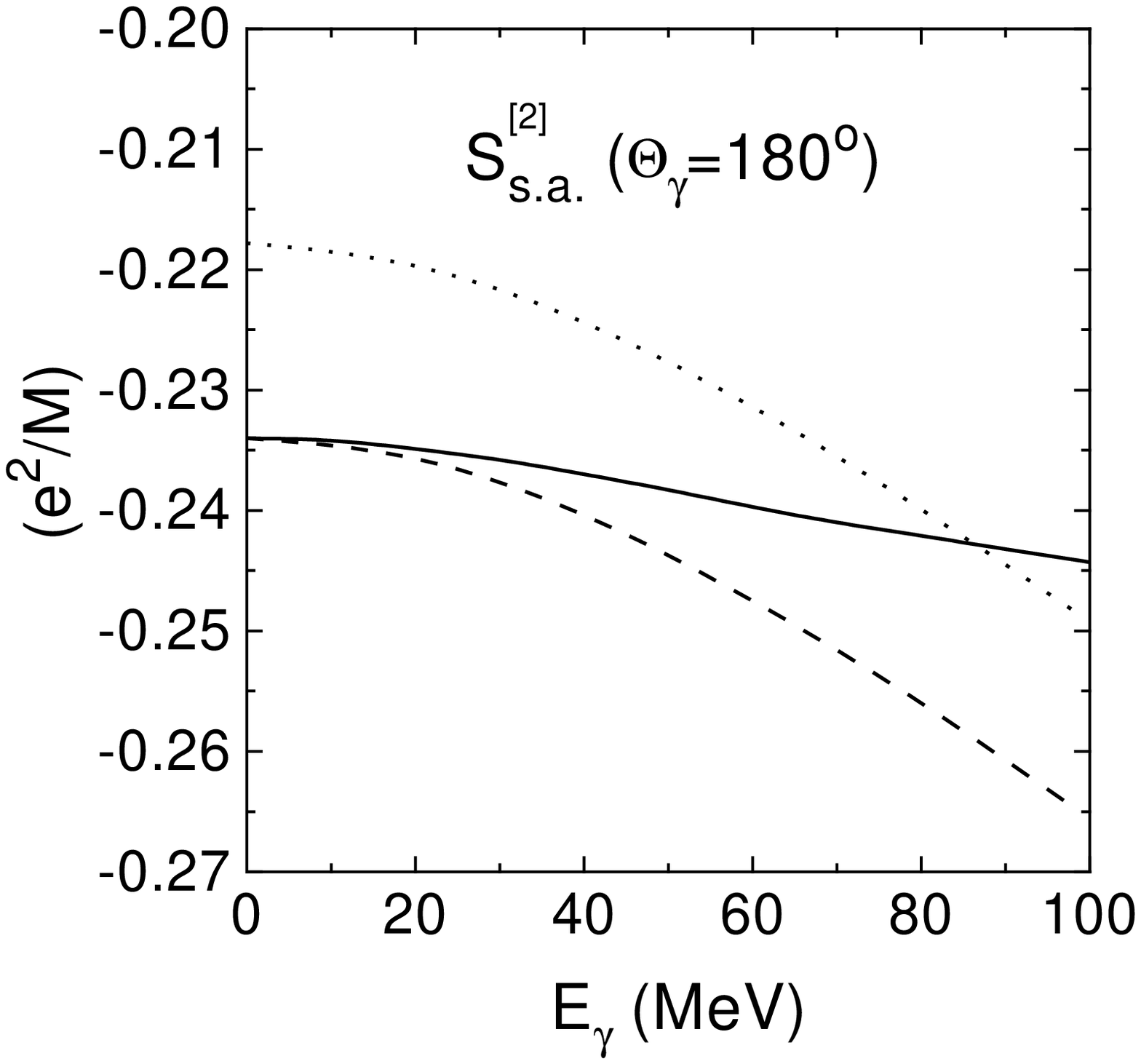}
            \epsfxsize=0.35\textwidth\epsfbox{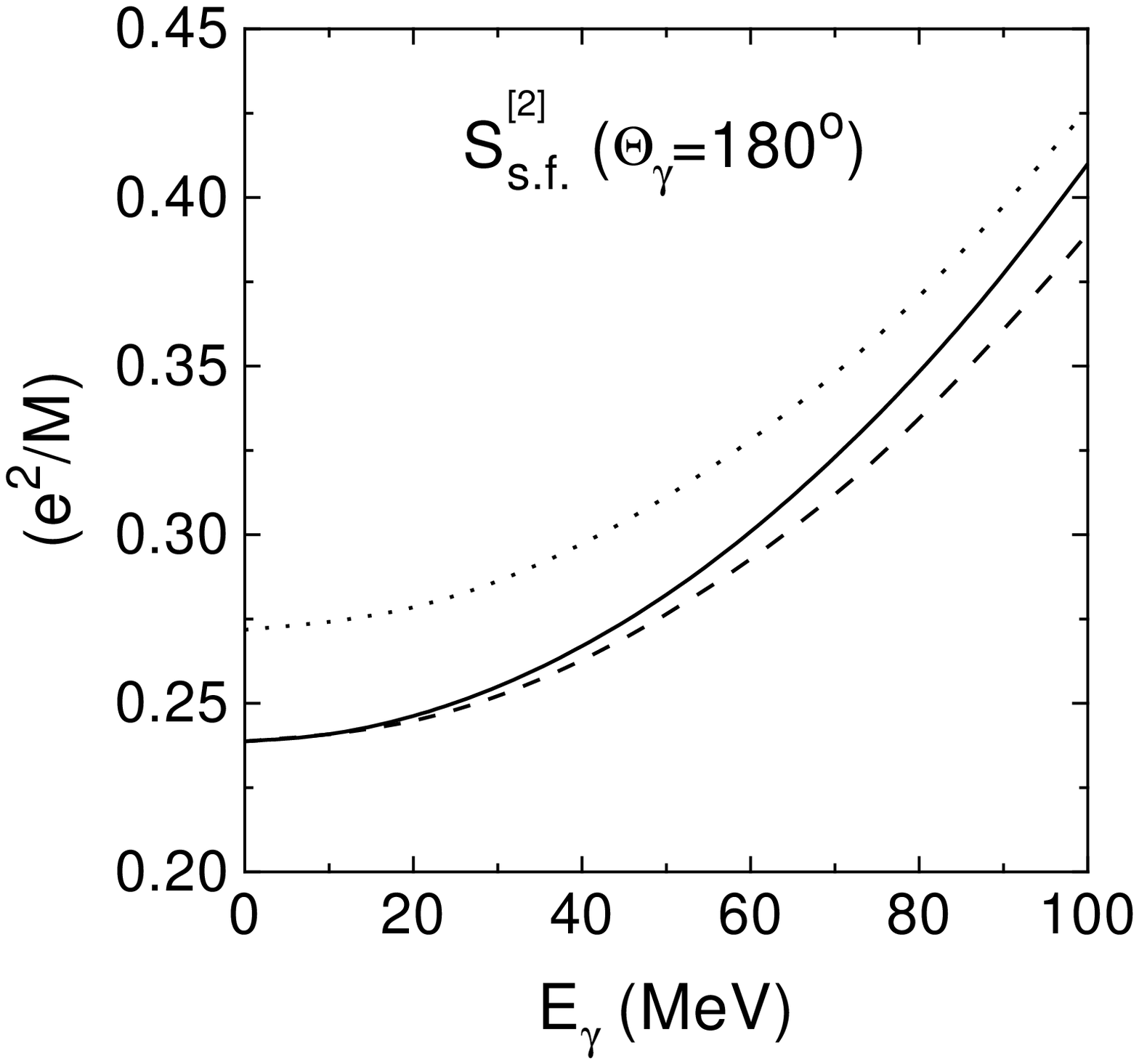}}
\caption{
The energy dependence of the spin-averaged two-body seagull amplitude
$S^{[2]}_{\rm s.a.}$ and the spin-flip amplitude
$S^{[2]}_{\rm s.f.}$. Units are $e^2/M$.
The (retarded) pion-exchange contribution is shown in dotted lines.
Successive additions of heavy mesons and the $\Delta$-isobar
leads to the dashed and solid lines, respectively.}
\label{fig:seagull}
\end{figure}

The energy dependence of the two-body seagull contribution $S^{[2]}$ at
forward scattering angle is illustrated in upper panels of
Fig.~\ref{fig:seagull}.  It is rather flat in the case of the
spin-averaged amplitude,
\beq
\label{S[2]-spin-av(theta=0)}
    S^{[2]}_{\rm s.a.}(E_\gamma,0) = \frac13 \sum_{m=-1,0,1}
   \langle 1, m| S^{[2]}(E_\gamma,0) | 1, m \rangle ,
   \qquad  S^{[2]}_{\rm s.a.}(0,0) = -\frac{e^2}{2M} \,\kappa \,,
\eeq
and it is more pronounced in the case of the spin-flip amplitude,
\beq
\label{S[2]-spin-flip(theta=0)}
    S^{[2]}_{\rm s.f.}(E_\gamma,0) =
   \langle -1, 1| S^{[2]}(E_\gamma,0) | 1, -1 \rangle ,
   \qquad  S^{[2]}_{\rm s.f.}(0,0) = \frac{e^2}{M} \,\kappa_T \,.
\eeq
The $\Delta$-resonance contribution becomes rather noticeable above 60
MeV, and it diminishes the energy dependence introduced by the pion
exchange. As the result, the energy dependence of the seagull amplitude
$S^{[2]}(E_\gamma,0)$ gives only a 2\% increase (in the absence of the
polarizability contribution from the free nucleon) in the differential
cross section of $\gamma d$ scattering at forward angle and
$E_\gamma=100$ MeV.

In the bottom panels of Fig.~\ref{fig:seagull} we show the
spin-averaged and spin-flip two-body seagull amplitudes $S^{[2]}$ in
the case of backward scattering.  They are defined by equations like
(\ref{S[2]-spin-av(theta=0)}) and (\ref{S[2]-spin-flip(theta=0)}), in
which $\Theta_\gamma=0$ is replaced by $\Theta_\gamma=180^\circ$ and
the helicity of the final photon is inverted.  We see that the energy
dependence of the spin-flip amplitude is rather steep in this case. It
increases the differential cross of deuteron Compton scattering at 100
MeV by 5\% (again, in the absence of the free-nucleon
polarizabilities).

In general, we have found that the effect of the $\Delta$ excitation
onto the seagull amplitude $S^{[2]\pi}$ is not large.  It does not
exceed $-2\%$ in the differential cross section of $\gamma d$
scattering, $d\sigma/d\Omega$, at all considered energies.  The
two-body contribution of the $\Delta$-isobar to the resonance amplitude
$R$ was found to be not large too.  For example, it changes the
differential cross section at 100 MeV by $+3\%$ at forward  angle and
by $-0.6\%$ at backward angle.  At this point we disagree with the
results of Ref.\ \cite{weyr83}, in which it was found that the effect
of the $\Delta$-excitation onto the resonance amplitude at 100 MeV is
negligible at small angles and very large at $\Theta_\gamma \ge
90^\circ$ giving a $+15\%$ increase in $d\sigma/d\Omega$.

\subsection{Model dependence}

\begin{figure}[hbt]
\centerline{
\leavevmode\epsfxsize=0.33\textwidth\epsfbox{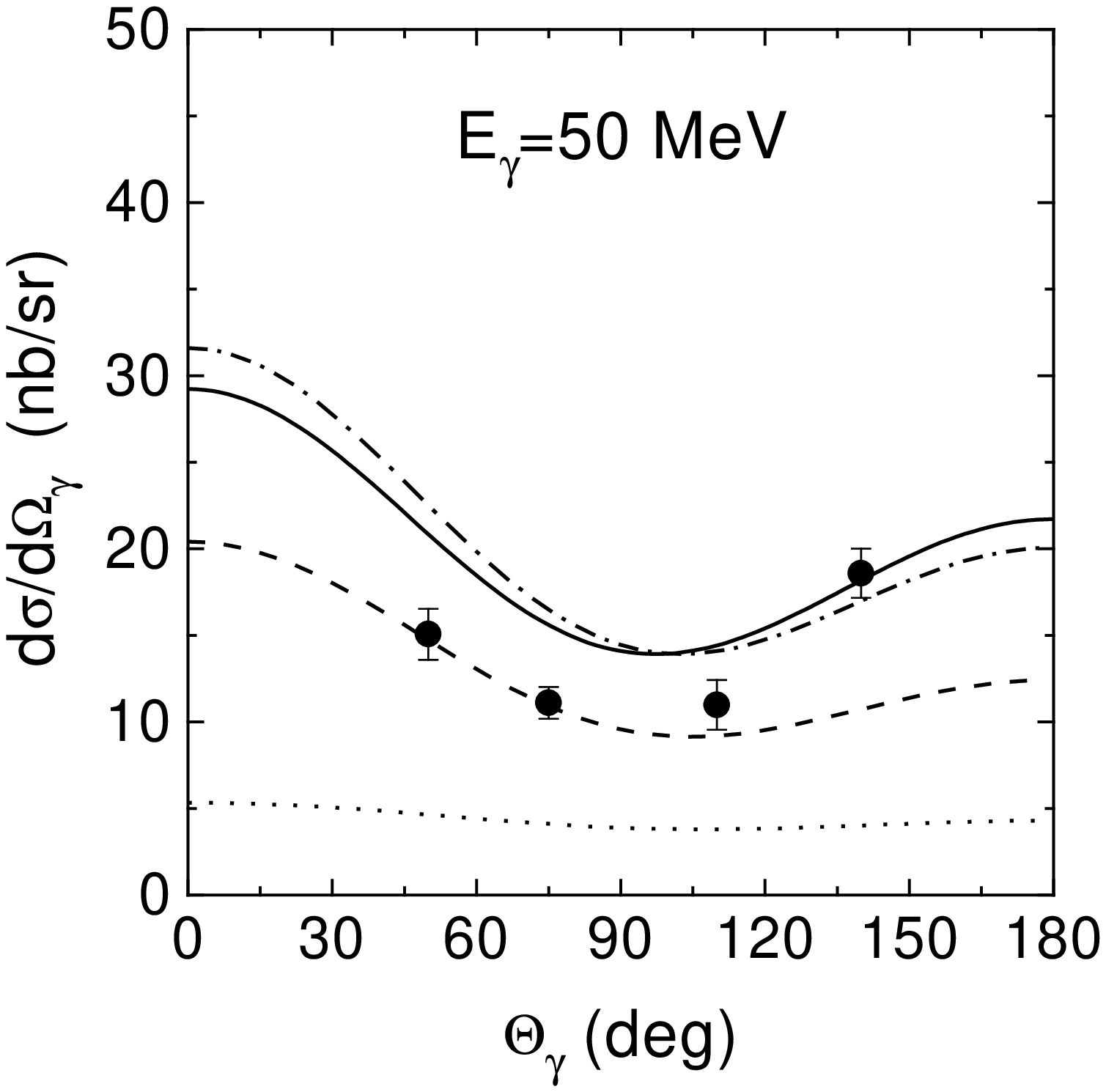}
\leavevmode\epsfxsize=0.33\textwidth\epsfbox{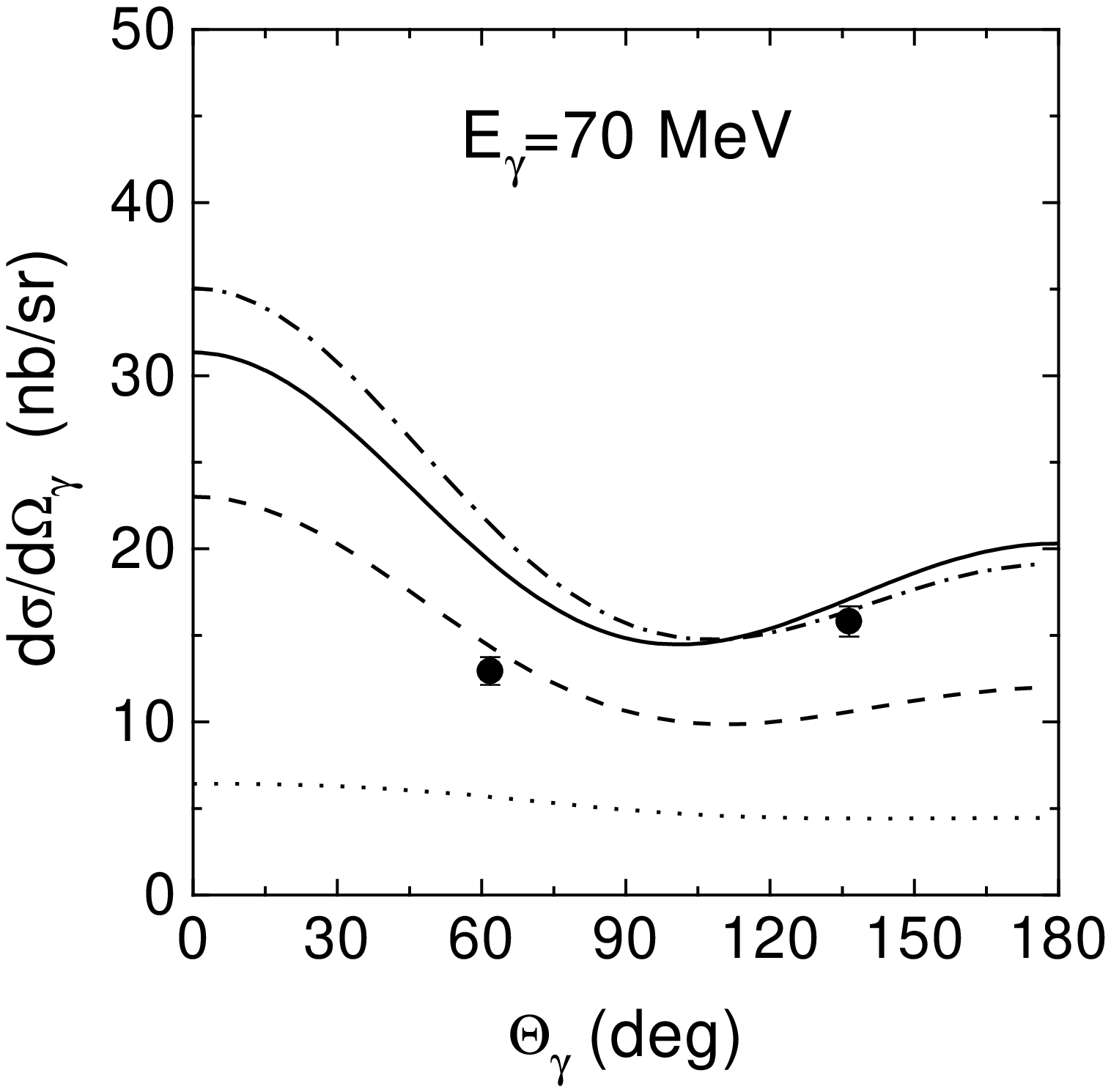}
\leavevmode\epsfxsize=0.33\textwidth\epsfbox{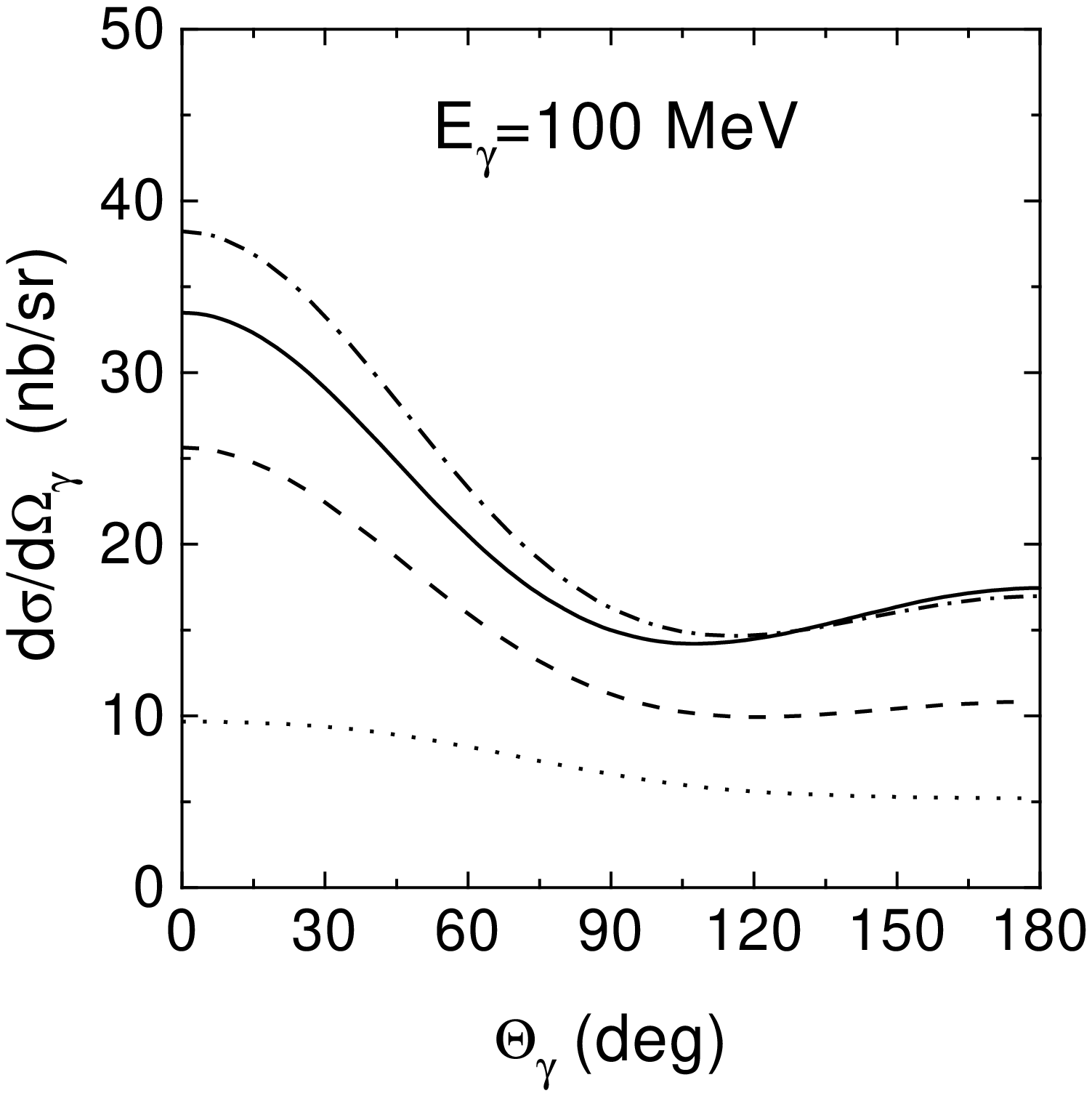} }
\caption{
Contributions of different parts of the amplitude
to the differential cross section (CM) of $\gamma d$ scattering
at 50, 70, and 100 MeV.
The contribution of the resonance amplitude $R^{\rm no\,rescat}$ alone
is shown in dotted lines. Successive additions of the one-body seagull
$S^{[1]}$, the two-body seagull $S^{[2]}$,
and the rescattering amplitude $R^{\rm rescat}$ give the
dashed, dash-dotted and solid lines, respectively.
Nucleon polarizabilities are not included into $S^{[1]}$.
Data are from Ref.\ \protect\cite{luca94}. }
\label{fig:s-r-tot}
\end{figure}

Contributions of the different components $R^{\rm no\,rescat}$,
$S^{[1]}$, $S^{[2]}$ and $R^{\rm rescat}$ of the total amplitude $T$ to
the differential cross section at a few selected energies are shown in
Fig.~\ref{fig:s-r-tot}.%
\footnote{Since the seagull and resonance contributions are not
gauge-invariant separately, we remind that we use the gauge
(\ref{gauge}) to calculate $R$ and $S$.}
It is seen that the effects of the resonance amplitude, the one-body
seagull as well as the two-body seagull are of similar scale, though
the total effect of the seagulls is more than 70\% in the energy region
under consideration. At the same time rescattering has a modest impact
on the differential cross section.  Our results confirm findings of
other approaches \cite{weyr90,wilb95,levc95} that the rescattering
decreases the forward differential cross section and that this decrease
is between 7\% to 12\% in the energy region of 50 to 100 MeV.

There was some discrepancy in the previous work concerning the role of
$R^{\rm rescat}$ at backward angles.  Weyrauch \cite{weyr90} found that
the rescattering does not contribute at backward angles at all, whereas
in the later calculations \cite{wilb95,levc95} a visible increase in
$d\sigma/d\Omega$ was claimed.  Our results agree with the latter
conclusions and suggest that the effect of
$R^{\rm rescat}(E_\gamma,180^\circ)$ ranges between $+7\%$ at 50 MeV to
$+3\%$ at 100 MeV.  Since an accurate calculation of $R^{\rm rescat}$
is a difficult problem for our computational scheme, this finding of a
relatively small effect of $R^{\rm rescat}$ at energies and angles
where experimental data are available provides some justification to
our approximate use of a separable potential \cite{haid84} for a
computation of the off-shell $T$-matrix of $NN$ rescattering.

An instructive feature of the calculation is that the spin-orbit (s.o.)
contributions to the electromagnetic current and seagull,
$\vec j^{[1]\,\rm s.o.}$ and $S^{[1]\,\rm s.o.}$, which are
relativistic corrections of order $1/M^2$ in Eqs.\ (\ref{j1}) and
(\ref{S1}), are rather essential.
Previously, it was found \cite{camb84} that the spin-orbit current
is responsible for the long-standing discrepancy between the theory and
data on deuteron photodisintegration at forward and backward angles.
In the reaction of Compton scattering, the spin-orbit interaction
is important not only at extreme angles, and this importance
increases with the photon energy.
Either $\vec j^{[1]\,\rm s.o.}$ or $S^{[1]\,\rm s.o.}$
leads to an approximately equal decrease in the differential cross
section at the forward angle, and the total effect of the spin-orbit
interaction on $d\sigma(E_\gamma,0)/d\Omega$ is $-4\%$ at 50 MeV and
$-15\%$ at 100 MeV.  A somewhat different situation happens at the backward
angle. The spin-orbit current $\vec j^{[1]\,\rm s.o.}$ still decreases
the differential cross section, but the spin-orbit seagull $S^{[1]\,\rm
s.o.}$ makes a bigger increase. The net effect of the spin-orbit
interaction at $180^\circ$ is $+4\%$ at 50 MeV and $+8\%$ at 100 MeV.
In the central angular region of $\Theta_\gamma \simeq 90^\circ$, the
spin-orbit interaction has a little impact on $d\sigma/d\Omega$ ranging
between $+0.8\%$ at 50 MeV and $-0.2\%$ at 100 MeV.

Staying within the Bonn-potential picture, we have checked how the
differential cross section depends on a specific choice of the
potential's parameters. The OBEPR and OBEPR(A) versions of the Bonn
potential give $d\sigma/d\Omega$ which are different at most by 1\% in
the energy range of $50{-}100$ MeV.  A bigger difference is found for
the OBEPR and OBEPR(B) versions, though it decreases with the photon
energy. For example, the OBEPR(B) potential gives $d\sigma/d\Omega$
which is bigger by 5\% (7\%) at 50 MeV and 0.5\% (5\%) at 100 MeV for
forward (backward) angles, respectively.  The main reason for such a
difference comes from a very large value for the cut-off parameter
$\Lambda_\pi=2$ GeV used in the $\pi$-exchange potential of the
OBEPR(B) version, whereas $\Lambda_\pi=1.3$ GeV for OBEPR.
Respectively, the seagull's parameters $\kappa$ and $\kappa_T$ are
bigger for OBEPR(B) too.%
\footnote{As a word of precaution we have to remind that we always
calculate the rescattering contribution $R^{\rm rescat}$ using a
fixed $T$-matrix obtained with the (separable-approximated) Paris
potential.  Therefore, the specific numbers indicated in the above
discussion do not show the full change in the theoretical predictions
when the potential is changed, although we assume that they are
qualitatively correct.}

We may note that the OBEPR(B) version of the Bonn potential does not
provide a satisfactory description of observables in deuteron
photodisintegration \cite{levc95a} and thus it is not fully realistic.
Therefore, one may conjecture that the sensitivity of the results on
$\gamma d$ scattering would not be so noticeable if one restricts
oneself to ``realistic" potentials only.  In this respect it is worth
mentioning that the use of different momentum-space versions of the
Bonn OBE potentials was found \cite{wilb95} to yield
version-independent results for $\gamma d$ scattering within 1\%.

The present results are in a qualitative agreement with our previous
calculation \cite{levc95} done in the framework of a ``minimal model",
in which MECs and MESs are evaluated through the minimal substitution
$\vec p \to \vec p - e\vec A$ in the (Paris) $NN$ potential.  As they
were published, those older results did not include the spin-orbit
interaction. After taking into account $\vec j^{[1]\,\rm s.o.}$ and
$S^{[1]\,\rm s.o.}$ the predictions of the minimal model become closer
to the results of the present work, especially as for the shape of the
angular dependence.  Still, the minimal model gives a lower
differential cross section:  by 6\% at 70 MeV and by 11\% at 100 MeV.
Such a difference can be traced in part to a weaker energy dependence
of the seagull amplitude found in the minimal model and to the absence
of the two-body $\Delta$-isobar effects in that model.

\begin{figure}[hbt]
\centerline{
\leavevmode\epsfxsize=0.33\textwidth\epsfbox{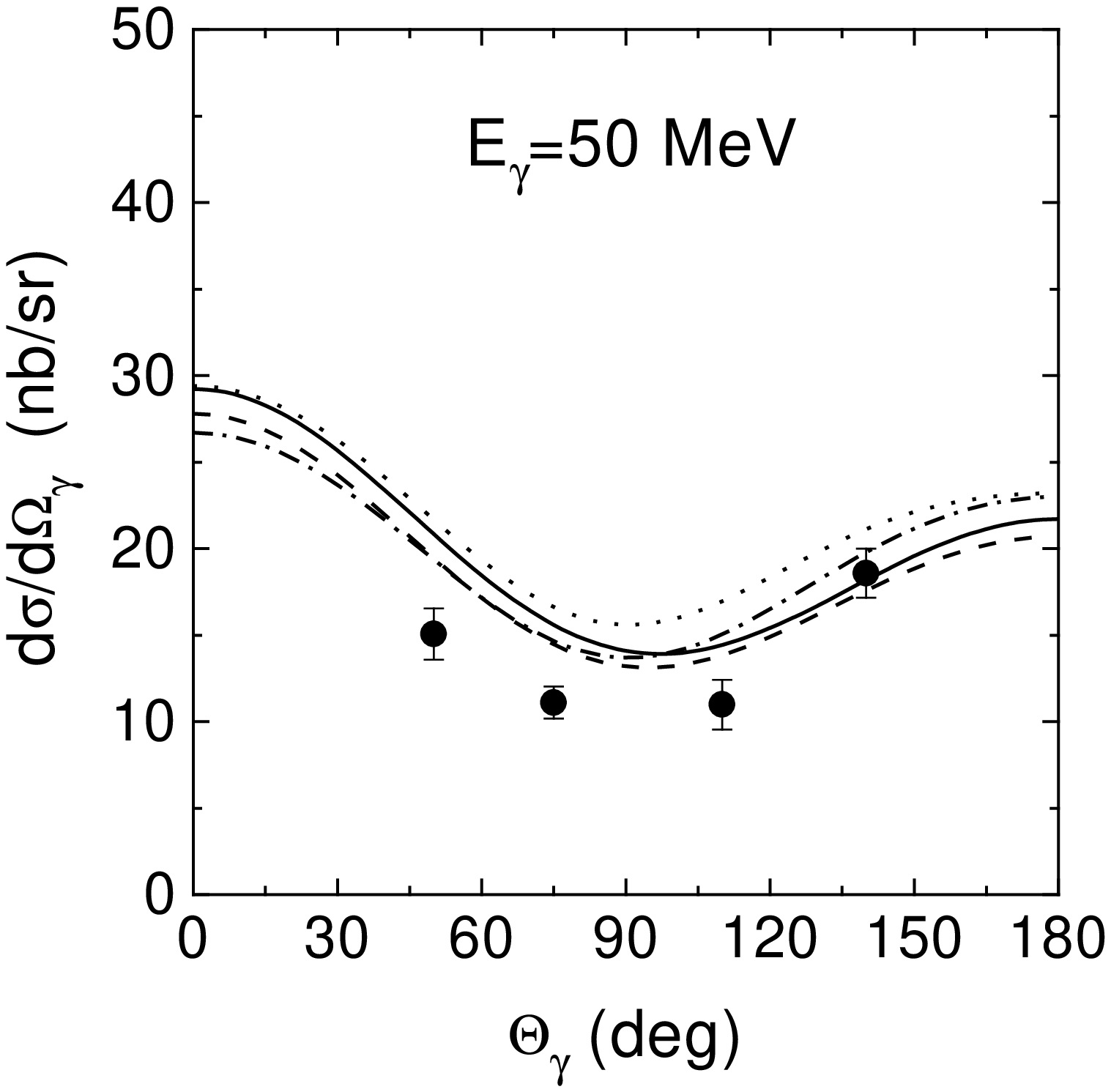}
\leavevmode\epsfxsize=0.33\textwidth\epsfbox{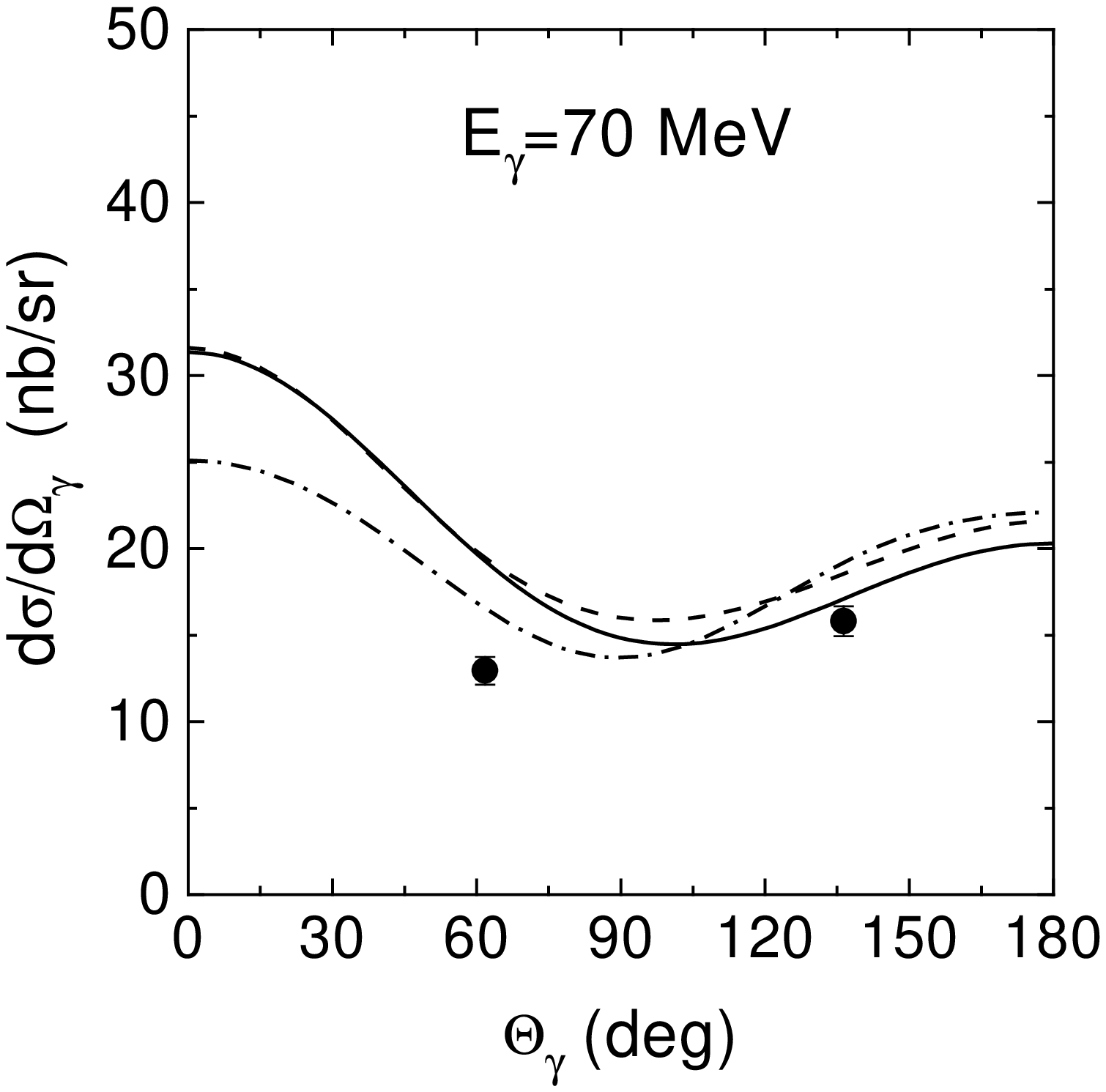}
\leavevmode\epsfxsize=0.33\textwidth\epsfbox{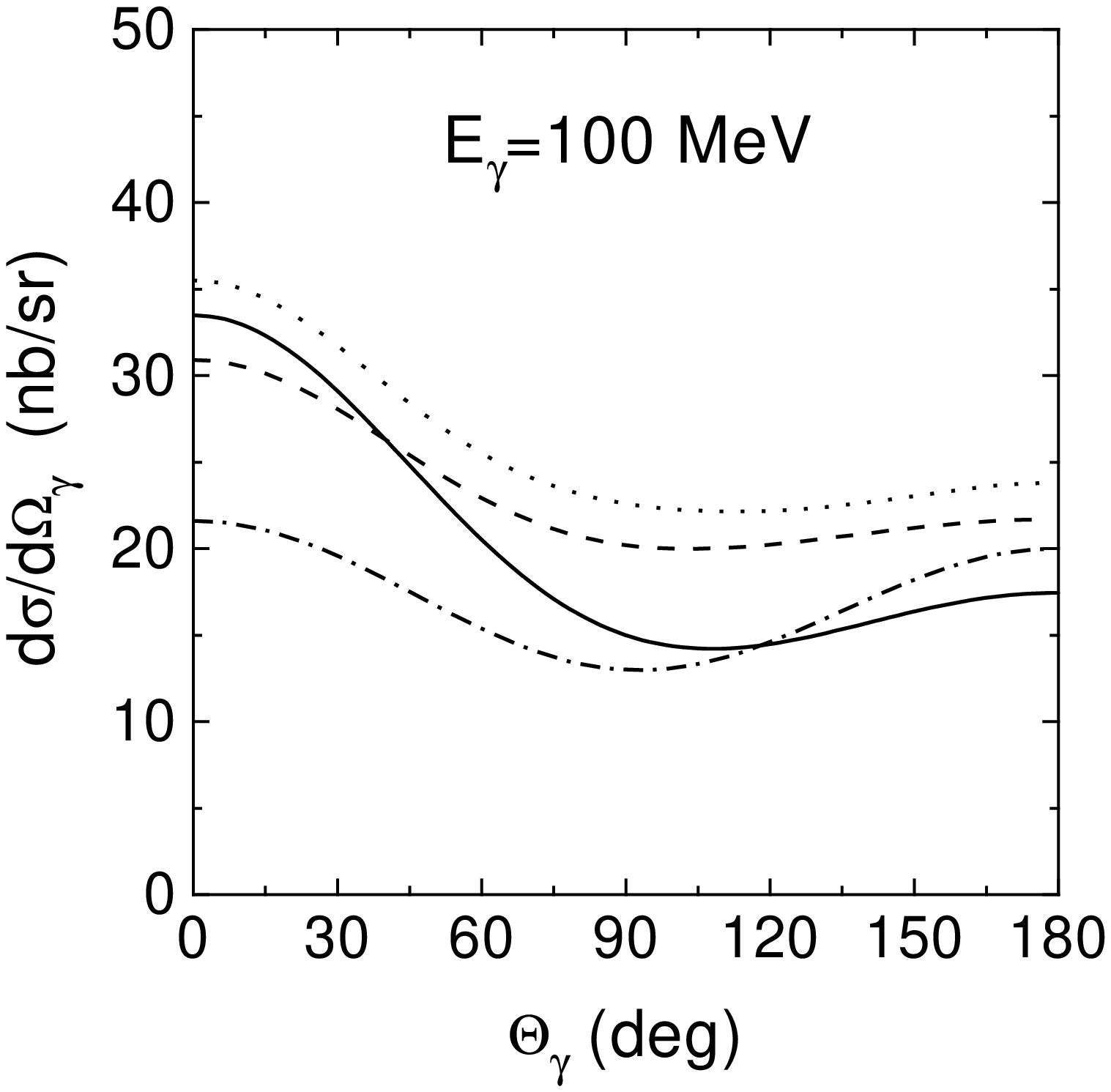} }
\caption{
Differential cross sections (CM) of $\gamma d$ scattering.
Dotted lines: Weyrauch and Arenh\"ovel \protect\cite{weyr83}.
Dashed lines: Weyrauch \protect\cite{weyr90}.
Dash-dotted lines: Wilbois \etal\ \protect\cite{wilb95}.
Solid lines: the present work.
Nucleon polarizabilities are turned off.
Data are from Ref.\ \protect\cite{luca94}.}
\label{fig:comparison1}
\end{figure}

A comparison of the present predictions with the results of other
calculations of 80's -- mid 90's \cite{weyr83,weyr90,wilb95} is shown
in Fig.\ \ref{fig:comparison1}.  There is a reasonable agreement
between all the predictions at low energies like 50 MeV, maybe with the
except for those of Ref.\ \cite{weyr83}.  When the energy increases up
to 100 MeV, we predict a bigger angular variation of $d\sigma/d\Omega$
that other works do.  In fact, the discrepancy between all the results
of different authors is dramatically large at 100 MeV.  Moreover, since
the spin-orbit interaction was not taken into account in Refs.\
\cite{weyr83,weyr90,wilb95} and since the spin-orbit effects decrease
the forward cross section by 15\% at 100 MeV, the genuine disagreement
of other components of the scattering amplitude with those of Ref.\
\cite{wilb95} is even more serious than Fig.\ \ref{fig:comparison1}
suggests.

\begin{figure}[hbt]
\centerline{
\leavevmode\epsfxsize=0.33\textwidth\epsfbox{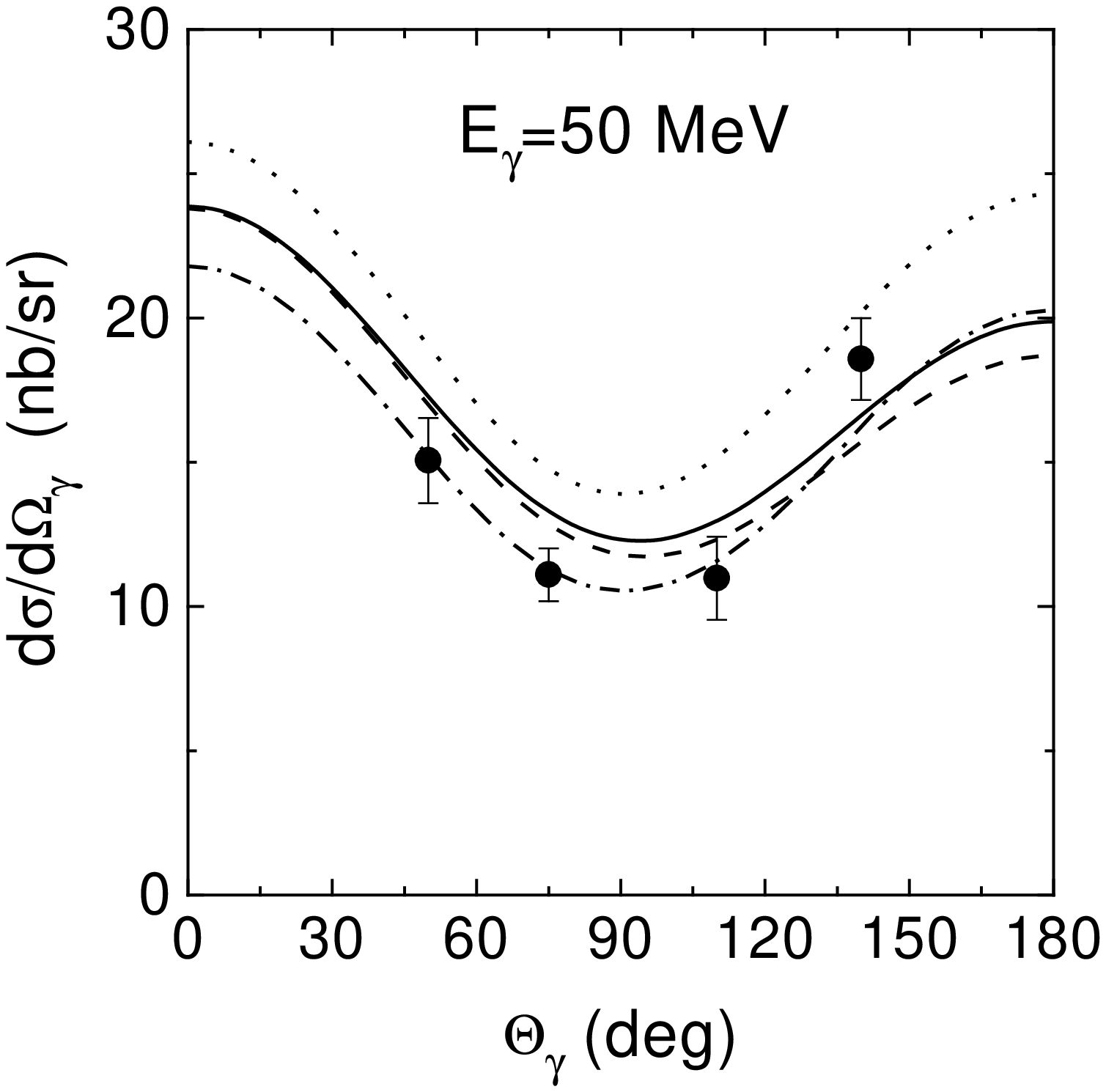}
\leavevmode\epsfxsize=0.33\textwidth\epsfbox{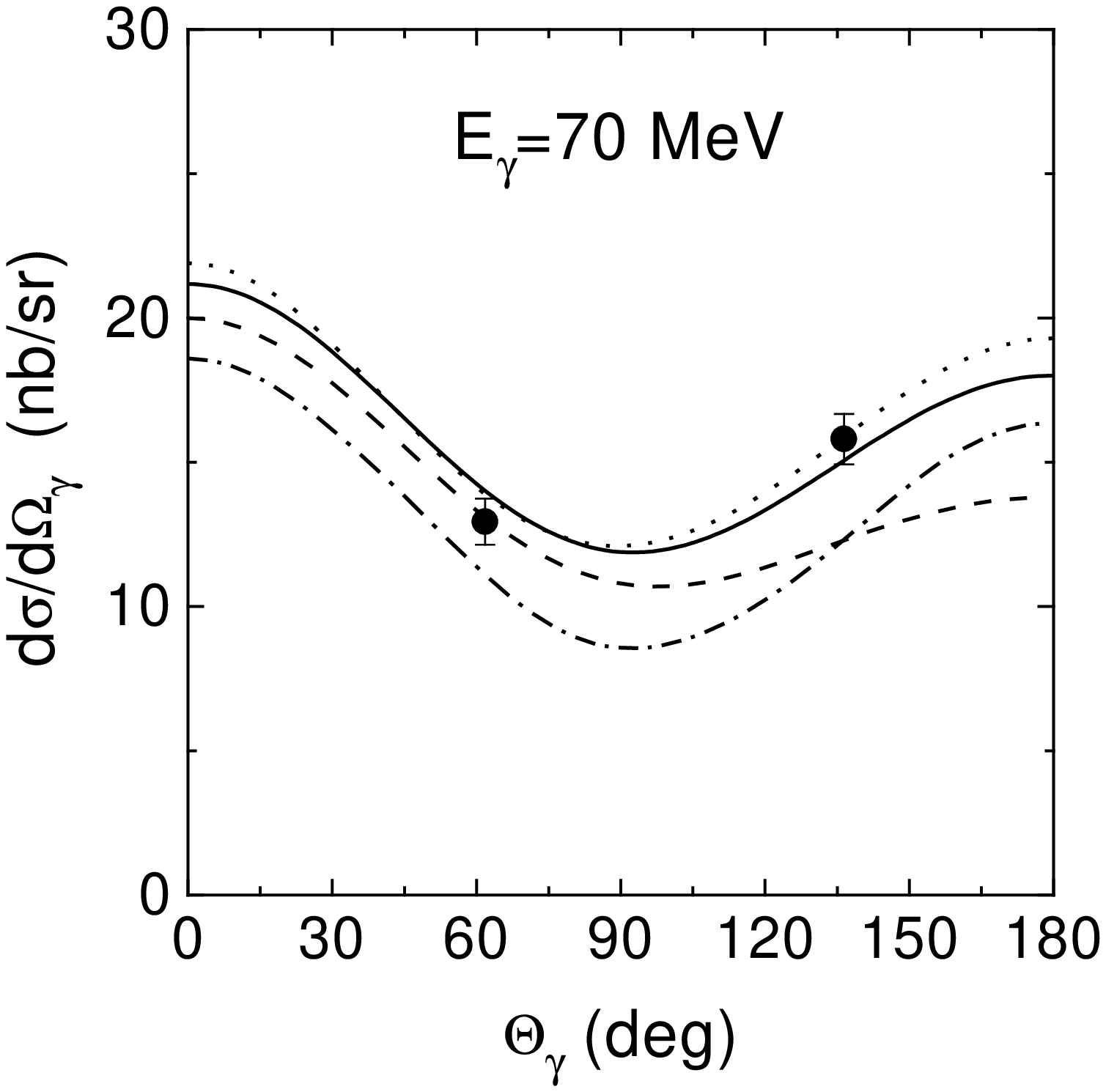}
\leavevmode\epsfxsize=0.33\textwidth\epsfbox{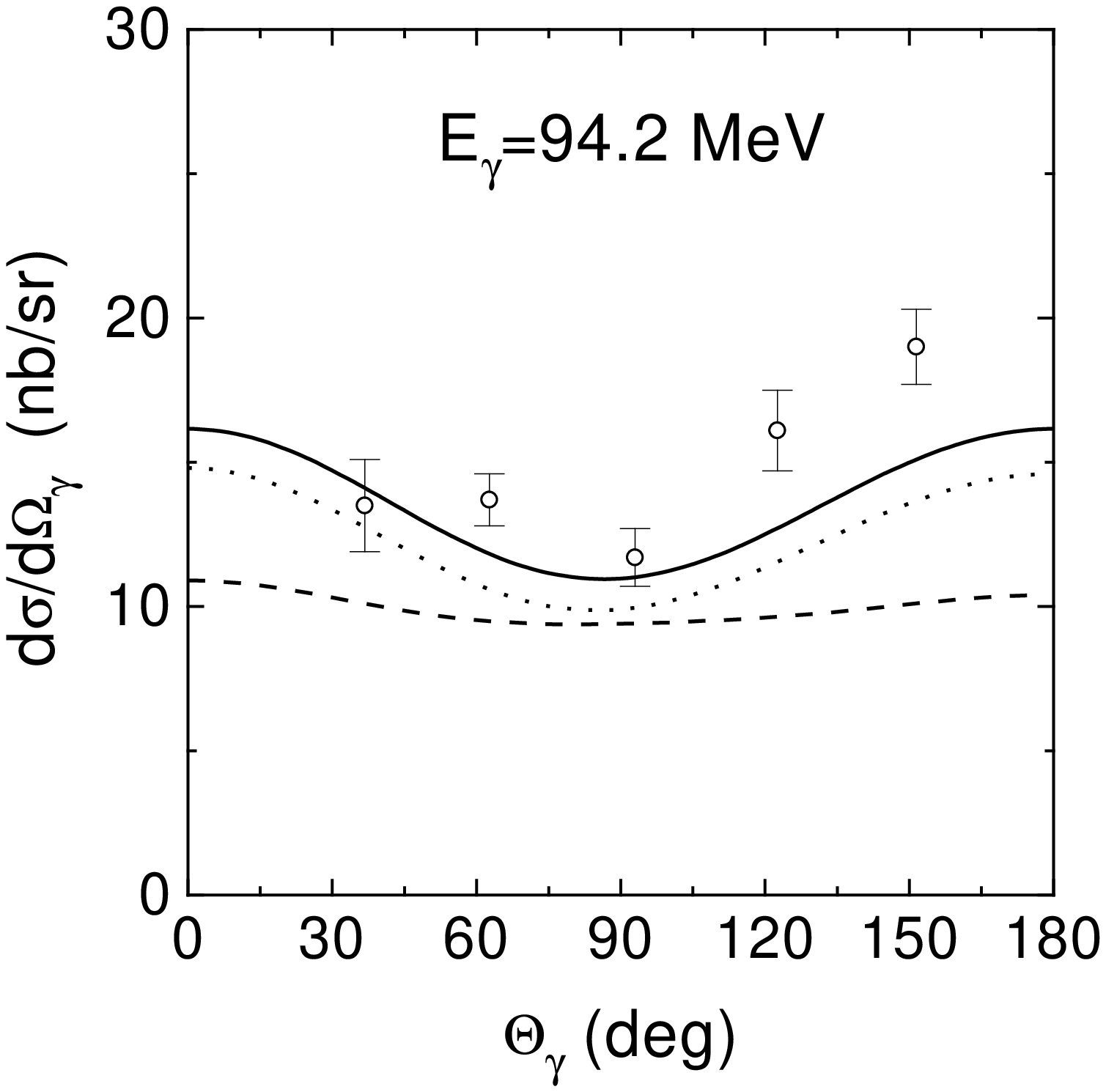} }
\caption{
Differential cross sections (CM) of $\gamma d$ scattering.
Dashed lines: Karakowski and Miller \protect\cite{kara99}.
Dotted lines: Beane \etal\ \protect\cite{bean99}.
Dash-dotted lines: Chen \etal\ \protect\cite{chen98}.
Solid lines: the present work.
Nucleon polarizabilities are included, and
$\bar\alpha_N - \bar\beta_N =9$ is used for drawing the solid curves.
Data are from Ref.\ \protect\cite{luca94} (solid circles) and
Ref.\ \protect\cite{horn99} (open circles).}
\label{fig:comparison2}
\end{figure}

It is interesting that the very low differential cross section obtained
by Wilbois \etal\ \cite{wilb95} at 100 MeV has found a support from a
recent work of Karakowski and Miller (KM) \cite{kara99}, see Fig.\
\ref{fig:comparison2}.  Below we mention a couple of possible reasons
why the KM model disagrees with our model and shows a wrong behavior at
high energies $E_\gamma \sim 100$ MeV.%
\footnote{We are indebted to Prof.\ G.A. Miller for useful comments
made on this point.}
First, we have rather different effects of the spin-orbit interaction.
In Ref.\ \cite{kara99}, this interaction was taken into account partly,
i.e.\ only through the electromagnetic seagull, not through the
electromagnetic current. However, it gave a much bigger decrease in the
differential cross section at forward angles than that we found (cf.\
Fig.~12 in Ref.\ \cite{kara99}).  The second reason, perhaps less
important numerically, might be that there is a mismatch between the
electromagnetic and strong-interaction parts of the Hamiltonian $H_{\rm
KM}$ of the KM model which destroys the gauge invariance and actually
signals that some electromagnetic charges or currents in the system are
missing in the theoretical formalism.  The mentioned mismatch is that
the electromagnetic two-body part of $H_{\rm KM}$ includes only the
point-like pion-exchange piece, whereas the wave function of the
deuteron is constructed using a more complicated (and more realistic)
Bonn potential.%
\footnote{Actually, the procedure of Ref.\ \cite{kara99} is even more
complicated in this respect, because one more potential (Reid93) is
used in another part of their computation, when the rescattering
correction is found.  To be honest, we have to say that we also do a
similar mixture in order to evaluate $R^{\rm rescat}$ (see
Section~\ref{sec:computation}).}
The violation of the gauge invariance in the KM model did not lead to
visible problems at low photon momenta $kr \ll 1$ owing to the use of
the Siegert transformation which ensured automatically the fulfillment
of the low-energy theorem (\ref{LET}).  However, when $kr$ becomes
large (this is the case for energies $E_\gamma \sim 100$ MeV), the
Siegert transformation does not help, and the missing charges and/or
currents can become important.

There is a simple way to verify the theoretical calculations
in the particular case of $\Theta_\gamma=0$ and to see that
the predictions of Ref.\ \cite{kara99} at high energies are invalid.
Using the Gell-Mann--Goldberger--Thirring dispersion relation
jointly with the optical theorem (cf.\ Eq.\ (\ref{sigma-tot-BP}))
for the spin-averaged amplitude $T_{\rm s.a.}$, we write
\beq
\label{DR}
     \Re T_{\rm s.a.}(\omega,0)  = - \frac{Z^2 e^2}{AM}
     + \frac{2\omega^2}{\pi}\, {\rm P} \int_0^\infty
      \frac{\sigma_{\rm tot}(\omega')}
       {{\omega'}^2 - \omega^2} \,d\omega',
\eeq
where $\sigma_{\rm tot}$ is the total photoabsorption cross section.
Keeping in mind that the total cross section of meson production off
the deuteron is dominated by meson production off quasi-free nucleons,
we see that this part of the photoabsorption cross section is
responsible for the component of the $\gamma d$ scattering amplitude
which is related with the polarizabilities of free nucleons (up to
relatively small effects due to medium modifications of these
polarizabilities).  Therefore, subtracting the meson-production part of
the photoabsorption cross section and keeping in Eq.\ (\ref{DR}) {\em
non-mesonic}, or photodisintegration part of the cross section, we can
approximately identify the resulting r.h.s.\ of Eq.\ (\ref{DR}) with
the $\gamma d$ scattering amplitude, in which the internal (mesonic)
structure of the nucleon is disregarded.  In other words, using the
deuteron photodisintegration cross section
$\sigma^{\gamma d\to pn}_{\rm tot}$ instead of $\sigma_{\rm tot}$ in
Eq.\ (\ref{DR}), we should obtain the $\gamma d$ scattering amplitude
with point-like nucleons having zero polarizability.  See Ref.\
\cite{huet00} for a more detailed discussion of these steps.

\begin{figure}[hbt]
\centerline{\epsfxsize=0.4\textwidth \epsfbox{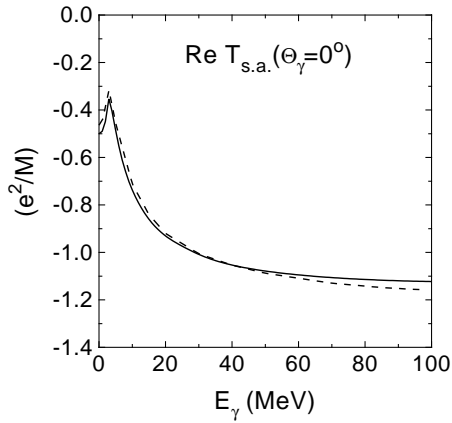}}
\caption{
The spin-averaged amplitude of forward $\gamma d$-scattering.
Solid line: the r.h.s.\ of Eq.\ (\ref{DR}).
Dashed line: results of the present work based on the Bonn potential.
Meson photoproduction and nucleon polarizabilities are disregarded.}
\label{fig:DR}
\end{figure}

We have evaluated the integral in Eq.\ (\ref{DR}) at energies $\omega
\alt 100$ MeV using:  (i) the effective-range parameterization of
$\sigma^{\gamma d\to pn}_{\rm tot}(\omega')$ at energies
$\omega'$ below 20 MeV (see Ref.\ \cite{aren91}, Eq.\ (2.18))
which gives an accurate description of
experimental data at low energies, and (ii) a phenomenological fit
\cite{ross89} to available experimental data between 20 and 440 MeV. At
higher $\omega'$, the photodisintegration cross section is
small and can be safely neglected in the integral.  The result of such
an evaluation of Eq.\ (\ref{DR}) is shown in Fig.\ (\ref{fig:DR}) by
the solid curve, together with our predictions (dashed line) based on
the Bonn-potential picture, in which the nucleon polarizabilities are
disregarded.  Generally, we find very good agreement between the two
curves.  Some disagreement of about 6\% at very low energies appears
due to an approximate way of finding the rescattering amplitude $R^{\rm
rescat}$, as it was already mentioned in Section~\ref{sec:computation}.
At energies above 10 MeV, the rescattering amplitude is less important,
and the agreement between the two calculations improves.  It is better
that 3\% even at 100 MeV.  It is needless to say that the
Bonn-potential picture nicely reproduces the experimental data on the
total cross section $\sigma^{\gamma d\to pn}_{\rm tot}$ at all energies
below pion threshold as well as the differential cross section of
deuteron photodisintegration and polarization observables
\cite{levc95a}.

Now let us compare our predictions with those obtained within two
different flavors of effective field theory (EFT) for few-nucleon
systems \cite{chen98,bean99} (a general review of the EFT approach to
nuclear problems can be found in Ref.\ \cite{vank99}).  At ``high"
energies $E_\gamma \sim 70{-}100$ MeV, we have a qualitative agreement
with the results of Beane \etal\ \cite{bean99}, who used the so-called
Weinberg formulation of the nuclear EFT.  See
Fig.~\ref{fig:comparison2} for a comparison.  In part, a proximity of
our and Beane \etal\ predictions is caused by their use of a realistic
wave function to evaluate matrix elements of the ChPT kernel, the
latter denoting the amplitude of $\gamma NN \to \gamma NN$ taken to
order $\O(Q^3)$ in the chiral perturbation theory expansion.  It was
actually the wave function of the Bonn potential.%
\footnote{The optional use of the Nijmegen-potential wave function (see
Fig.~15 in Ref.\ \cite{bean99}) leads to bigger deviations from the
dispersion predictions and from the Urbana data \cite{luca94}.  It was
claimed in Ref.\ \cite{bean99} that any wave function of the
deuteron with the correct binding energy, thus including for instance
$\Psi(\vec p)$ given by Eq.\ (\ref{Psi0}), could also be used
within the considered order $\O(Q^3)$ of the power expansion.
Being valid theoretically, this makes, however, a big numerical difference!}
Furthermore, the dominating part of the two-body seagull operator is
the same in our two approaches. It comes from the retarded pion
exchange.  Despite the Bonn-potential picture uses form factors in the
$\pi NN$ vertices and additional heavier mesons (to make improvements
to the potential at small distances), the effects of the form factor
and the heavy mesons onto the seagull amplitude are reduced to some
changes of $\kappa$ which almost cancel each other.  Specifically, we
have found $\kappa = 0.48 - 0.04 + 0.03$, where the three numbers are
the contribution of the pure pion exchange, the contribution from the
fictitious $\Lambda$ particle simulating the $\pi NN$ form factor (see
Section~\ref{sec:MES}), and the contribution of $\rho$, $\omega$,
$\delta$ and $\sigma$ mesons, respectively.

We can notice that {\em all} the contributions counted in Ref.\
\cite{bean99} have been taken into account in our calculation as well.
Beyond that, we included other corrections which formally belong to
higher orders in $Q$ in the power counting scheme of the EFT but are
large numerically.  An instructive example is the amplitude $R^{\rm
no\, rescat}$ given by Eq.\ (\ref{R1}), in which each of the amplitudes
of the $\gamma d \to pn$ transitions is dominated, at high energies, by
the pion-exchange current $\vec j^{[2]\,\pi}$, as shown in
Fig.~\ref{fig:T-gd}.  This rather large contribution formally appears
only in order $\O(Q^5)$ of the power counting scheme of Ref.\
\cite{bean99}.

A resemblance of the two predictions seems to be lost at lower
energies, where the scattering amplitude of Beane \etal\ begins to
deviate from the correct value fixed, for instance, by the dispersion
relation (\ref{DR}) at the forward angle.  Such a failure is not a
surpise and it was anticipated in Ref.\ \cite{bean99} as a result of
contributions from the $NN$ intermediate states which break down the
used power counting at low energies.

It is worth mentioning that the evaluation of matrix elements of the
ChPT kernel between {\em phenomenological} (Bonn or Nijmegen) wave
functions, which are not consistent with the one-pion-exchange dynamics
of the $NN$ interaction incorporated into the ChPT kernel,
automatically means the absence of the gauge invariance in the scheme
of Ref.\ \cite{bean99}.  This itself is a sufficient reason for a
failure of such a theory at very low energies where the gauge
invariance is crucial.

In view of close magnitudes of $\kappa$ arising in our approach and in
that of Ref.\ \cite{bean99}, we can {\em conjecture} that the main
difference between the two predictions at energies of about 100 MeV is
related with our taking into account the $\Delta$-resonance excitation
(both in one-body and two-body operators) and with our taking into
account the contribution $R^{\rm no\, rescat}$ (of the correct
magnitude) and the contribution $R^{\rm rescat}$.

The problem with the gauge invariance and with the region of very low
energies does not exist in the version of EFT used
by Chen \etal\ \cite{chen98}.  Their work is based on the so-called
Kaplan--Savage--Wise (KSW) regularization which successfully resolves
the problem of a poor power-series convergence in the case of large
$s$-wave $NN$-scattering lengths \cite{kapl98}.  $NN$ rescattering
contributions are accurately taken into account in that approach.
However, even being quite accurate at low energies, this approach
becomes inapplicable when the momenta of nucleons in the rescattering
diagrams exceed the range of convergence of power series which is about
$\Lambda_{NN} = 16\pi M/g_{\pi NN}^2 \simeq 300$ MeV.  This makes
predictions of Ref.\ \cite{chen98} not well controlled at energies
$\agt 70{-}90$ MeV.  Therefore, there is no surprise that these
predictions at 70 MeV lie visibly lower than our predictions
(and those of Beane \etal\ \cite{bean99}), including
the $\Theta_\gamma=0$ point, where the dispersion relation (\ref{DR})
strongly favors our calculation.  It is worth mentioning that neither
$\Delta$-isobar excitation nor the spin-orbit current and seagull are
taken into account in Ref.\ \cite{chen98} since these pieces appear
only in higher orders of the used expansion.

\subsection{Determination of the nucleon polarizabilities}

\begin{figure}[hbt]
\centerline{
\leavevmode\epsfxsize=0.33\textwidth \epsfbox{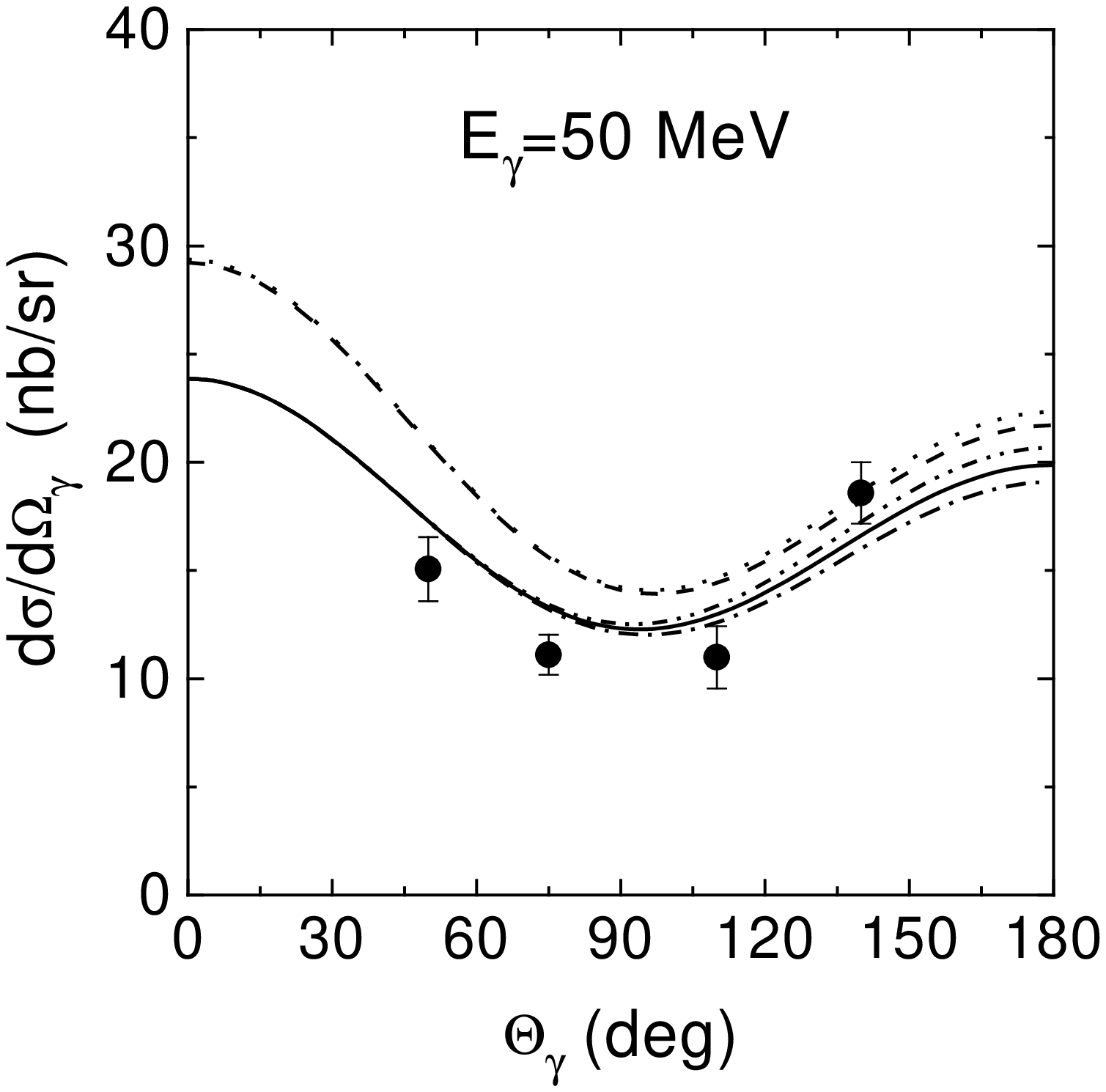}
\leavevmode\epsfxsize=0.33\textwidth \epsfbox{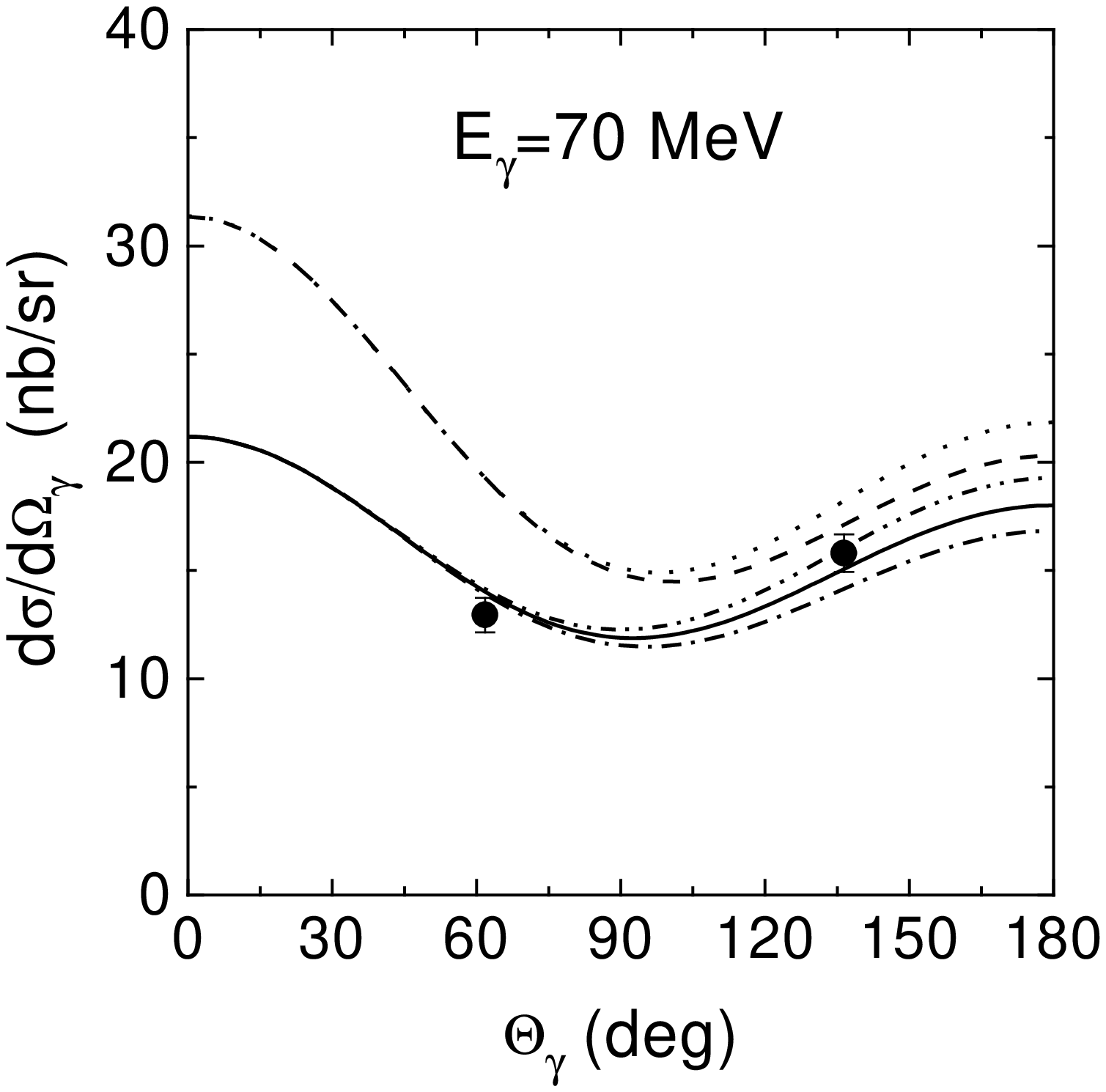}
\leavevmode\epsfxsize=0.33\textwidth \epsfbox{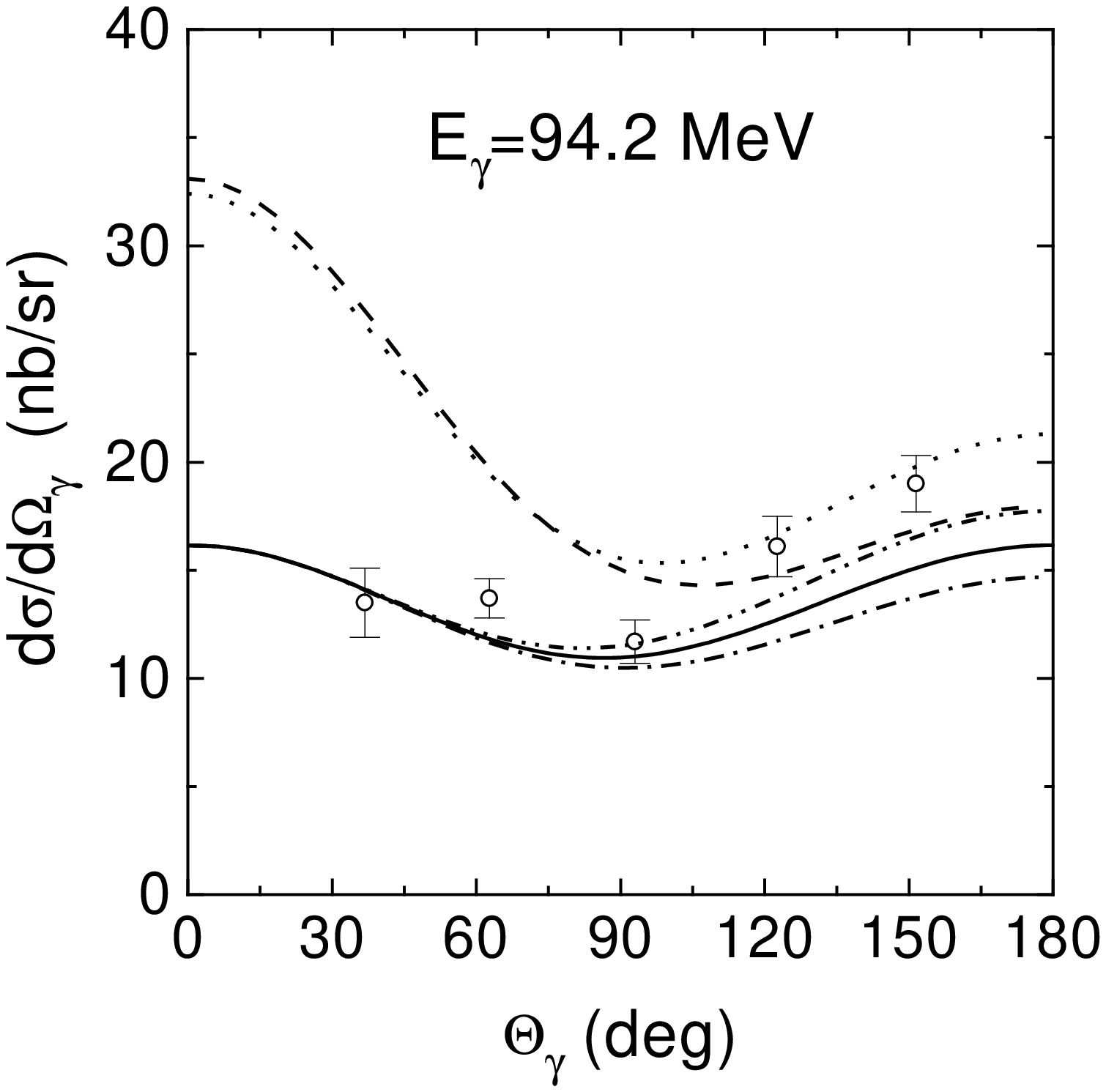} }
\caption{
Dependence of the differential cross section (CM) of $\gamma d$
scattering on the nucleon-averaged dipole polarizabilities
$\bar\alpha_N$ and $\bar\beta_N$.
Dashed lines: all the polarizabilities
(including those of higher order) are turned off.
Dotted lines: only higher-order polarizabilities,
Eq.\ (\protect\ref{H1add}), are included.
Dashed-double-dotted, solid, and dashed-dotted lines:
$\bar\alpha_N - \bar\beta_N = 6$, 9 and 12, respectively,
$\bar\alpha_N + \bar\beta_N = 14.6$ is fixed, and
the higher-order polarizabilities are included.
Data are from Ref.\ \protect\cite{luca94} (solid circles) and
Ref.\ \protect\cite{horn99} (open circles).}
\label{fig:s-polar}
\end{figure}

Considering the nucleon dipole polarizabilities $\bar\alpha_N$ and
$\bar\beta_N$ in the electromagnetic seagull operator (\ref{S1}) as
free parameters, we can check the sensitivity of the differential cross
section $d\sigma/d\Omega$ with respect to variation of these
parameters.  Our results are shown in Fig.\ \ref{fig:s-polar} together
with a few experimental data available from Urbana ($E_\gamma = 49$ and
69 MeV) \cite{luca94} and Saskatoon ($E_\gamma=94$ MeV) \cite{horn99}.
We do not show how the differential cross section depends on
the sum of the electric and magnetic polarizability, because
this sum is reasonably-well fixed by the Baldin sum rule, Eq.\
(\ref{a+b}).  As for the difference of $\bar\alpha_N$ and $\bar\beta_N$
which is not well-known theoretically, it can be determined from data
at large scattering angle.

Of course, the highest sensitivity is observed at the highest photon
energy. Nevertheless, we believe that data at medium energies
like 70 MeV are also quite useful, because theoretical uncertainties in
our computation related, for example, with omitted
relativistic corrections or with omitted dispersion effects
due to two-pion exchanges are expected to be
smaller at lower energies.

Dotted lines in Fig.\ \ref{fig:s-polar} illustrate the fact
that the higher-order polarizabilities (\ref{H1add})
are in no way negligible
when the nucleon dipole polarizabilities are determined from
$\gamma d$ scattering at energies $\agt 70$ MeV.
This feature was paid attention to also in Ref.\ \cite{bean99}, in which
the higher-order contributions appeared as an intrinsic part
of the one-pion-loop diagrams of the ChPT kernel.
Since, however, we use the higher-order polarizabilities
which are given by dispersion relations \cite{babu98,drec98,lvov99,hols99}
and which are very different from those suggested
by the one-pion-loop mechanism
(in part, due to the $\Delta$-contribution, see Ref.\ \cite{babu98}),
we predict a much bigger effect at backward angles.

A straightforward two-parameter fit of the Urbana data \cite{luca94}
gives $\bar\alpha_N =14.5\pm 2.7$ and $\bar\beta_N= 6.6 \pm 2.7$,
whereas a similar fit of the Saskatoon data \cite{horn99} gives
a lower value of the electric polarizability:
$\bar\alpha_N = 8.4 \pm 1.8$ and $\bar\beta_N= 6.2 \pm 1.8$.
Making a combined fit of all the data, we obtain
\beq
\label{a+b:exp}
   \bar\alpha_N + \bar\beta_N = 17.1 \pm 1.6,
\eeq
what is in agreement with the theoretical expectation (\ref{a+b}), and
\beq
\label{a-b:exp}
   \bar\alpha_N - \bar\beta_N = 4.0 \pm 1.5,
\eeq
though with a poor $\chi^2/N_{\rm d.o.f.} = 21/9$.  Systematic
uncertainties of the experimental data are included into the obtained
estimates (\ref{a+b:exp}) and (\ref{a-b:exp}).  However, it is not so
easy to estimate uncertainties introduced by the theoretical model.
Certainly, they are not less than the experimental uncertainties.

Taken as they are, these numbers, together with the experimental data
on the polarizabilities of the proton (\ref{alpha-p-exp}) can be considered
as an indication that the electric polarizability of the neutron is
$\bar\alpha_n = 9 \pm 3$, and the neutron magnetic polarizability is
$\bar\beta_n =  11 \pm 3$.
While the obtained sum $\bar\alpha_n + \bar\beta_n = 20 \pm 3$
reasonably agrees with the theoretical estimate (\ref{(a+b)lvov-n}),
the obtained difference $\bar\alpha_n - \bar\beta_n = -2 \pm 3$ is
rather far from both the similar difference in the proton case
found experimentally,
$\bar\alpha_p - \bar\beta_p = 10 \pm 2$ \cite{macg95,bara99}, and from
theoretical estimates based on dispersion relations which predict
roughly $\bar\alpha_n - \bar\beta_n \simeq \bar\alpha_p - \bar\beta_p$
(see, e.g., Refs.\ \cite{petr81,lvov93,hols94}).
It is clear that a further experimental and theoretical work is needed
to reduce the uncertainties. New data can appear from Lund \cite{lund99}.

\begin{figure}[hbt]
\centerline{
\leavevmode\epsfxsize=0.33\textwidth \epsfbox{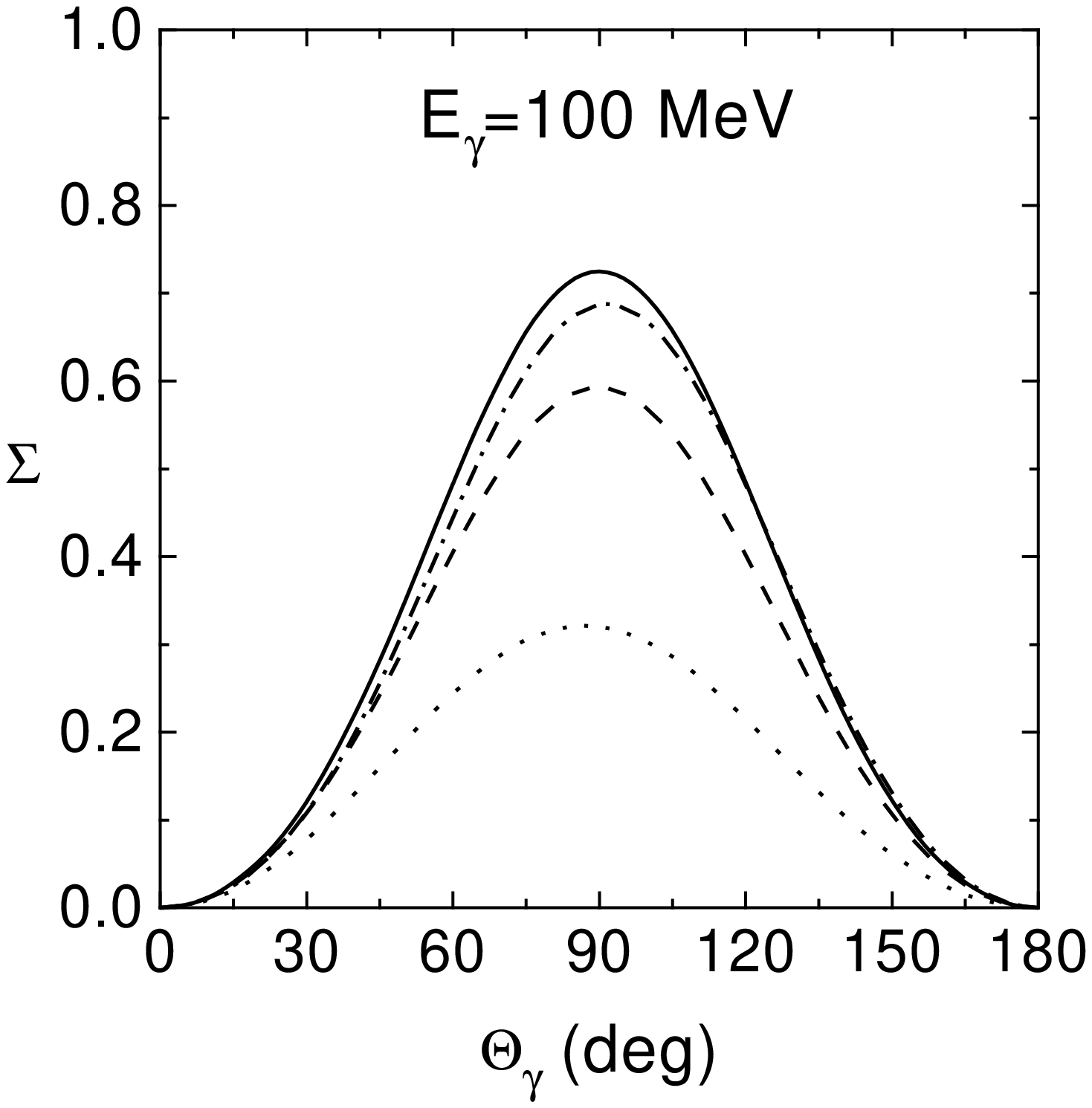}
\leavevmode\epsfxsize=0.33\textwidth \epsfbox{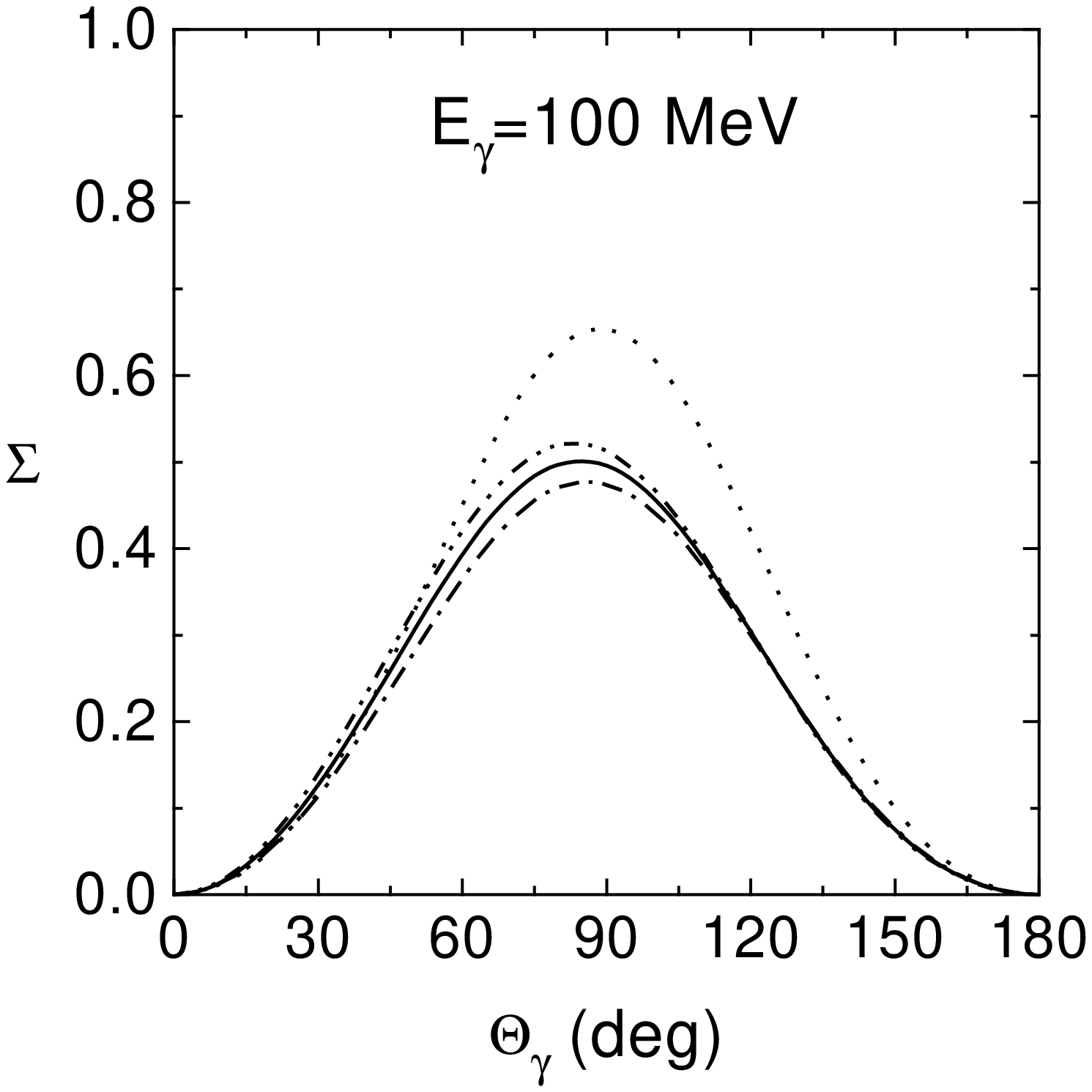} }
\caption{
Left panel: Contributions of different parts of the Compton scattering
amplitude to the beam asymmetry at 100 MeV.
Notation of curves is as in Fig.~\protect\ref{fig:s-r-tot}.
Right panel: Dependence of the beam asymmetry
on the nucleon-averaged dipole polarizabilities
$\bar\alpha_N$ and $\bar\beta_N$.
Notation of curves is as in Fig.~\protect\ref{fig:s-polar}. }
\label{fig:a-polar}
\end{figure}

Among other observables of $\gamma d$ scattering
which are sensitive to the nucleon polarizabilities too,
we briefly discuss the beam asymmetry $\Sigma$, Eq.\ (\ref{Sigma}).
In Fig.\ \ref{fig:a-polar} we show how different components of the
Compton scattering amplitude affect $\Sigma$
(this is helpful for imagining a possible scale of model uncertainties)
and how $\Sigma$ is sensitive to the nucleon polarizabilities.
One can notice a strong dominance of the one-body seagull amplitude,
whereas the role of the two-body seagull contribution is smaller than
that in the case of the differential cross section.
The role of $NN$ rescattering is again small.

The spin-orbit interaction essentially affects $\Sigma$ and gives a
10\% increase at central angles.  It mainly comes through the one-body
seagull amplitude $S^{[1]}$.
The contribution of the $\Delta$ excitation to the two-body seagull
amplitude $S^{[2]\pi}$ changes the beam asymmetry by less than 1\%,
but the $\Delta$ contribution into the resonance amplitude $R$
is rather visible, reducing $\Sigma$ by 6\% at 100 MeV.
Pion-retardation effects have only a tiny impact on $\Sigma$.
The use of the potential OBEPR(B) instead of OBEPR has a big effect
and reduces $\Sigma$ by 14\% at 100 MeV.

It looks like experiments with the linearly-polarized photon beam
can also be useful for measuring the nucleon polarizabilities,
provided the accuracy of measurements is better than $\sim 5{-}10\%$.

We conclude saying that the reaction of deuteron Compton scattering
at energies of about 50--100 MeV has a great potential
for a determination of the electromagnetic polarizabilities of
the neutron. Currently, the available theoretical models
show a big divergence in their results, in part because they
do not take into account all important contributions.
So, a further theoretical work is needed to improve the accuracy
of the models before any firm conclusions
could be inferred about the values of $\bar\alpha_N$ and $\bar\beta_N$.
Better experimental data are also needed to this aim.

\acknowledgments
We would like to thank N.V.~Maksimenko, M.~Schumacher and T.~Wilbois
for useful discussions, and D.~Hornidge and N.~Kolb for supplying us with
the new experimental data.  This work was supported by Advance Research
Foundation of Belarus and by the
Russian Foundation for Basic Research, grant No.\ 98-02-16534.

\newpage
\appendix


\section{Electromagnetic seagulls from heavy mesons of the OBE potential}
\label{sec:MES-other}

In this appendix we give explicit formulas for the seagulls
$S_{ij}^\alpha$ produced by the meson exchanges $\alpha=\eta$,
$\sigma$, $\delta$, $\omega$ and $\rho$ of the Bonn potential (OBEPR).
All of them are obtained through a direct evaluation of the diagrams
shown in Fig.\ \ref{fig:V-j-S}{\it c}.  The electromagnetic effective
meson-nucleon vertices $\gamma\alpha NN$ and $\gamma\gamma\alpha NN$ in
these diagrams arise from the relativistic boson-nucleon effective
Lagrangian of Refs.\ \cite{mach87,mach89}, in which a nonrelativistic
reduction is done and terms up to order $\O(M^{-2})$ are only retained.
Such a procedure is consistent with the whole construction of the
OBEPR, because this potential itself is built through the truncation of
the relativistic Feynman diagrams of the one-boson exchanges to order
$\O(M^{-2})$ (see  Ref.\ \cite{mach89}, Appendix A.3).

\begin{figure}[htb]
\epsfxsize=0.8\textwidth
\centerline{\epsfbox[35 665 480 800]{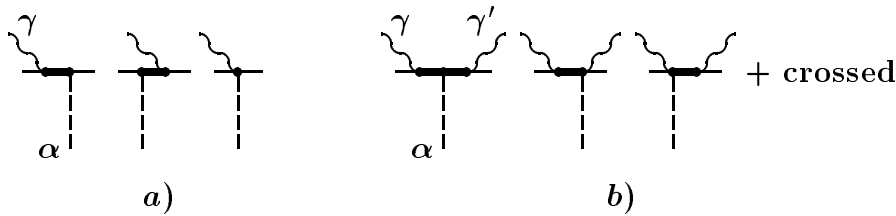}}
\caption{Effective contact meson-nucleon vertices
$\gamma\alpha NN$ (diagrams {\it a}) and $\gamma\gamma\alpha NN$
(diagrams {\it b}) arising from the antinucleon degrees of freedom.
Thick lines denote the negative-energy part $P_-$ of the nucleon
propagator, Eq.\ (\protect\ref{P-}).}
\label{fig:anti-N}
\end{figure}

Technically, the nonrelativistic reduction can be conveniently
performed \cite{adam89} by considering appropriate relativistic Feynman
diagrams (see Fig.\ \ref{fig:anti-N}) and keeping only the
negative-energy part $P_-$ of the nucleon propagators. To leading order
in $1/M$, it is sufficient to take $P_-$ in the static limit, i.e.\
\beq
\label{P-}
   P_- = -\frac{1}{2M} \pmatrix{ 0 & 0 \cr 0 & 1}.
\eeq
It is worth noticing that the contact vertex $\gamma\alpha NN$ in Fig.\
\ref{fig:anti-N} appears only in the case of $\alpha=\rho$, being
caused by the tensor coupling of the charged $\rho$-meson to the
nucleon. In the formalism of the pseudo-scalar $\pi NN$ coupling used
in Refs.\ \cite{mach87,mach89}, the contact vertex $\gamma\pi NN$ is
absent.

We use the same notation as in Section~\ref{sec:MES}.  In particular,
the momenta $\vec q_1$, $\vec q_2$, $\vec K_1$ and $\vec K_2$ are
defined by Eqs.\ (\ref{q_i}) and (\ref{K_i}).
We introduce also the vectors
\beq
   \vec P_1 = \vec p_1 + \vec p'_1, \quad
   \vec P_2 = \vec p_2 + \vec p'_2.
\eeq
The functions $G_\alpha(\vec q)$, $G_{1\alpha}(\vec q_1,\vec q_2)$,
$G_{2\alpha}(\vec q_1,\vec q_2,\vec K)$, and
$D_\alpha(\vec q_1,\vec q_2,\vec K_1,\vec K_2)$ used below are defined
as in Section~\ref{sec:MES} with the evident replacement
$\pi \to \alpha$ in all masses $m_\alpha$ and cutoff parameters
$\Lambda_\alpha$.  The constants $\kappa_\alpha$ denote the ratios of
the tensor and vector coupling constants for the $\omega NN$ and
$\rho NN$ vertices, $\kappa_\alpha = f_\alpha/g_\alpha$.

Evaluating the diagrams in Figs.\ \ref{fig:V-j-S}{\it c} and
\ref{fig:anti-N}, we obtain the following results.

\vspace{1ex}\noindent$\bullet$ Isoscalar exchanges:
\beq
      \epsilon_i^{\prime *} \epsilon_j S_{ij}^\eta
      (-\vec k',\vec k; \vec p'_1, \vec p'_2; \vec p_1, \vec p_2) = 0,
\eeq
\beq
      \epsilon_i^{\prime *} \epsilon_j S_{ij}^\sigma
      (-\vec k',\vec k; \vec p'_1, \vec p'_2; \vec p_1, \vec p_2) =
      -\e\cdot\E \, \frac{e^2 g_\sigma^2}{2M^2}
       \Big[ Z_1 G_\sigma(\vec q_2) + (1 \leftrightarrow 2) \Big],
\eeq
\beq
      \epsilon_i^{\prime *} \epsilon_j S_{ij}^\omega
      (-\vec k',\vec k; \vec p'_1, \vec p'_2; \vec p_1, \vec p_2) =
      -\e\cdot\E \, \frac{e^2 g_\omega^2}{2M^2}
       \Big[ Z_1 G_\omega(\vec q_2)
       - 2Z_1 Z_2 G_\omega(\vec K_1) + (1 \leftrightarrow 2) \Big].
\eeq
Writing the last equation, we have used that $\kappa_\omega=0$ for the
Bonn potential.

\vspace{1ex}\noindent$\bullet$ Isovector exchanges:
\beqn
\label{S-delta}
    &&   \epsilon_i^{\prime *} \epsilon_j S_{ij}^\delta
      (-\vec k',\vec k; \vec p'_1, \vec p'_2; \vec p_1, \vec p_2) =
      -\e\cdot\E \, \frac{e^2 g_\delta^2}{2M^2}
       \Big[ Z_1 \tau_2^z G_\delta(\vec q_2) + (1 \leftrightarrow 2) \Big]
\nn && \qquad {}
    + \frac{e^2 g_\delta^2}{4M^2}
       \Big\{ \Big( \vec q_2\cdot\E G_{1\delta}(\vec q_2,\vec K_1)
     \Big[ T_{12} (\vec q_1\cdot\e + i\vec\sigma_1\times\vec P_1\cdot\e)
\nn && \qquad\qquad {}
    + i(\vec\tau_1\times\vec\tau_2)^z \,
         (\vec P_1\cdot\e + i\vec\sigma_1\times\vec q_1\cdot\e) \Big]
      + (1 \leftrightarrow 2) \Big)
\nn && \qquad\qquad\qquad {}
   + (\e \leftrightarrow \E,~ \vec K_1 \leftrightarrow -\vec K_2) \Big\}
\nn && \qquad {}
   - e^2 g_\delta^2 \, T_{12} D_\delta(\vec q_1, \vec q_2, \vec K_1, \vec K_2)
    \Big[ 1 -
    \frac{\vec P_1^2 + \vec P_2^2 - \vec q_1^2 - \vec q_2^2} {16M^2}
\nn && \qquad\qquad\qquad\qquad {}
    - i \frac{\vec\sigma_1\times\vec q_1\cdot\vec P_1
         +    \vec\sigma_2\times\vec q_2\cdot\vec P_2} {8M^2} \Big],
\eeqn
\beqn
\label{S-rho}
    &&   \epsilon_i^{\prime *} \epsilon_j S_{ij}^\rho
      (-\vec k',\vec k; \vec p'_1, \vec p'_2; \vec p_1, \vec p_2) =
      -\e\cdot\E \, \frac{e^2 g_\rho^2}{2M^2}
       \Big[ (Z_1 \tau_2^z - \kappa_\rho T_{12}) G_\rho(\vec q_2)
            + (1 \leftrightarrow 2) \Big]
\nn && \qquad {}
      + \frac{e^2 g_\rho^2}{4M^2} \Big\{ G_\rho(\vec K_1)
         \Big[ \e\cdot\E (T_{12} - 4Z_1 Z_2)
      + (1+\kappa_\rho)^2 \, T_{12} \,
          \vec\sigma_1\times\e \cdot \vec\sigma_2\times\E
\nn && \qquad\qquad\qquad {}
      - (1+\kappa_\rho) (\vec\tau_1\times\vec\tau_2)^z \,
          (\vec\sigma_1 + \vec\sigma_2)\cdot\E\times\e \Big]
            + (1 \leftrightarrow 2) \Big\}
\nn && \qquad {}
    + \frac{e^2 g_\rho^2}{4M^2}
       \Bigg[ \Big( \vec q_2\cdot\E G_{1\rho}(\vec q_2,\vec K_1)
     \Big\{ T_{12} \Big[ (1+4\kappa_\rho)\vec q_1\cdot\e
        + i(1+2\kappa_\rho)\vec\sigma_1\times\vec P_1\cdot\e
\nn && \qquad\qquad\qquad\qquad {}
        - 2i(1+\kappa_\rho)\vec\sigma_1\times\vec P_2\cdot\e
        - 2(1+\kappa_\rho)^2 \,  \vec\sigma_1\times\e \cdot
                             \vec\sigma_2\times\vec q_2 \Big]
\nn && \qquad\qquad {}
    + i(\vec\tau_1\times\vec\tau_2)^z \,
         \Big[ (\vec P_1 - 2\vec P_2)\cdot\e
    + i(1+2\kappa_\rho)\vec\sigma_1\times\vec q_1\cdot\e
    - 2i(1+\kappa_\rho)\vec\sigma_2\times\vec q_2\cdot\e \Big] \Big\}
\nn && \qquad\qquad\qquad\qquad {}
    + (1 \leftrightarrow 2) \Big)
    + (\e \leftrightarrow \E,~ \vec K_1 \leftrightarrow -\vec K_2) \Bigg]
\nn && \qquad {}
   + e^2 g_\delta^2 \, T_{12} D_\rho(\vec q_1, \vec q_2, \vec K_1, \vec K_2)
    \Bigg[ 1 +
    \frac{\vec P_1^2 + \vec P_2^2 - 4\vec P_1\cdot\vec P_2} {16M^2}
   + (1 -4\kappa_\rho) \frac{\vec q_1^2 + \vec q_2^2} {16M^2}
\nn && \qquad\qquad\qquad {}
    + i(1 + 2\kappa_\rho)\,  \frac{\vec\sigma_1\times\vec q_1\cdot\vec P_1
         + \vec\sigma_2\times\vec q_2\cdot\vec P_2} {8M^2}
\nn && \qquad\qquad\qquad {}
    - i(1 + \kappa_\rho) \, \frac{\vec\sigma_1\times\vec q_1\cdot\vec P_2
         + \vec\sigma_2\times\vec q_2\cdot\vec P_1} {4M^2}
\nn && \qquad\qquad\qquad\qquad {}
    + (1 + \kappa_\rho)^2 \, \frac{\vec\sigma_1\times\vec q_1
        \cdot \vec\sigma_2\times\vec q_2} {4M^2} \Bigg].
\eeqn
Writing Eqs.\ (\ref{S-delta}) and (\ref{S-rho}), we used the radiation
gauge (\ref{gauge}).

With the help of Eqs.\ (\ref{G-G1}) and (\ref{G1-G2}), one can verify
that thus constructed seagull operators $S_{ij}^\alpha$ satisfy the
equation (\ref{S2-conservation-p}), provided the electromagnetic
currents $\vec j^\alpha$ are taken as obtained \cite{levc95a} from the
same boson exchanges to order $\O(M^{-2})$.

Numerical values of the masses $m_\alpha$, the couplings
$g_\alpha$ (as well as $\kappa_\alpha$ for the vector mesons), and the
cutoff parameters $\Lambda_\alpha$ for different bosons
are taken exactly the same as for the Bonn potential (OBEPR) itself
\cite{mach87,mach89}.  The only exception concerns the
$\sigma$-exchange.  The matter is that the Bonn parameterization
of the $\sigma$-exchange suggests to use a different mass $m_\sigma$ and
the coupling $g_\sigma$ for different $NN$ channels with the total
isospin $I=0$ or $I=1$. We found this feature inconvenient for building
electromagnetic operators which mix the isospin.  Since we noticed no
visible distinction between our predictions using $\sigma$-MEC and
$\sigma$-MES with either of the two sets of ($m_\sigma$, $g_\sigma$),
we took for the sake of simplicity the $\sigma$-meson parameters
proposed by the Bonn group for the $I=0$ channel.

As a final comment we have to mention that, strictly speaking, the
nonrelativistic reduction of the Feynman diagrams to order $\O(M^{-2})$
considered as a method of a determination of the OBE potentials
$V^\alpha$, the OBE electromagnetic currents $j_\mu^\alpha$, and the OBE
electromagnetic seagulls $S_{\mu\nu}^\alpha$ may need a further
refinement. The matter is that the operators obtained in this way are
manifestly frame-dependent.  Specifically, they explicitly depend on
the individual momenta of the nucleons, $\vec P_1$ and $\vec P_2$,
rather than on the relative variable $\vec P_1 - \vec P_2$  (see, e.g.,
Eqs.\ (\ref{S-delta}), (\ref{S-rho}) and also Ref.\ \cite{levc95a}).
This is not what is expected for potentials, MEC, and MES in the
nonrelativistic framework.

It is possible to propose a modification of the diagrammatic
representation of the boson exchanges which leads to Galilei-invariant
results and to some changes in the above equations for MEC and MES
\cite{levc00}.  We checked, however, that such a modification has only
a minor numerical effect and can be neglected in the present context.

\end{document}